\documentclass[longauth]{aa}   

\usepackage{graphicx}
\usepackage{txfonts}
\usepackage{lipsum}
\usepackage{subcaption}         
\usepackage{booktabs}
\usepackage{lscape}             
\usepackage{placeins}           
                                
\usepackage[colorlinks=true,linkcolor=bluea,allcolors=blue]{hyperref}

\newcommand{\kms}{km\,s$^{-1}$}
\newcommand{\cmm}{cm$^{-3}$}
\newcommand{\oiii}{[O\:{\sc iii}]}
\newcommand{\oiiib}{[O\:{\sc iii}]$4959\AA$}
\newcommand{\oiiir}{[O\:{\sc iii}]$5007\AA$}

\begin{document}
\title{The Quasar Feedback Survey: Ionised outflows in type 2 {QSOs} with MUSE data}
\titlerunning{QFeedS: Ionised outflows in type 2 QSOs}



   \author{M. Bianchin\inst{1,2,3}\thanks{\email{marina.bianchin@iac.es}}
   \and C.~M. Harrison\inst{4}
   \and T. Costa\inst{4}
   \and V. Mainieri\inst{5} 
   \and R.~A. Riffel\inst{3} 
   \and G. Venturi\inst{6,7}
   \and D. Kakkad\inst{8}
   \and C. Ramos Almeida\inst{1,2}
   \and P.~H. Cezar\inst{1,2}
   \and S. Ward \inst{9}
   \and A. Girdhar\inst{10}
   \and S. Molyneux\inst{11} 
   \and L. Ulivi\inst{12}
   \and E.~P. Farina\inst{13,14}
   \and J. Mullaney\inst{15}
   \and F. Arrigoni Battaia\inst{16}
}

   \institute{
    Instituto de Astrof\' isica de Canarias, Calle V\'ia L\'actea, s/n, E-38205, La Laguna, Tenerife, Spain
    \and Departamento de Astrof\'isica, Universidad de La Laguna, E-38206, La Laguna, Tenerife, Spain
    \and Universidade Federal de Santa Maria, Departamento de Física, Centro de Ciências Naturais e Exatas, 97105-900, Santa Maria, RS, Brazil
    \and School of Mathematics, Statistics and Physics, Newcastle University, NE1 7RU, UK
    \and European Southern Observatory, Karl-Schwarzschild-Strasse 2, Garching bei München, D-85748, Germany
    \and INAF – Osservatorio Astrofisico di Arcetri, Largo E. Fermi 5, I-50125, Firenze, Italy
    \and Scuola Normale Superiore, Piazza dei Cavalieri 7, I-56126 Pisa, Italy 
    \and Centre for Astrophysics Research, Department of Physics, Astronomy and Mathematics, University of Hertfordshire, Hatfield AL10 9AB, UK
    \and Center for Computational Astrophysics, Flatiron Institute, New York, NY 10010, USA
    \and Ludwig-Maximilians-Universit{\"a}t, Professor-Huber-Platz 2, D-80539 M{\"u}nchen, Germany
    \and Institute of Theoretical Astrophysics, University of Oslo, P.O. Box 1029, Blindern, 0315 Oslo, Norway
    \and Centro de Astrobiolog\'ia (CAB), CSIC--INTA, Cra. de Ajalvir Km.~4, 28850 -- Torrej\'on de Ardoz, Madrid, Spain
\and International Gemini Observatory/NSF NOIRLab, 670 N A’ohoku Place, Hilo, Hawai'i 96720, USA
\and INAF -- Osservatorio di Astrofisica e Scienza dello Spazio di Bologna, via Gobetti 93/3, I-40129, Bologna, Italy
\and Astrophysics Research Cluster, School of Mathematical and Physical Sciences, University of Sheffield, Sheffield, S3 7RH, UK
\and Max Planck Institut f\"{u}r Astrophysik, Karl Schwarzschild Stra{\ss}e 1, D-85741 Garching, Germany
}
\date{Received June 25, 2026}

  \abstract
{The impact of feedback from active supermassive black holes on their host galaxies is a key phenomenon in understanding galaxy evolution. We study ionised gas properties in a sample of 18 luminous (log L$_{\rm bol}\geq45$\:erg\:s$^{-1}$) obscured {Quasi-stellar objects (QSOs)} in the nearby Universe ($0.085<z<0.2$) observed with VLT/MUSE as part of the multi-wavelength Quasar Feedback Survey (QFeedS) and with VLA at low- and high-resolution data{, 1\arcsec and 0.3\arcsec respectively}. Our main goal is to characterise the disturbed gas kinematics in this galaxy sample. We compare the outflow-related quantities with values reported in the literature and investigate possible correlations between the outflows and other properties of the ionised gas, such as the electron density. We adopt a non-parametric approach to characterise the ionised gas kinematics, traced by the \oiiir~ emission line, and use $W_{80}$ measured from the unresolved nuclear spectra as a threshold to identify outflow-dominated spaxels. A qualitative comparison between the radio and the $W_{80}$ maps shows that, in general, the peak of the radio coincides with the peak of the $W_{80}$. \oiiir~ is used as a proxy for the ionised gas mass, and the outflow-related properties are obtained in two different ways: the global, using flux-weighted velocities and radii and the peak, which relies on radial profiles. We correct the \oiiir~ flux for extinction using the Balmer decrement. We take advantage of the MUSE sensitivity and use the [Ar\:{\sc iv}]4711,4740 doublet, only observable in the integrated $r=1.5\arcsec$ spectra, to measure the electron density in each galaxy.  We obtain $500<n_{\rm e}{\rm[ cm^{-3}]}<27700$, leading to 0.03$<\dot{M}_{\rm out}/{\rm [M_{\odot} yr^{-1}]}<$4.17 and $39<\dot{E}_{\rm kin}/{\rm [erg\:s^{-1}]}<42$.  Our results emphasize the importance of accurate electron density determination, leading to a difference of ~1 dex in the outflow rates and kinetic powers. We observe a positive correlation between $n_e$ and $V_{\rm out}$, interpreted as compression of the gas at higher velocities. This correlation should be further investigated at other luminosity and redshift ranges, but it can inform future outflow simulation studies.}

   \maketitle
   \nolinenumbers

\section{Introduction}
Accreting supermassive black holes (SMBH, $M_{\rm BH}\gtrsim10^6$\,M$_{\odot}$) are the powerhouses of active galactic nuclei (AGN). Diverse processes drive the gas towards the centre and trigger the AGN \citep[see ][for reviews on SMBH feeding]{sb19, alexander25}. Subsequently, the phenomena associated with the so-called AGN feedback can quench or even enhance star formation in the host galaxy, thus having a profound impact on the evolution of their host galaxies \citep[e.g.][] {dimatteo05,hopkins_elvis10,fabian12, kormendy13, king15, harrison24}.
The AGN-driven galactic-scale feedback is associated with accretion disc winds or jets of relativistic particles, and the energy released by the AGN during its lifetime can even be higher than the host galaxy binding energy \citep{begelman06}.

Observational studies reveal that turbulent motions of the ionised gas in the AGN narrow-line region (NLR) are often associated with radio emission \citep[e.g.][]{heckman81, mullaney13, zakamska14, riffel26}. This is also consistent with simulations showing that radio jets can disturb and increase the turbulence of the surrounding gas, producing outflows \citep{mukherjee18, tanner22, audibert23, mukherjee25, riffel26}. The AGN feedback modes, quasar/radiative and radio, do not seem to be two completely distinct phenomena \citep{wylezalek18} as radio emission associated with outflows and/or feedback is observed in radiatively efficient AGN \citep[e.g.][]{zakamska16,jarvis19, audibert25}.

Theoretical models predict that AGN-driven gas outflows can suppress the star formation of the host \citep{granato04, zubovas17,costa20} and affect the luminosity function of galaxies at the high-mass end \citep[e.g.][]{bower06}. Such ionised gas outflows have been observed not only in nearby ($z<0.4$) quasar hosts \citep{rupke11, greene12, mullaney13, harrison14, somalwar20, bruno21,trindade-falcao21, girdhar22, ulivi24, speranza22, speranza24, bessiere24, kakkad26} but also in quasars at cosmic noon and beyond ($z>2$) \citep{kakkad16,kakkad20, vayner24, venturi26, bertola25}. However, the necessary kinetic coupling efficiency, a key ingredient in theoretical predictions \citep{hopkins_elvis10}, remains an open question in the literature \citep{harrison18, ward24}. 

One of the key factors in determining the mass carried by the outflows, and therefore their loss rates and powers, is the electron density ($n_{\rm e}$).
At rest-frame optical wavelengths, there are different approaches to measure such a quantity, which in many cases can be underestimated when obtained from the [S\:{\sc ii}] lines \citep{holden26}, therefore leading to overestimated masses. One of the main sources of such underestimation is due to the fact that this transition is being produced in the partially ionised zone of the narrow line region, {where there is a lower number of free electrons, which collisionally excite the [S\,{\sc ii}], leading to an underestimate of the electron density \citep{revalski22}.} Attempts to obtain more precise measurements have been performed in the literature \citep[e.g.][]{rose18, davies20, holden24, bessiere24, ramos-almeida25}, but they rely on the detection of emission lines that may lie outside the typical coverage of the optical spectrographs.

Quasars at $z\sim0.1$ are located at the sweet spot for observing these phenomena as they have high bolometric luminosities ($L_{\rm bol}\sim 10^{45}$erg~s$^{-1}$) and the gas structures can still be resolved at sub-kiloparsec resolutions at seeing-limited ground-based observations. The Quasar Feedback Survey (QFeedS) is designed to study low redshift ($z<0.2$) quasars that are analogues of high-z sources \citep{jarvis21}. In this work, we analyse the ionised gas properties in 18 QSOs from QFeedS observed with the Multi Unit Spectroscopic Explorer (MUSE) at the Very Large Telescope (VLT). We aim at studying the \oiiir-based ionised gas kinematics and derive properties such as electron density, mass, and mass loss rates linked to kinematically perturbed motions in a consistent manner for this particular sample. 
This work complements previous observational papers on subsamples of QFeedS \citep{jarvis19,jarvis20,jarvis21,silpa22,molyneux24,girdhar24,njeri25, njeri26} and on single objects \citep{harrison15,lansbury18,girdhar22}.

The paper is organised as follows. In Section \ref{sec:sample_obs}, we present the sample selection, based on the parent QFeedS sample \citep{jarvis21, njeri26}, and describe the MUSE and VLA datasets. In Section\,\ref{sec:medidas} we describe our emission-line fitting procedure and the non-parametric measurements that we use in the analysis of the gas kinematics. In Section\:\ref{sec:results} we present the properties of the ionised gas, including electron density, extinction, and estimates of outflow-related quantities. We discuss our results in Section\:\ref{sec:discussion}, putting them into context with predictions from simulations. Section\:\ref{sec:conclusions} presents our conclusions.
We adopt a cosmology of H$_{0} = 70$~\kms Mpc$^{-1}$, $\Omega_m=0.3$, and $\Omega_{\lambda}=0.7$ throughout the paper. 

\begin{figure}[ht!]
    \centering
\includegraphics[width=0.87\linewidth]{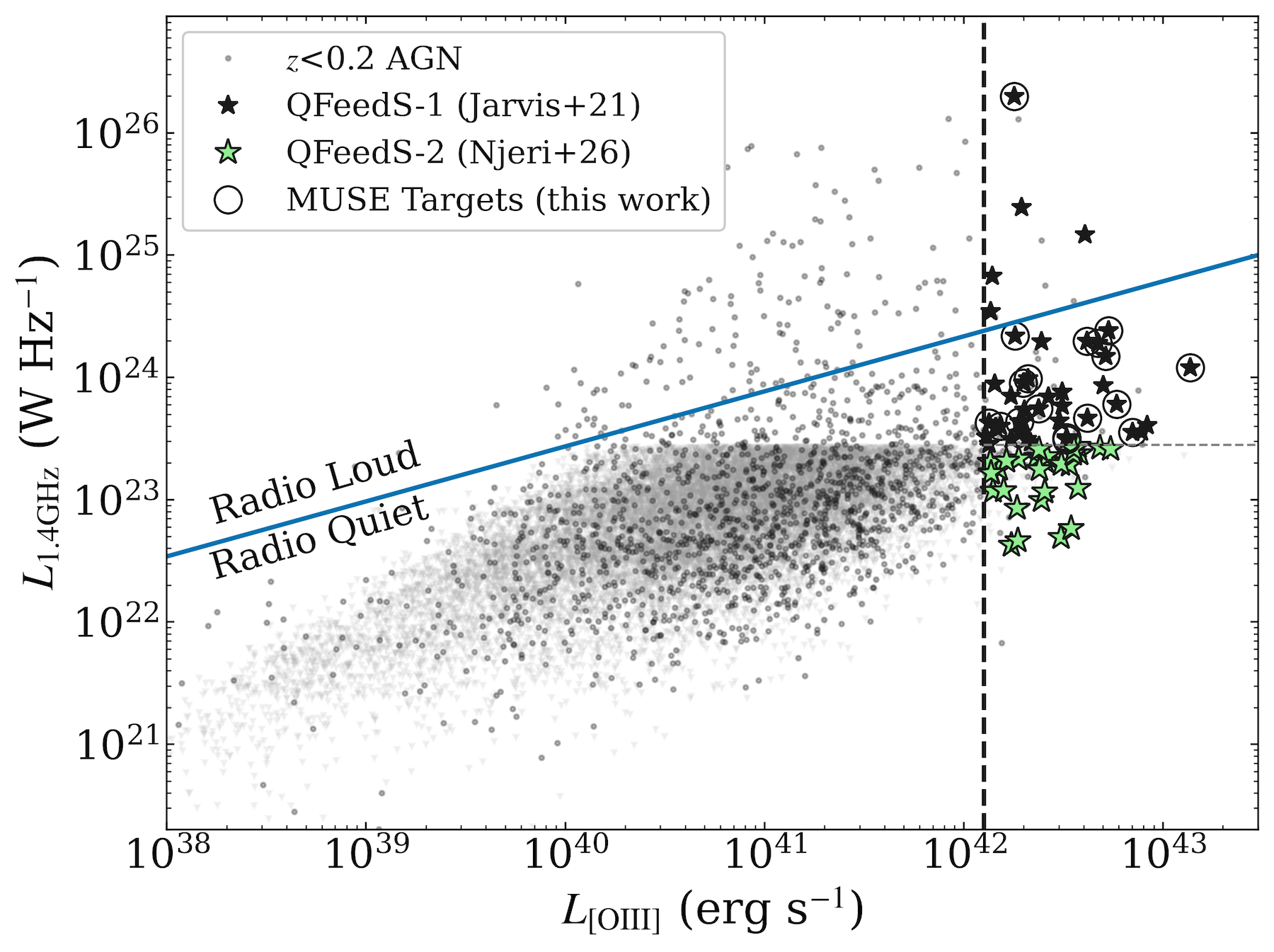}
    \includegraphics[width=0.87\linewidth]{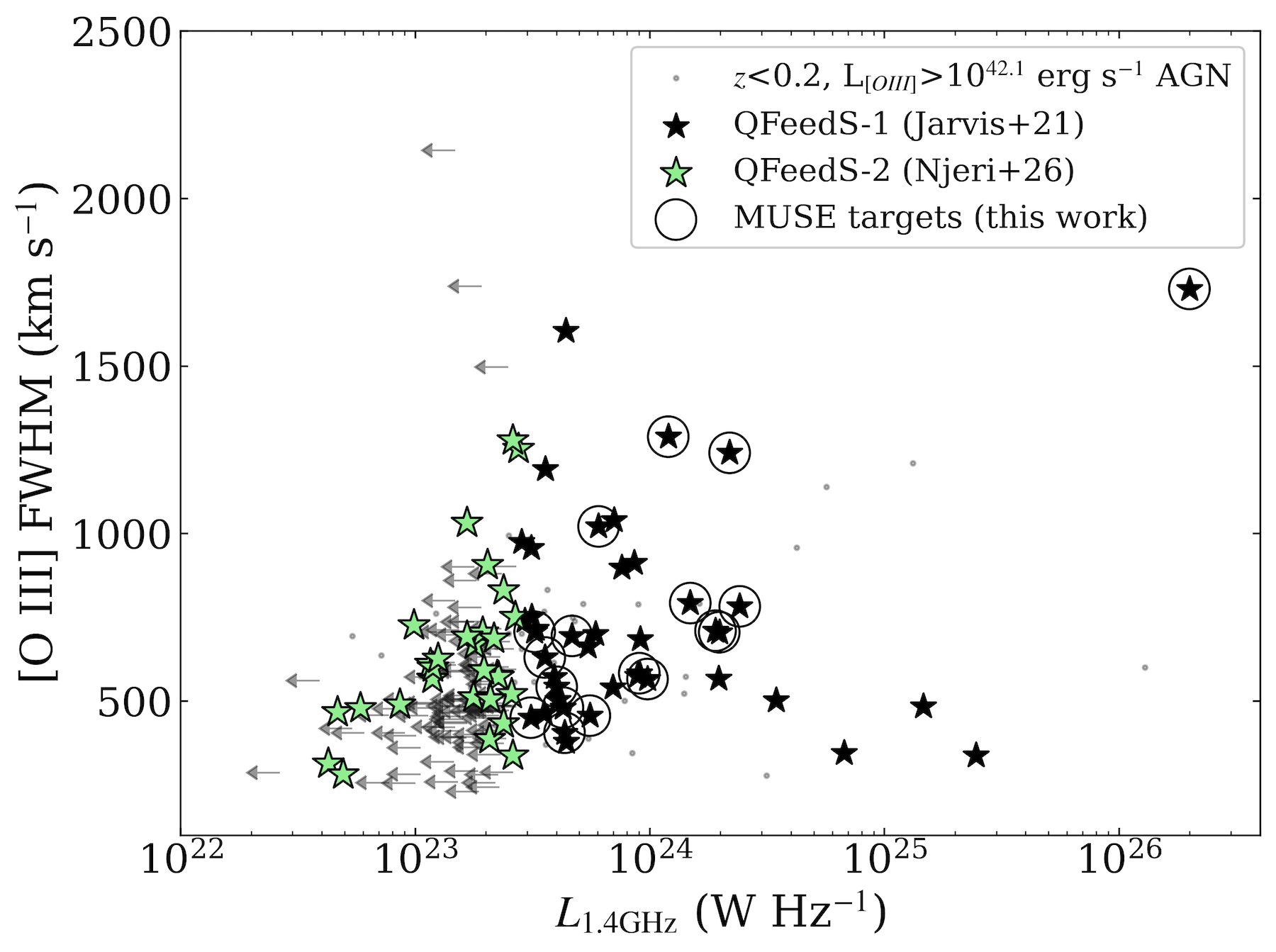}
    \caption{{\em Upper panel:} 1.4~GHz luminosity vs \oiii~luminosity as the grey points (light grey triangles for the upper limits) for the $z<0.2$ quasars in the \citet{mullaney13} sample. The solid blue line indicates the radio loudness criteria from \citet{xu99}. The dashed black line indicates the QFeedS selection, with the black stars representing the \citet{jarvis21} sample and the green stars from the sample studied in \citet{njeri26}. The 18 targets we study here are highlighted as the black circles. {\em Bottom:} \oiii~luminosity vs 1.4\:GHz Luminosity for the nearest ($z<0.2$) and brightest ($L_{\rm [O III]}>10^{42.1}$\:erg\:s$^{-1}$) galaxies in the \citet{mullaney13} sample. The symbols are the same as the upper plot, with the upper limits represented by arrows. The galaxies studied here populate the high radio luminosity space.}
    \label{fig:sample}
\end{figure}
\section{The sample and observations}
\label{sec:sample_obs}
\subsection{The type 2 QSO sample}
The original 42 objects of QFeedS-1 are drawn from the 24\,264 local galaxies ($z<0.4$) in the Sloan Digital Sky Survey data release 7 (SDSS DR7) sample \citep{mullaney13}. The QFeedS objects are located at $z<0.2$, allowing them to be resolved within a few kiloparsec, have quasar-like luminosities \citep[$L_{\rm [O III]}>10^{42}$\,erg\,s$^{-1}$,][]{reyes08}, and radio luminosities of $L_{\rm 1.4\,GHz}>10^{23.45}$\,erg\,s$^{-1}$. Detailed sample selection and description are provided in \citet{jarvis21}. We note that this is the parent sample of 42 sources that we focus on for this work (now called QFeedS-1), although the parent sample was recently expanded to 71 sources, now including quasars with lower radio luminosities \citep[QFeedS-2;][]{njeri26}.

From the selection in \citet{jarvis21}, we select the 18 galaxies classified as type 2 \citep{mullaney13} and observable from Cerro Paranal ($-80^{\circ}<\delta<40^{\circ}$) for which MUSE data were obtained. 
The choice of focusing on obscured quasars in this work is due to the lack of contamination from the broad component in the permitted lines, simplifying the study of the gas distribution and kinematics in the narrow-line region. Most of the previous work focused on these sources \citep[e.g.][]{harrison14,harrison15,lansbury18, jarvis19, jarvis20, girdhar22, girdhar24}, facilitating the comparison and interpretation of the results. Table\,\ref{tab:obs} presents the coordinates of the galaxies, redshifts, and [O{\sc iii}]5007 luminosities \citep[3\arcsec diameter aperture SDSS DR7 photometry;][]{mullaney13, jarvis21}.  The table also showcases information on the MUSE project ID, exposure time, and the spatial resolution of each galaxy observation. 

{The top panel of Figure~\ref{fig:sample} shows the galaxies in the parent sample of \citet{mullaney13}, the QFeedS targets (black and green stars), and the targets selected for this work highlighted as black circles, in the $L_{\rm 1.4~GHz}$ versus $L_{\rm [O III]}$ plane.  All galaxies studied here are classified as radio quiet, except J1347+1214 (also known as 4C+12.50 or PKS 1345+125).
The bottom panel shows only galaxies within $z<0.2$ and $\log L_{\rm [O III]}>42.1$\:erg\:s$^{-1}$ in the \oiii~full width at half maximum (FWHM) versus $L_{\rm 1.4 GHz}$ plane. The \citet{jarvis21} sample, thus by definition the sample studied here, occupies the highest radio luminosities compared to the QFeedS-2 sample \citep{njeri26}, although with a similar \oiii~ line width (see also Appendix\:\ref{app:maps} for contextualisation with the literature for individual sources).
 }

\label{sec:sample}

\begin{figure*}[ht!]
    \centering
    \includegraphics[width=0.89\textwidth]{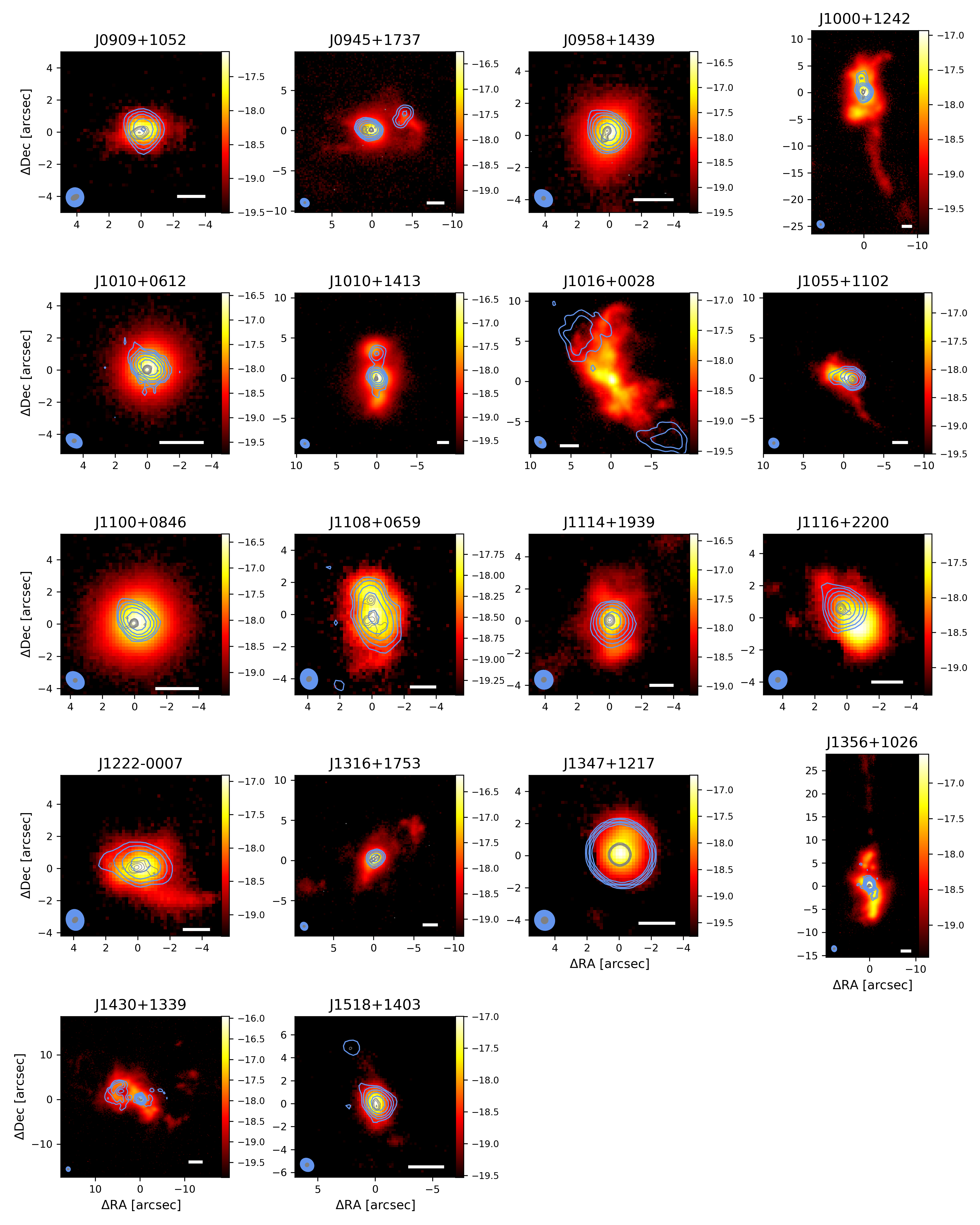}
    \caption{Pseudo narrow-band image constructed from the MUSE data cube accounting for the emission of H$\beta$, \oiiib,\ and \oiiir. The white bar at the bottom of each panel corresponds to a physical size of 5~kpc.
     The contours indicate the VLA radio data. In light blue, the $\sim 1\arcsec$  and in grey the $\sim 0.3\arcsec$ resolution observations (see Sect.~\ref{sec:vla} for details about the frequencies), and the circles at the bottom left of each plot are the beam sizes. The images are in log-scale and in units of erg\:s$^{-1}$\:\AA$^{-1}$ as indicated in the colour bar. In most galaxies, there is a clear correspondence between the radio and optical emission morphologies.}
    \label{fig:flux+radio}
\end{figure*}

\subsection{Observations and data}
\subsubsection{MUSE data}
\label{sec:muse}
The galaxies were observed {with the} MUSE instrument \citep{bacon10}, mounted at the Nasmyth B focus of the 8\,m UT4 of the VLT and operated by the European Southern Observatory (ESO) using, when possible, the adaptive optics (AO) mode.  MUSE is an integral field spectrograph operating in the optical ($465-930$\,nm) with a spectral resolution of $R\approx3000$ ($\approx 100$~\kms) and a spectral pixel of $1.25$\AA. The galaxies in our sample were observed in the Wide Field Mode, which has a field of view (FoV) of $1\arcmin\times1\arcmin$ ($\approx 150\times 150$\:kpc$^2$ at the distance of the galaxies) with a spatial sampling of 0.2\arcsec. 
The white bar in each panel of Figure \ref{fig:flux+radio} corresponds to 5\:kpc in the galaxies.

The data cubes analysed here are provided by the ESO Science Portal\footnote{\url{https://archive.eso.org/scienceportal/home}}. 
Data reduction follows the standard procedure of the MUSE data reduction pipeline \citep{weilbacher20}. The main data reduction steps include bias subtraction, division by flat field, determination of the wavelength solution, geometric and astrometric calibrations, correction for the atmospheric refraction, flux calibration, and sky subtraction.  
Table\,\ref{tab:obs} shows the project ID, the exposure times, and the spatial resolution, obtained from the \texttt{sky\_res} keyword in the header of the cubes. At the distances of the galaxies, this corresponds to physical spatial resolutions of $\approx 1-3$~kpc. The data for two of the galaxies (J1010+0612 and J1010+1413) correspond to the combination of two different programs, identified as MUSE-DEEP in the ESO Science Portal, as opposed to the single program observations identified as MUSE. This is indicated in the Project ID columns in Table~\ref{tab:obs}. 

Figure~\ref{fig:spec_sample} shows the extracted nuclear spectra for the 18 galaxies in our sample. We adopted an aperture with a diameter of 3\arcsec~, which is the same as the SDSS optical fibres and larger than the spatial resolution, thus not requiring aperture corrections. 
In some galaxies, which were observed using the MUSE adaptive optics (AO) mode, the sodium emission ($582-597$\,nm) from the AO laser contaminates the observed spectra and is masked out during the reduction. This region is indicated as 'Na mask' in the figure. In the case of the galaxies J0909+1052, J1222-0007, and J1108+0659, the \oiiir~ lies in this region, and thus the kinematics associated with this line will not be analysed here. We keep these three galaxies in the sample as we provide density and extinction measurements for them, despite the lack of the \oiiir~line. We indicate the observed emission lines, highlighting the [Ar\,{\sc iv}]$4711,4740$\:\AA~doublet used as an electron density tracer in the outflow component \citep[see Section \ref{sec:density} and][]{binette24}.

\begin{table*}[ht!]
\caption{Type 2 MUSE QSO sample: Observations.}
    \centering
    \begin{tabular}{lccccccccc}
        \hline\hline
        (1)& (2) & (3) & (4) & (5) & (6) & (7) & (8)\\
        Galaxy &  RA & Dec  &$\log{L_{\rm[O\,III]}}$& $z$ & Project ID & Exposure Time& FWHM \\
         & J2000 & J2000 & [erg\,s$^{-1}$] & & &[sec] &[\arcsec] \\
       \hline 
  J0909+1052$^{\star}$ &09:09:35.49 & +10:52:10.53 & 42.28&0.166& 0103.B-0071& $11\times612$&0.74\\ 
  J0945+1737 &09:45:21.33 &  +17:37:53.20  & 42.67& 0.128& 0103.B-0071 &  $4\times550$&0.93 \\ 
  J0958+1439 &09:58:16.88 &  +14:39:23.70  &42.52&  0.109& 0103.B-0071&  $4\times550$& 1.04\\
  J1000+1242 &10:00:13.14  & +12:42:26.20  & 42.62& 0.148& 0104.B-0476&  $12\times700$&1.09\\
  J1010+0612 &10:10:43.36 &  +06:12:01.40  & 42.26& 0.098& 0103.B-0071\& 0104.B-0476&  $12\times650$&0.93\\
  J1010+1413 &10:10:22.95  & +14:13:00.90   &43.14& 0.199& 0103.B-0071\& 0104.B-0476& $16\times625$&0.71 \\
  J1016+0028 &10:16:53.82   &+00:28:57.15 &42.18&0.116 & 0103.B-0071 &  $8\times550$ &0.71\\
  J1055+1102 &10:55:55.34 &+11:02:52.21 &42.52&0.145 & 0103.B-0071 &  $10\times550$&0.72\\
  J1100+0846& 11:00:12.38  & +08:46:16.30  &42.71&  0.100& 0104.B-0476 & $6\times450$&1.06 \\
  J1108+0659$^{\star}$ &11:08:51.04 & +06:59:01.44&42.32& 0.181 & 0103.B-0071 &  $9\times550$& 0.88\\
  J1114+1939 &11:14:23.81& +19:39:15.89 &42.30&0.199 & 0103.B-0071 & $4\times550$& 1.0\\
  J1116+2200 &11:16:25.34 &+22:00:49.37&42.38& 0.143 & 0103.B-0071 &  $4\times550$& 1.17\\
  J1222-0007$^{\star}$ &12:22:17.85 & -00:07:43.76& 42.85&0.173 & 0103.B-0071 & $8\times550$&  0.81\\
  J1316+1753& 13:16:42.90  & +17:53:32.50  & 42.77 &0.150  & 0103.B-0071 & $4\times550$&0.75\\
  J1347+1217& 13:47:33.36&   +12:17:24.30 &42.25& 0.121 & 0103.B-0391 &  $11\times612$&1.03\\
  J1356+1026& 13:56:46.10 &  +10:26:09.00&  42.73&  0.123  & 0103.B-0071 &  $4\times550$& 1.04\\
  J1430+1339& 14:30:29.88  & +13:39:12.00&  42.62&  0.085 & 0102.B-0107 &  $6\times950$&0.74\\
  J1518+1403& 15:18:56.27 &+14:03:19.05 &42.13&0.139  & 0103.B-0071 &  $8\times550$ &0.86\\
       \hline
    \end{tabular}
    \tablefoot{(1) Galaxy name; (2) Right ascension and (3) Declination ; (4) \oiiir~ luminosity from \citet{jarvis21} based on the \oiiir~ line fluxes reported by \citet{mullaney13}; (5) Galaxy redshift from SDSS DR7; (6) Muse project ID; (7) Exposure times; (8) Spatial resolution of the MUSE data. The galaxies marked with $\star$ are the objects where the \oiiir~region is masked out due to the sodium emission from the AO laser.}
    \label{tab:obs}
\end{table*}

 \subsubsection{VLA data}
\label{sec:vla}
We used the observations from the Karl Jansky VLA to compare the morphology of the radio emission with the distribution and kinematics of the ionised gas derived from MUSE. For the nine galaxies which are also part of the \citet{jarvis19} sample, namely J0945+1737, J0958+1439, J1000+1242, J1010+0612, J1010+1413, J1100+0846, J1316+1753, J1356+1026, and J1430+1339, we used the VLA observations in the C-band ($4-8$\:GHz) with A-array and B-array configurations \footnote{Data available at \url{https://doi.org/10.25405/data.ncl.c.5203919}.} with 0.3\arcsec and 1.0\arcsec resolution, respectively.
For the remaining nine galaxies, we used the A-array VLA observations taken in the L- ($1-2$\,GHz)\footnote{L-band data available at \url{https://data.ncl.ac.uk/articles/dataset/M_E_Jarvis_et_al_2021_L-band_VLA_images/13416164}} and C-bands ($4-8$\:GHz) data \footnote{C-band data available at \url{https://data.ncl.ac.uk/articles/dataset/M_E_Jarvis_et_al_2021_-_C-band_VLA_images/13702021}} with nominally the same resolution as the data from \citet{jarvis19}. 
The reduction performed in \citet{jarvis21} was done consistently for the entire sample, thus some fainter features are not well observed in some sources, while in \citet{jarvis19} the particularities of each galaxy were taken into account in the reduction. Thus, we adopted the two datasets for comparison with our work here.

In Figures \ref{fig:flux+radio} and \ref{fig:w80+radio} we show the VLA contour levels in light blue (1.0\arcsec resolution) and in grey (0.3\arcsec resolution) overlaid on the ionised gas emission and \oiiir~$W_{80}$ maps. The contours and noise levels are the same as those published by \citet{jarvis19} and \citet{jarvis21} for the respective galaxies. {However, we do not plot the lowest contour level for some galaxies, especially in J1430+1339, as they are closer to the noise level and introduce noisier features in the outer parts of the images.} We note the morphological correspondence between the extended ionised gas and the low resolution radio features \citep[see][]{jarvis21}. {In the case of $W_{80}$, its highest values are perpendicular to the radio emission} in some galaxies, notably J1010+1413, J1000+1242, and J1316+1753 \citep[see][]{ulivi24, girdhar22}.  A detailed comparison between the orientation of the radio and ionised gas emission is beyond the scope of this paper, and we limit ourselves to a qualitative analysis. Appendix\:\ref{app:maps} expands the discussion of the radio features in context with the MUSE data and the existing literature.

\subsection{Trimming and alignment of MUSE and VLA data}
The MUSE FoV is $\approx1$\,arcmin$^2$, and comprises a large field of view at the projected distances of the galaxies ($\approx100\times100$\,kpc$^2$ to $\approx200\times200$\,kpc$^2$). The area with ionised gas emission is much smaller, as shown in Figure~\ref{fig:flux+radio}. In order to make the emission line fitting procedure faster and more effective, that is,  not fit spaxels without detected emission lines, we trimmed the data cubes to account for the regions where the ionised gas is detected within $3\sigma$. First, we built pseudo H$\beta+$[O{\sc iii}] narrow-band images using the software QFitsView \citep{qfitsview}. We summed the flux in a window centred at the H$\beta$+[O{\sc iii]}4959,5007 emission lines and subtracted the continuum measured in 100\AA\ windows close to the lines. This resulted in the images in  Figure\,\ref{fig:flux+radio}.
The angular sizes covered by the galaxies are indicated in the figure. In the cases of J1000+1242 and J1356+1026, which have very extended tidal tails, almost the entirety of the north-south direction of the original data cube is present in the trimmed cubes. 
Since the astrometry solution delivered by the phase III MUSE pipeline is not always reliable, we assigned the correct coordinates to each data cube. We follow a similar procedure as presented in \citet{girdhar22}, where we first find the peak of the emission in a pseudo {\em r-band} image
using the \texttt{find\_peaks} function in the \texttt{photutils} package of Astropy. This pixel is then linked to the SDSS coordinates of the sources, quoted in Table\,\ref{tab:obs}.
The VLA data already has an accurate \texttt{wcs} solution, and we use the reference pixels on the image's header to align them to the \texttt{wcs}-corrected MUSE data.

\section{Measurements}
\label{sec:medidas}
We performed emission-line fitting for all spaxels in each trimmed data cube using the Python-based package \texttt{IFSCUBE} \citep{ifscube}.  {Due to the wings and complexities in the quasar emission lines, we adopted up to three Gaussian curves to reproduce each emission line. We used the {\em refit} parameter of \texttt{IFSCUBE}, which assumes the best fit parameter of the spaxels located in a radius of 0.6\arcsec from the current spaxel as its initial guess. Therefore, the amplitude of unnecessary components tends to zero, and the code automatically chooses the best number of Gaussians to reproduce a certain emission-line profile \citep[see][for more detail]{rogemar_N1275}. Our strategy is to best reproduce the emission-line profiles without attributing a physical meaning to individual components of the fit \citep[e.g.][]{zakamska14, ruschel-dutra21, rogemar_agnifs, rogemar_agnifskin, harrison14, girdhar22, girdhar24, ulivi24, speranza24, bessiere24}. Therefore, since we adopt non-parametric measurements (see below) to analyse the line kinematics and are interested in the best reproduction and minimisation of the noise of the overall line profile, the ultimate number of components does not affect our analysis.} We limited our analysis to the range of $4400-5500$\,\AA\ (in the rest-frame) since the \oiiir~is strong and a well-known outflow tracer \citep[e.g.][]{storchi-bergmann18, kakkad18, mullaney13, fischer13, bruno21}. A linear polynomial was used to reproduce the underlying continuum, as the spectral range {is} small. 
We tied the kinematics of the \oiiib\ to the \oiiir~ lines and also assumed a theoretical line ratio \oiiir/\oiiib$=$3 for each component.
Since our goal {is} to measure the outflow properties traced by the \oiiir~line, our fitting procedure did not include the modelling of the underlying stellar continuum. The stellar absorptions observed in the optical are not a major contaminant to the \oiii~emission line fluxes \citep[e.g.][and see also Fig.\:\ref{fig:spec_sample}]{kakkad22} as they are for H$\beta$, and thus our procedure is adequate for the proposed analysis.

\begin{figure}
    \centering
\includegraphics[width=0.9\linewidth]{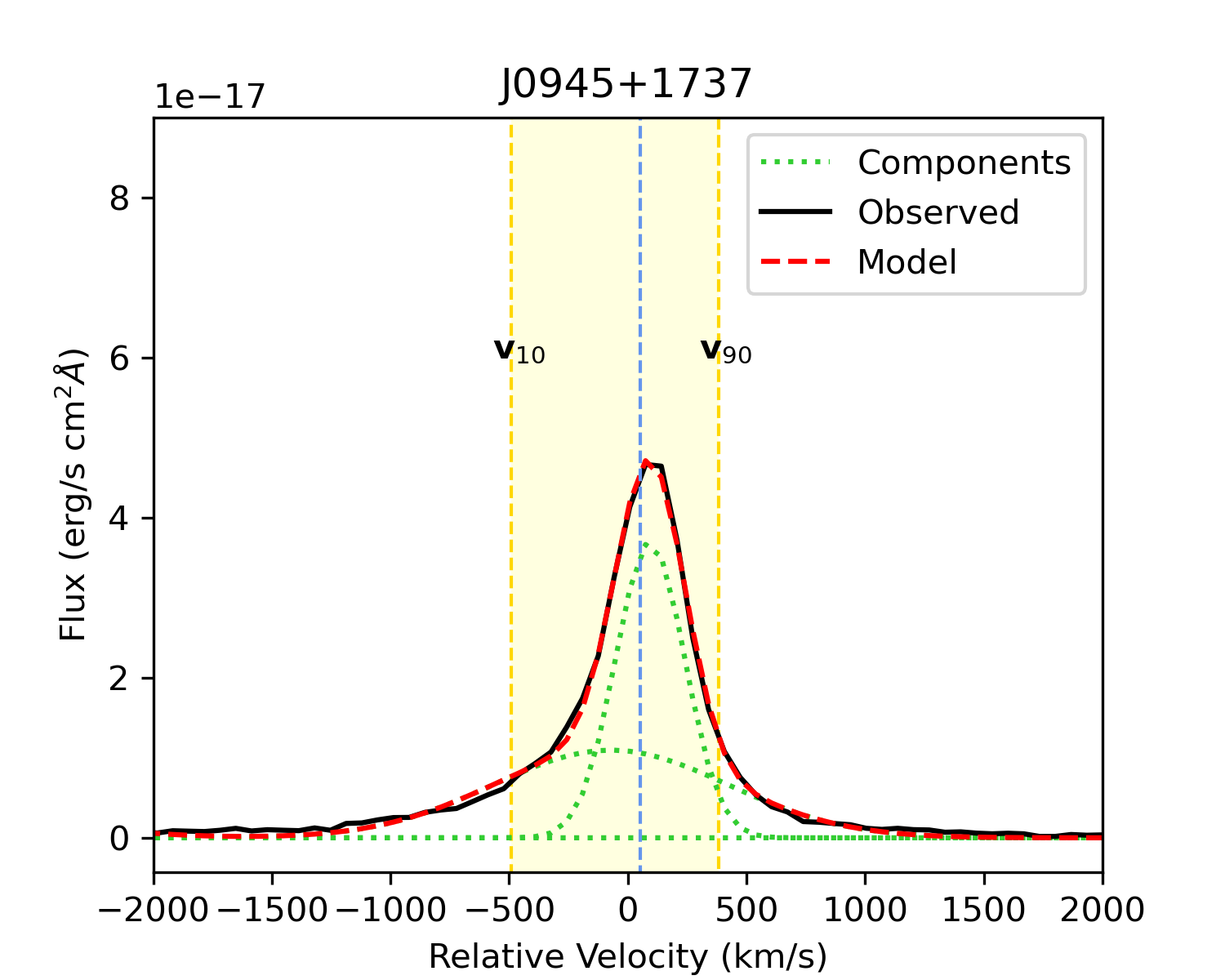}
    \caption{Fitting example of the \oiiir\ for the nuclear spaxel of J0945+1737. The linear continuum fit is subtracted from both modelled and observed spectra. The green-dotted lines represent the individual Gaussian components. The shaded yellow area corresponds to the $W_{80}$ where the $v_{10}$ and $v_{90}$ are the velocities encompassing 10 and 90\,percent of the line flux. The dashed blue line corresponds to the mean line velocity, $V_{\rm cen}$.}
    \label{fig:example_fit}
\end{figure}

Here we used a non-parametric approach for analysing the emission-line kinematics following the strategy described in \citet{liu13}. This method is not dependent on the shape or number of functions used to fit the emission lines. Figure \ref{fig:example_fit} shows an example of the fitting to  \oiiir\ in a nuclear spaxel of the galaxy J0945+1737. The green lines represent the Gaussian components, while the red curve corresponds to the modelled spectra. We also followed \citet{liu13} in adopting the modelled spectra instead of the data to obtain the non-parametric measurements, as it is less noisy in the areas with lower signal-to-noise ratios. The yellow area in the figure corresponds to the parameter $W_{80}\equiv v_{90}-v_{10}$, where $v_{90}$ and $v_{10}$ are the velocities comprising 90 and 10 percent of the emission line flux. We also calculated the centroid velocity, $V_{\rm cen}$, which corresponds to the 50th velocity percentile, indicated by the blue vertical line in the figure, corresponding to the mean line velocity. This procedure was applied to all spaxels in every trimmed data cube. Figures \ref{fig:w80+radio} and \ref{fig:maps_J0945}-\ref{fig:maps_J1518} show the corresponding line emission and underlying continuum maps. Appendix \ref{app:maps} presents a discussion of the maps for individual galaxies. 

Besides the data cubes, we also relied on two sets of extracted spectra, centred at the nuclei of the galaxies, to perform some of the analysis. 
The first with $r=1.5\arcsec$ (see Fig.\ref{fig:spec_sample}) was used for the estimations of extinction (see Sec. \ref{sec:av}), where up to three Gaussian curves plus the underlying continuum were fitted to the emission lines, and electron density (see Sec. \ref{sec:density}) where just one Gaussian function was enough due to the low signal-to-noise ratio of the lines.  The second extraction had $r=0.3\arcsec$ and we also fitted up to three Gaussians to the \oiii~to obtain the unresolved nuclear velocity dispersion, which is the basis for defining outflow-dominated spaxels (see Sec. \ref{sec:outflowcalc}).

\section{Properties of the ionised gas}
\label{sec:results}

\subsection{Nebular extinction}
\label{sec:av}
We calculated the interstellar dust extinction following Equation 7 in \citet{rogerio21}. This prescription adopts an extinction curve as described in \citet{CCM89}, $R_V=3.1$, and an intrinsic H$\alpha$ to H$\beta$ line ratio of $2.86$ assuming an excitation temperature of 10\,000 K, electron density of 100 \cmm, and case B recombination \citep{Osterbrock06}. 
Due to complexities in the line profiles, especially the blending between the emission lines of H$\alpha$ and  [N{\sc ii}], we chose to calculate the extinction based on the integrated spectrum with $r=1.5\arcsec$.

Table \ref{tab:dens} presents the optical extinction ($A_{\rm V})$ in magnitudes for all galaxies except J1114+1939, where the H$\beta$ emission line cannot be observed due to the sodium emission from the AO laser (see {Sect. \ref{sec:muse}} and Fig. \ref{fig:spec_sample}). In this particular case, no correction was applied to the observed line fluxes. For the subsequent outflow analysis (Sec. \ref{sec:outflowcalc}), we corrected the \oiiir\ fluxes by the extinction at the observed wavelength of the emission line. The mean value of our $A_V$ measurements is $\sim 1.3$\:mag {with a mean uncertainty of 0.6\,mag}. \citet{ulivi24} derived a median $A_V$ of $1-1.8$\:mag on a spaxel-by-spaxel basis for four of the galaxies that are in our sample J1010+1413, J1010+0612,  J1100+0846, and J1000+1242. The mean $A_V$ value we obtain for these four galaxies is $\sim 1.6$\,{mag}, which is within the values obtained by \citet{ulivi24}.

\subsection{Electron density}
\label{sec:density}
The electron density can be estimated by different methods, each of which has its own uncertainties and limitations. The ratio of the [S\:{\sc ii}]6717,6730 emission line doublet is one of the most common methods to derive the electron density using optical spectra. However, the range of densities probed by this line ratio is limited to $50\lesssim n_e [{\rm cm^{-3}}]\lesssim5000$\,\cmm \citep{Osterbrock06}. \citet{holden26} argue that this indicator probes an even stricter range of densities ($100\lesssim n_e [{\rm cm^{-3}}]\lesssim3000$\,\cmm). The energy necessary to create  [S\:{\sc ii}] is also low ($10$\,eV) and this transition is produced in the partially ionised region, while the emission\oiii, our adopted tracer of outflows here, originates in areas where the gas is fully ionised. The transauroral lines [O\:{\sc ii}]\,7320,7330 and [S\:{\sc ii}]4069,4076 have also been used as density estimators {\citep{holt11, rose18, davies20, bessiere24}}, giving rise to the so-called {TR} method. The range of densities probed depends on the extinction model adopted and other factors, for example \citet{davies20} found $1200<n_e[{\rm cm^{-3}}]<3000$ for a sample of local Seyferts, while \citet{holden26} derived $700\lesssim n_e[{\rm cm^{-3}}]\lesssim28000$ for a sample of type 2 QSOs at $z\sim0.1$. 
This method is not applicable to our dataset as the doublets [S\:{\sc ii}]4069,4076 and  [O\:{\sc ii}]3726,3729 are not observable by MUSE at the redshift of the galaxies. Photoionisation models can also provide gas electron densities \citep[e.g.][]{baron19,davies20,trindade-falcao21, revalski22, revalski25}, which is usually called the ionisation parameter method. It relies on the estimate of the number of hydrogen-ionising photons via a combination of observed line ratios with photoionisation models and can probe densities of $\sim 10^{4.5}$\cmm \citep[see][for a detailed discussion]{baron19}.

We used the [Ar\,{\sc iv}]$4711, 4740$ emission lines as tracers of the ionised gas electron density. The parent ion of these transitions, Ar$^{3+}$ and O$^{++}$ have similar ionisation potentials --- 40\,eV and 35\,eV, respectively. Moreover, this emission line ratio can probe a wide range of densities, from $\sim100$ to $\sim60~000$~\cmm~ at excitation temperatures of 15\:000~K. The range of densities probed is slightly more restrictive ($\sim300 -57\:000$\:\cmm) if we assume a temperature of 10\:000~K. {Direct measurement of the densities} can provide more robust estimates of the mass of ionised gas and outflow properties, mitigating some of the intrinsic uncertainties involved in these calculations. 

The [Ar\,{\sc iv}] lines are usually weak in AGN spectra \citep{rogemar_abundances, binette24}. In our dataset, these lines are  not detected in the individual 
spaxels. Thus, we used the $r=1.5\arcsec$ integrated spectra where these lines are detected (see Fig.~\ref{fig:spec_sample}, where the insets show the region of the [Ar\:{\sc iv}]$4711,4740$ lines), where it is possible to observe that the doublet is detected in all but two galaxies, J1114+1939  and J1347+1217. In the sources where it was detected, we fitted each [Ar\:{\sc iv}] line by a single Gaussian curve and used their flux ratio as an input for the {\tt getTemDen} routine in the Pyneb package \citep{pyneb}, assuming a temperature of $15\,000$\,K \citep[typical value for AGN,][]{revalski22} to derive the electron density. In Table\,\ref{tab:dens} we present the electron densities measured for 16 galaxies.  Since the [Ar\:{\sc iv}] lines are faint and only detected in the integrated spectra, no broad and/or disturbed gas components are observed (see the insets in Fig.\:\ref{fig:spec_sample}). Consequently, the derived electron density is likely tracing the overall density of the galaxy, not specifically the outflow density.

The $n_e$ we measure here is in the range $2.67<\log n_e/{\rm cm^{-3}}<4.44$ with a median value of $\approx5900$~\cmm. 
Assuming an excitation temperature of $10\:000$\:K, we find that the densities tend to be lower, with a median difference of $\sim 500$\:\cmm, than the values reported in Table\,\ref{tab:dens}. It should be noted that the greatest differences arise from galaxies with densities $\sim10^4$\,\cmm.  Lower excitation temperatures have been found to result in lower densities in previous work in the literature {\citep[e.g.][]{bianchin26, cezar26}}. 
Previous works focused on the study of QSO2s also estimated $n_e$ through various approaches. \citet{bruno21} finds $2.08 \leq \log n_e/{\rm cm^{-3}}\leq2.85$, $2.93 \leq \log n_e/{\rm cm^{-3}}\leq3.16$, and $2.56 \leq \log n_e/{\rm cm^{-3}}\leq4$ using [S\:{\sc ii}], [Ar\:{\sc iv}], and the ionisation parameter methods, respectively, derived from spatially resolved optical data. A wider range of values ($1.21\leq \log n_e/{\rm cm^{-3}} \leq 4.22$) is measured by \citet{holden26} for a sample of 48 QSO2s using the [S\:{\sc ii}]. These authors also use the {TR} method ($2.94\leq \log n_e/{\rm cm^{-3}} \leq 4.79$) that provides a range closer to ours. In a subset of the same sample, \citet{cezar26} measured $2.6 \leq \log n_e/{\rm cm^{-3}}\leq3.1$ using the [S{\sc ii}] method and
$3.0 \leq \log n_e/{\rm cm^{-3}} \leq 4.1$ using the {TR} method. We find good agreement between the densities we estimate and the values obtained by \citet{ramos-almeida25} using the [Ne{\sc v}] emission lines observed in the rest-frame mid-infrared through JWST/MIRI observations. This sample and ours overlap for three galaxies: J1010+0612, J1356+1026, and J1430+1339.
These authors adopt two excitation temperatures (10$^4$ and $2\times10^4$~K) and find $2.8\leq \log n_e/{\rm cm^{-3}} \leq 4.07$ and $3.1\leq \log n_e/{\rm cm^{-3}} \leq 4.33$, respectively.

\subsection{Mass of ionised gas} 

\label{sec:outs}
The mass of the ionised gas can be estimated from the \oiiir\ emission line following 

\begin{equation}
    M_{\rm [O III]}=8\times10^{7}{\rm M_{\odot}}\left(\frac{1}{10^{\rm[O/H]-[O/H]_{\odot}}}\right)\left(\frac{L_{\rm[OIII]}}{10^{44}\rm erg\,s^{-1}}\right)\left(\frac{n_e}{500\,{\rm cm^{-3}}}\right)^{-1},
\label{eq:moiii}
\end{equation}
where [O/H] is the oxygen abundance adopted to be equal to the solar abundance,  $L_{\rm[OIII]}$ is the luminosity of the [O{\sc iii}]5007 emission line, and $n_e$ is the electron density of the gas \citep{carniani15,kakkad16,kakkad20}. Eq.\,\ref{eq:moiii} is derived assuming a gas temperature of 10$^4$\,K in a fully ionised gas \citep{rupke05, genzel11, carniani15}. We also use this relation to estimate the mass of the outflows ($M_{\rm out}$) when taking into account only the outflow-dominated spaxels (see Sec.\,\ref{sec:outflowcalc}). 

Other studies use the H$\beta$ line as a proxy for gas masses \citep[e.g.][]{harrison14, storchi-bergmann18, bruno21}. According to the comparison presented by \citet{carniani15}, the masses calculated using the H$\beta$ line luminosity are higher than the masses obtained from \oiiir by a factor of $\sim 2$ because H$\beta$ has a lower IP and therefore a larger emitting volume $\sim 2$  \citep[see also][]{venturi23}. Thus, the gas masses, mass outflow rates, and kinetic powers we present here should be interpreted as being underestimated by at least a factor of 2. We also assumed that the mass outflow rate is constant over time \citep[see Sec.\,\ref{sec:outflowcalc} and][]{lutz20}, which leads to a factor of 3 that is not accounted for in our calculations. 
The other main source of uncertainty in the mass estimates lies in the electron density (see the discussion above). We mitigate this by adopting our measurements from the [Ar\:{\sc iv}] doublet (Table\:\ref{tab:out_props}). Furthermore, comparing with measurements from the literature, we assumed a typical density of $10^3$\,cm$^{-3}$ (Table\:\ref{tab:out_103}) since most of the data points from other works were normalised to this common density. 

\begin{table}
   \centering
    \caption{Gas properties.}
    \begin{tabular}{cccc}
    \hline\hline
    (1) & (2) & (3) & (4)\\
    Galaxy & $n_e$ & $A_{\rm V}$ & $W_{80}$\\
           &  [\cmm] & [mag] & [\kms]\\
    \hline
J0909+1052 & $3030\pm132$ & $1.06$ & $-$\\
J0945+1737 & $3888\pm64$ & $0.13$ & 384.3\\
J0958+1439 & $2159\pm109$& $2.50$&  412.3\\
J1000+1242 & $470\pm121$& $1.72$&277.1\\
J1010+0612 & $21516\pm 1788$& $2.60$& 258.2\\
J1010+1413 & $27690\pm2141$& $1.35$& 757.9\\ 
J1016+0028 & $7610\pm304$& $0.55$& 346.0\\  
J1055+1102 & $25714\pm214$& $0.94$&  180.4\\  
J1100+0846 & $5949\pm908$& $0.91$& 159.4\\     
J1108+0659 &$5103\pm11$& $1.42$& $-$\\ 
J1114+1939  & $-$ & $-$ & 255.1\\      
J1116+2200  &$5127\pm351$& $1.31$ &270.7\\   
J1222-0007 &$5763\pm100$& $0.11$& $-$\\  
J1316+1753 &$6076\pm41$& $1.07$& 307.5 \\    
J1347+1217  & $-$& $0.74$ & 671.7\\        
J1356+1026  &$2043\pm119$& $0.96$ & 544.5\\   
J1430+1339  &$2879\pm204$& $1.80$& 586.2\\  
J1518+1403 & $7240\pm281$ & $1.47$& 239.5\\  
    \hline
    \end{tabular}
    \tablefoot{(1) Galaxy name; (2) Electron density estimated from the [Ar\,{\sc iv}] and assuming $T_{\rm e}=15\:000$\:K;  (3) Visual extinction estimated from the Balmer decrement {of extracted nuclear spectra with $r=1.5\arcsec$} (see Sec.\:\ref{sec:av}). (4) $W_{80}$ threshold. In the galaxy J1114+1939 the sodium mask is located in the region that includes the [Ar\,{\sc iv}] and H$\beta$ lines. For J1347+1217, the [Ar\,{\sc iv}] doublet is not detected within 3$\sigma$. }
    \label{tab:dens}
\end{table} 

\subsection{Outflow geometry, mass, rates and energetics}
\label{sec:outflowcalc}
The geometry adopted for the outflow depends on the assumed outflow history \citep{lutz20}. We assume that the mass outflow rate is constant and that the average mass density decreases over time. We estimate both the {\em global} and {\em peak} outflow properties. In the latter case,
the mass outflow rate is assumed to be constant within each shell but variable along the extension of the outflow.

Regardless of the chosen geometry, it is challenging to determine the velocity at which the gas can be considered as part of an outflow. Different definitions have been adopted throughout the recent literature. They include multi-Gaussian decomposition and the attribution of the broadest Gaussian as due to the outflow \citep[e.g.][and references therein]{bianchin22, riffeln3884, ramos-almeida19, ramos-almeida22, speranza22, ramos-vieira25, revalski25}, the use of the maximum velocity of the line profile \citep{rupke11, bruno21},
or assuming $W_{80}$ as a proxy \citep[e.g.][]{liu13, harrison16, rogemar_agnifskin, ruschel-dutra21}.  Two works determined $W_{80}$ thresholds to define outflows in samples of low-luminosity AGN \citep[500~\kms]{wylezalek20} and type-2 quasars at cosmic noon \citep[600~\kms]{kakkad20} for the \oiii\,line, arguing that the gas is no longer gravitationally bound at these velocities. However, \citet{gatto24} found that the average $W_{80}$ has a range of $200-900$\kms, and adopted a threshold of 315\,\kms\ to define the outflow-dominated spaxels based on the distribution $W_{80}$ in the MaNGA AGN sample. These authors argue that using $W_{80}$ as a proxy for kinematically disturbed regions accounts for turbulence or radiation heating on the gas, in addition to the typical bipolar outflow structures. Such turbulent mechanisms are known to contribute to the disturbance of the gas, especially in type 2 QSOs \citep[see][]{ulivi24}. 
Alternatively, \citet{bessiere24} used the stellar velocity dispersion and compared it with \oiii~velocity dispersion to define the outflow velocity. Also, \citet{hervella23} compared the parametric and non-parametric approaches to estimate the outflow-related properties, finding considerable differences between those approaches. From the simulation point of view, an increase in the outflow velocity can result in underestimation of the outflow mass by a factor of 8, implying that the slow-moving gas carries a lot of mass content despite being at lower energies \citep{ward24}. Thus, if adopting only the more conservative single threshold in $W_{80}$ can result in the outflow mass being underestimated.

Here, we also adopt a variable $W_{80}$ threshold to define the outflow-dominated spaxels in the galaxies, following the procedure described in \citet{rogemar_agnifskin}. First, we extract the spectra with an aperture with $r=0.3\arcsec$, which at the resolution of our data is unresolved. Then, we fit the \oiiir~ line with up to three Gaussian functions. Finally, we assume $W_{80}=2.563\sigma_n$ \citep[see][for the $W_{80}$]{liu13}, where $\sigma_n$ is the velocity dispersion of the narrowest component, as the threshold to define the outflow-dominated spaxels in each galaxy. Table \ref{tab:dens} includes the thresholds we adopt to define the outflow-dominated spaxels.  The average of the $W_{80}$ values is $376$~\kms with a standard deviation of 176~\kms, similar to the value determined by \citet{gatto24}. The mass of the outflowing gas, and therefore the quantities derived from it, is determined by considering only spaxels with $W_{80}$ higher than the thresholds presented in Table \ref{tab:dens} in Equation \ref{eq:moiii}.

\subsubsection{Global properties}
\label{sec:global}
 For the spherical geometry, the mass of the outflow ($M_{\rm out}$) is given by the sum of the ionised gas mass in each spaxel with $W_{80}$ higher than the threshold presented in Table \ref{tab:dens}. Following our previous assumption of a constant mass outflow rate over time, the mass outflow rate $\dot{M}_{\rm out}^{\rm g}$ in a sphere of radius $R_{\rm out}^{\rm g}$ is given by 

\begin{equation}
         \dot{M}_{\rm out}^{\rm g}=\frac{M_{\rm out} V_{\rm out}}{R_{\rm out}^{\rm g}}, 
     \end{equation}
     with $R_{\rm out}^{\rm g}$ defined as 
\begin{equation}
     R_{\rm out}^{\rm g} = \frac{\langle R_{\rm out} F_{\rm [O III]} \rangle}{\langle F_{\rm [O III]} \rangle},
 \end{equation}
where $R_{\rm out}$ is the radius of each outflow-dominated spaxel measured from the nucleus, that is,  the peak of the continuum emission, in each galaxy,
$F_{\rm [O III]}$ is the \oiiir\ line flux in each spaxel, and ${V_{\rm out}}$ is the velocity of the outflow given by
\begin{equation}
     V_{\rm out} = \frac{\langle W_{\rm 80} F_{\rm [O\,III]} \rangle}{\langle F_{\rm [O III]} \rangle},
 \end{equation}
where $W_{80}$ is also the value measured in each outflow-dominated spaxel.

The kinetic power of the outflows is calculated by 
  \begin{equation}
         \dot{E}_{\rm out}^{\rm g}=\frac{1}{2}\dot{M}_{\rm out}^{\rm g} V_{\rm out}^2.
         \label{eq:ekin}
     \end{equation}
     
Here we use a definition of $R_{\rm out}^{\rm g}$ and $V_{\rm out}$ similar to those presented in \citet{rogemar_agnifskin}.  We estimate $R_{\rm out}^{\rm g}$ and $V_{\rm out}$ using the flux average values as presented in the equations above, meaning that spaxels with higher fluxes have a larger contribution to the radius and velocity. This strategy aims to reduce the contribution of spurious features that may still be present on the kinematic maps. We estimate flux, therefore mass, uncertainties by adopting the root mean square of the residuals from the fitting in each spaxel and using it within the formal error calculation. For the velocity uncertainties, we obtain the uncertainties in $W_{80}$ by calculating its standard deviation from $W_{80}$ and then dividing it by the number of data points with valid measurements. 

{We emphasize that the adoption of flux-weighted $W_{80}$ is just one among the different definitions of outflow velocity.  In the literature, one of the most common definitions involves attribution of the centroid velocity of the broadest Gaussian component to the outflow \citep[e.g.][]{fischer18, bianchin22, cezar26}. }

\subsubsection{Peak properties}
\label{sec:peak}
In the spherical shell geometry, we estimate the mass outflow rate and kinetic power at individual rings of size $\Delta R_{\rm out}=0.6$\arcsec. With such a choice, the innermost extraction is a circle with a diameter of 1.2\arcsec, which is larger than the spatial resolution of our data (see Table \ref{tab:obs}). This choice of radius is motivated by the size of the spaxel ($0.2$\arcsec) and the angular resolution of the data, which means that the first ring has a diameter larger than the highest data angular resolution (see Table\,\ref{tab:obs}). These rings, or shells, are concentric with the nucleus position. We follow the definitions of \citet{Shimizu19}, \citep[see also][]{bruno21,rogemar_agnifskin, venturi23} to derive the outflow-related quantities. 
In this geometry, the mass outflow rate in each radial bin is given by

\begin{equation}
         \dot{M}_{\rm out}=\frac{M_{\rm out}^{\rm ring} V_{\rm out}}{\Delta R_{\rm out}}, 
     \end{equation}
where $M_{\rm out}^{\rm ring}$ is the mass encompassed in each radial bin and the velocity $V_{\rm out}$ is the azimuthal average of $W_{80}$ in each radial bin. The kinetic power of the outflow is calculated by Eq.\,\ref{eq:ekin} for each radial bin. 

\begin{figure*}
\sidecaption
\includegraphics[width=0.7\textwidth]{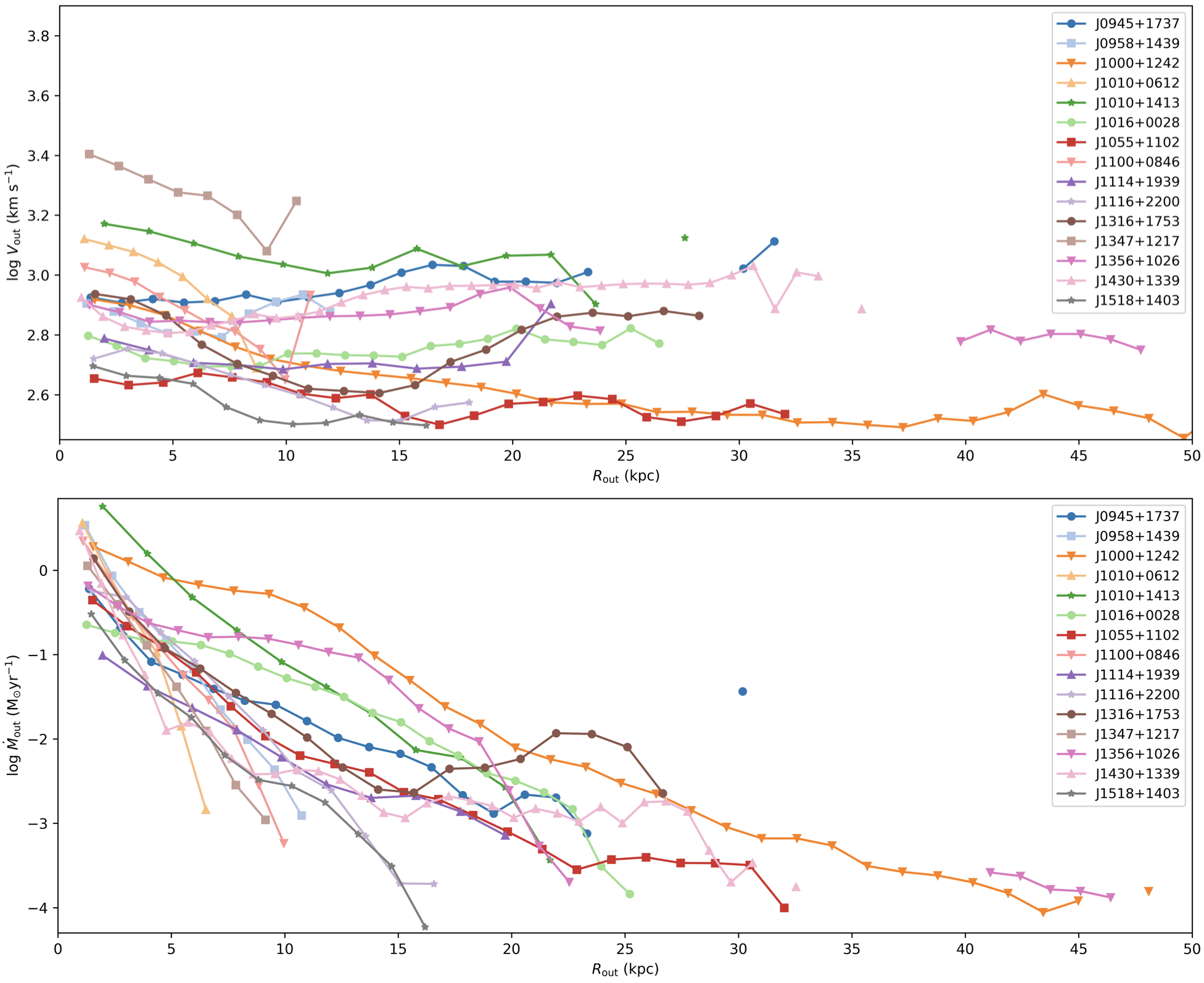}
    \caption{{\em Top panel:} Outflow velocity radial profile.  {\em Bottom panel:} Mass of the gas in the outflow radial profile assuming a density of $n_e=10^3$\,\cmm. The velocities are insensitive to our choice of electron density.
    }
    \label{fig:vel_mass_prof}
\end{figure*}

Figure \ref{fig:vel_mass_prof} presents the outflow velocity (top panel) and the mass outflow rate (bottom panel) for all galaxies. 
From the mass rate profiles, we define the peak radius of the outflow ($R_{\rm out}^{\rm p}$) as the radius where $\dot{M}_{\rm out}$ reaches its maximum, hereafter $\dot{M}^{\rm p}_{\rm out}$. The peak kinetic energy (radial profiles not shown here) is also defined as the energy at which $\dot{M}_{\rm out}$ is maximum. In all cases, the peak coincided with the innermost ring, which suggests that the nuclear region ($<2$\,kpc) contributes the most to the outflows. We also note that using these assumptions, $V_{\rm out}^{\rm peak}$ does not necessarily correspond to the highest value in the velocity profile (see Figure \ref{fig:vel_mass_prof}) but rather to where the mass outflow rate is the highest, which is ultimately governed by the mass ratio profile of the outflowing gas.

\begin{table*}[ht!]
    \centering
    \caption{Outflow properties considering the {\em global} and {\em peak} methods.}
    \label{tab:out_props}
    \begin{tabular}{lccccccccc}
    \hline
    \hline
          &  \multicolumn{5}{c}{Global}  &  \multicolumn{4}{c}{Peak}\\
          \cmidrule(l){3-6} \cmidrule(l){7-10}
        (1) & (2) & (3) & (4) & (5) & (6) & (7) & (8) & (9) & (10)\\
          Galaxy &  $\log M_{\rm out}$ &$R^{\rm g}_{\rm out}$ & $V^{\rm g}_{\rm out}$ & $\log \dot{M}^{\rm g}_{\rm out}$ & $\log \dot{E}^{\rm g}_{\rm kin}$ & $R^{\rm p}_{\rm out}$ & $V^{\rm p}_{\rm out}$  & $\log \dot{M}^{\rm p}_{\rm out}$ & $\log \dot{E}^{\rm p}_{\rm kin}$\\
          & [M$_{\odot}$]& [kpc]&[\kms]& M$_{\odot}$yr$^{-1}$ &[erg\:s$^{-1}$]& [kpc]&[\kms]& M$_{\odot}$yr$^{-1}$ &[erg\:s$^{-1}$]\\
          \hline
          J0945+1737 & 5.94$\pm$0.31 & 6.31$\pm$4.02 & 965$\pm$82 &  -0.87$\pm$0.61 & 40.60$\pm$0.61 & $1.37$ & 842.41$\pm$0.50 &-0.81$\pm$0.25 & 40.54$\pm$0.25\\
          J0958+1439 & 6.84$\pm$0.02 & 1.45$\pm$0.02&813$\pm$2 & 0.59$\pm$0.02 & 41.91$\pm$0.02 & $1.19$ & 802.53$\pm$0.62&0.20$\pm$0.02 & 41.51$\pm$0.02 \\
          J1000+1242 &  7.55$\pm$0.11 & 5.97$\pm$0.13 & 677$\pm$5 & 0.62$\pm$0.11 & 41.78$\pm$0.12 & $1.55$ & 828.7$\pm$0.27& 0.61$\pm$0.12 &41.95$\pm$0.12\\
          J1010+0612 & 5.58$\pm$0.06 & 1.25$\pm$0.10 & 1342$\pm$6 & -0.38$\pm$0.09 & 41.38$\pm$0.10 & $1.09$ & 1319.9$\pm$0.68& -0.77$\pm$0.06 & 40.97$\pm$0.06\\
          J1010+1413 & 5.85$\pm$0.29 & 2.57$\pm$1.00 & 1460$\pm$46 & -0.38$\pm$0.46 & 41.45$\pm$0.48 & $1.97$ &1483.81$\pm$0.77& -0.69$\pm$0.34 & 41.15$\pm$0.34 \\
          J1016+0028 & 5.61$\pm$0.02 & 5.59$\pm$0.11 & 586$\pm$3 & -1.36$\pm$0.02 & 39.67$\pm$0.02 & $1.26$ & 626.6$\pm$0.26 &-1.53$\pm$0.03 & 39.57$\pm$0.03\\
          J1055+1102 & 5.25$\pm$0.01 & 2.90$\pm$0.06 & 449$\pm$1 & -1.55$\pm$0.01 & 39.25$\pm$0.01 & $1.52$ & 626.6$\pm$0.26& -1.77$\pm$0.01 & 39.04$\pm$0.01\\
          J1100+0846 & 5.98$\pm$0.12 & 1.78$\pm$0.24 & 1057$\pm$14 & -0.24$\pm$0.17 & 41.31$\pm$0.18 &$1.11$ &1062.25$\pm$0.57 & -0.43$\pm$0.15 & 41.12$\pm$0.15 \\
          J1114+1939$^*$ &6.07$\pm$0.12 & 3.31$\pm$0.90 & 626$\pm$27 & -0.64$\pm$0.18 & 40.45$\pm$0.21 & $1.97$ & 613.75$\pm$0.41 & -1.01$\pm$0.24 & 40.06$\pm$0.24\\
          J1116+2200 & 5.93$\pm$0.03 & 3.27$\pm$0.04 & 485$\pm$2 & -0.89$\pm$0.03 & 39.98$\pm$0.03 & $1.51$ &525.39$\pm$0.32 & -0.96$\pm$0.03 & 39.99$\pm$0.03\\
          J1316+1753 & 6.11$\pm$0.03 & 2.34$\pm$0.15 & 930$\pm$18 & -0.28$\pm$0.06 & 41.16$\pm$0.08 & $1.57$ &864.87$\pm$0.51& -0.64$\pm$0.04 & 40.73$\pm$0.04\\
          J1347+1217$^*$ &6.21$\pm$1.15 & 1.74$\pm$3.00 & 2466$\pm$333 & 0.37$\pm$1.91 & 42.65$\pm$2.01 & $1.31$ & 2535.93$\pm$1.99 & 0.10$\pm$1.10 & 42.40$\pm$1.10\\
          J1356+1026 & 6.38$\pm$0.06 & 5.41$\pm$0.32 & 741$\pm$12 & -0.48$\pm$0.08 & 40.76$\pm$0.09 & $1.33$ & 794.17$\pm$0.24& -0.50$\pm$0.09 & 40.80$\pm$0.09\\
          J1430+1339 & 6.51$\pm$0.06 & 1.18$\pm$0.05 & $774\pm20$ & 0.34$\pm$0.06 & 41.62$\pm$0.08 & $0.96$ & 840.41$\pm$0.48& 0.01$\pm$0.07 &  41.35$\pm$0.07\\
          J1518+1403 & 5.53$\pm$0.02 & 2.08$\pm$0.07 & 496$\pm$2 & -1.08$\pm$0.04 & 39.81$\pm$0.04 & $1.47$ &496.34$\pm$0.33& -1.38$\pm$0.03 & 39.51$\pm$0.03\\
          \hline
    \end{tabular}
    \tablefoot{
    (1) Galaxy name; (2) Mass of the outflowing gas; (3)-(6) {\em global} outflow radius, velocity, mass rate, and kinetic power; (7)-(10) same as (3)-(6) for the {\em peak properties}. 
    The values displayed here use the [Ar\,{\sc iv}] 4711,4740 derived electron densities to estimate the mass and derived quantities.
    $^*$ The outflow properties in these galaxies are derived assuming $n_e=10^3$\cmm~as the [Ar\,{\sc iv}] lines are not detected (see Sec.~\ref{sec:density}.)}
    \label{tab:placeholder}
\end{table*}

\section{Discussion}
\label{sec:discussion}
\subsection{Outflowing gas properties in context}
\label{sec:out_context}
In Table\,\ref{tab:out_props}, we present radius, velocity, mass, mass outflow rate, and kinetic powers derived using the two methods, {global} and {peak} (see Sec.\ref{sec:global} and \ref{sec:peak}), and considering the densities estimated from the [Ar\,{\sc iv}] lines. The mass outflow rates and powers assumed by $n_e=10^3$\cmm, used for comparison with the results in the literature, are presented in {Table\,\ref{tab:out_103}}. For the two galaxies where[Ar\,{\sc iv}] is not detected, we include the values measured using $n_e=10^3$\cmm~in Table\,\ref{tab:out_props}. 

{When adopting the electron densities derived from [Ar\:{\sc iv}], we obtain {global} masses, mass outflow rates, and kinetic powers in the ranges $5.25<{\log{M}_{\rm out}^{\rm g}/{\rm M_{\odot}}}
<7.55$, $0.03<\dot{M}_{\rm out}^{\rm g}/{\rm M_{\odot}~yr^{-1}}<4.17$, 
and $39.25<\log\dot{E}_{\rm kin}^{\rm g}/{\rm erg~s^{-1}}<41.91$. 
By assuming $n_e=10^3$\cmm\ we obtain $6.07<\log{{M}_{\rm out}^{\rm g}/{\rm M_{\odot}}}<7.3$, $0.23<{\dot{M}_{\rm out}^{\rm g}/{\rm M_{\odot}~ yr^{-1}}}<11.48$ and $40.55<\log\dot{E}_{\rm kin}^{\rm g}/{\rm erg~s^{-1}}<42.25$, which are systematically larger than the values derived using the [Ar\:{\sc iv}] density. A similar comparison can be made when looking at the same adopted density but with the outflow properties calculated using different methods. Figure~\ref{fig:rates_lbol} illustrates this behaviour, where the filled circles are obtained adopting the $10^3$\cmm~density and the open points the [Ar\:{\sc iv}] density, and the different colours correspond to the {global} and {peak} values (see Sect.~\ref{sec:origin_outs} for more details).
{\citet{ulivi24} measured mass outflow rates that are approximately one order of magnitude higher than the values we obtain here for four galaxies in our sample. We attribute this discrepancy mainly to the lower $n_e$ they report for three of the galaxies, derived from the [S{\sc ii}] doublet. These authors also find larger mass outflow rates and kinetic powers.} An interesting case of close agreement is J1010+1413, which has a similar outflow velocity, mass loss rate, and powers when we assume $n_e=10^3$\cmm, 
despite the authors adopting different prescriptions for the mass calculation and an electron density of $n_e=360$\,\cmm~estimated from the [S\:{\sc ii}] lines. \citet{bessiere24} obtained the mass outflow rates and kinetic powers for a sample of 48 type 2 local ($z<0.2$) QSOs using SDSS spectra. They obtained electron densities, which are in close agreement with ours, via the TR method (see Sec.\:\ref{sec:density}). The mass outflow rates and powers derived by these authors fall within the range we measured here, although the mass outflow rates tend to be higher. \citet{speranza24} conducted a spatially resolved study of a subsample of five QSOs from \citet{bessiere24}, three of which overlap with our sample, using GTC-MEGARA. These authors adopted electron densities from the TR method and also obtained values that were in good agreement with our measurements. 
\citet{cezar26} obtained a similar range of mass and mass outflow rates from the ionised gas for a sample of six type 2 QSOs observed with the long slit spectograph GTC-EMIR (see the discussion on the densities adopted in Sec.\ref{sec:density}).
Such matches, even with different mass and velocity prescriptions, highlight the uncertainties inherent in the calculations of outflow properties, with one of the main contributors being the electron density.

We also emphasize that even though the mass outflow rates and powers are not always in agreement with the literature values at the individual source level, 
as mentioned above, they agree with other works designed to study type 2 quasars, and are on the order of magnitude of the results in samples of $z\sim0.3$ luminous ($L_{\rm bol}\sim10^{46}$erg\,s$^{-1}$) quasars \citep{bruno21, hervella23}, and with cosmic noon sources at similar luminosities \citep{kakkad20}. For luminous $L_{\rm bol}\sim10^{45-46}$ at $z>3$, \citet{bertola25, venturi26} also find higher mass outflow rates and powers, reaching up to $\sim500$\:M$_{\odot}$yr$^{-1}$ and $\sim10^{43}$erg\:s$^{-1}$, derived by adopting a common $n_e=10^3$\cmm throughout the whole sample.
The values we derive here are also higher than those derived for a sample of local Seyfert galaxies ($\log L_{\rm bol}\sim 41-45$\:erg\:s$^{-1}$) using rest-frame optical and near-IR observations \citep{ruschel-dutra21, rogemar_agnifskin}, highlighting how the outflows carry more mass and energy when driven by more luminous AGN \citep[e.g.][]{fiore17}.

\begin{figure}
    \centering
    \includegraphics[width=0.37\textwidth, trim= 0.2cm 1cm 1.2cm 2cm, clip=True]{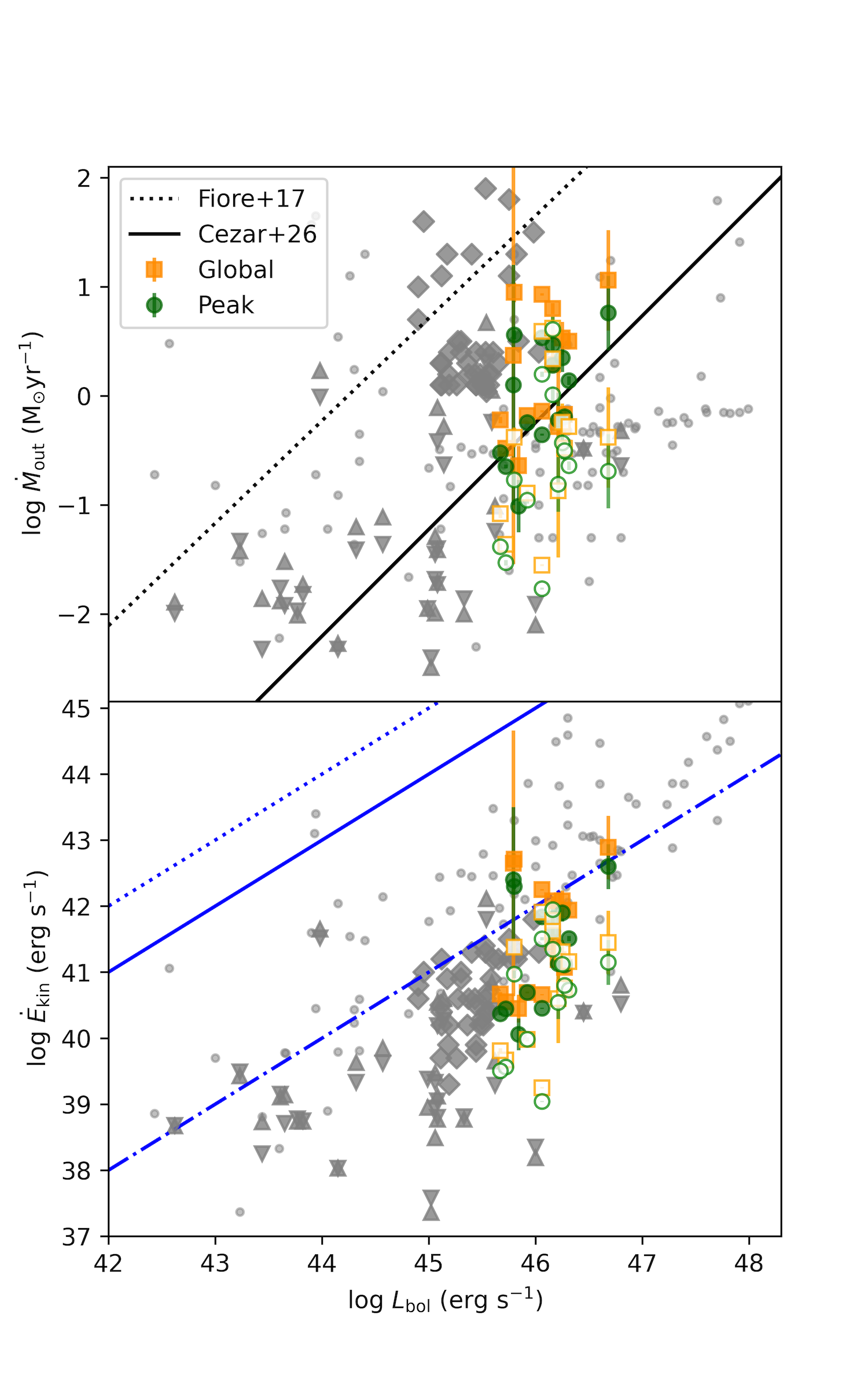}
    \caption{Mass outflow rates and kinetic power versus AGN $L_{\rm bol}$ on top and bottom panels, respectively. The filled symbols correspond to the values obtained assuming $n_e=10^3$\cmm\ and the open symbols to the ones obtained with the [Ar\:{\sc iv}] $n_e$. The grey dots are values from the literature, and up and down triangles from \citet[][and references therein]{rogemar_agnifskin} indicating the {\rm global} and {\rm peak} values in that work, respectively. The diamonds indicate the values from the QSO2s from \citet{bessiere24}.
    The dotted and black lines indicate the relation derived by \citet{fiore17} and \citet{cezar26}, respectively. The blue lines on the bottom plot correspond to $\dot{E}_{\rm kin}=L_{\rm bol}$ (dotted), $\dot{E}_{\rm kin}=0.1\times L_{\rm bol}$ (solid), and $\dot{E}_{\rm kin}=0.0001\times L_{\rm bol}$ (dot-dashed).}
    \label{fig:rates_lbol}
\end{figure}

\subsection{Outflow radial profiles}
Figure \ref{fig:vel_mass_prof} presents the radial profiles of the velocity and mass outflow rate. As a general trend, the mass outflow rate decreases from the centre towards the outer regions of the galaxies. We also observe the same behaviour in the mass radial profiles (not shown here). Simulations from \citet{costa18} showed that the clouds of cool emitting gas are eventually destroyed or slowed down in the halo. We interpret that our observations are consistent with  i) the destruction of \oiii~emitting clouds; ii) differences in the mass-to-light ratios used to estimate the gas masses, or even iii) the fact that the AGN cannot excite the \oiii~at such large radii.

The velocity profiles are complex, suggesting acceleration of the outflow in different regions, different outflow episodes, or varying density along the outflow \citep{zubovas25}. The peak of the mass outflow rate coincides with the central radial bin. \citet{venturi23} also identified this decreasing mass outflow rate profile in a detailed analysis of the J1430+1339 MUSE data. However, the mass outflow rates determined by those authors are higher than the values measured here, mainly due to the choices of electron density estimators ([S\:{\sc ii}] in their case), their outflow velocity definition, and the use of the H$_{\alpha}$ flux to obtain the mass of the outflow (see Sec. \ref{sec:outflowcalc}).  The values obtained by these authors using the \oiii~flux to estimate the mass are also higher than ours.
This trend of decreasing mass outflow is also seen in local Seyferts \citep[e.g.][]{venturi18, revalski18, revalski21}, while others, with a more limited field of view and higher spatial resolution, did not identify this trend \citep[e.g.][]{Shimizu19,bruno21,rogemar_agnifskin}. 
We emphasize that the radius of the peak outflow is simply the radius at which the mass outflow rate peaks, which is smaller than the global radius (Table \ref{tab:out_props}).  Given our choice of the shell radii and the results from previous works \citep{venturi23}, we interpret the outflows to be accelerated in the innermost regions of the galaxies ($\lesssim1$kpc), not probed here due to the limited spatial resolution. \citet{marconcini25} identified this small-scale acceleration in a sample of local Seyfert galaxies, where they were able to probe regions $<1$\:kpc.

Some galaxies display distinct features on the mass outflow rate radial profiles: 
J1316+1753 has a discontinuity on the radial profile, which is due to the ionised blueshifted cloud to the south-east of the centre of the galaxy. \citet{girdhar22} also note the presence of this cloud in the MUSE data but they do not analyse its kinematics and point out that this cloud can be reminiscent of a previous outflow event \citep{lintott09, keel12, keel19}. J1356+1026 has [O{\sc iii}] flux associated with high $W_{80}$ in a small blob to the east of the tail that connects the main galaxy and its companion. In the radial profiles, this region corresponds to the points between 45 and 50\,kpc. J1356+1026 is known to host powerful quasar-driven outflows \citep{greene12,somalwar20} {and is one of the most extended high-ionisation emitting regions \citep{bianchin26}}, so this feature could be associated with a previous outflow event as in J1316+1753.

\subsection{Origin of the outflows and influence on the host galaxy}
\label{sec:origin_outs}

In Figure\,\ref{fig:rates_lbol}, we present a comparison of our mass outflow rates and kinetic power with the literature. We adopt $n_e=10^3$\,\cmm as the standard electron density for our measurements (filled orange and green symbols) and use the compilation from \citet{rogemar_agnifskin} with scaled values to this common density.  \citet{rogemar_agnifskin} scaled the values presented in \citet{fiore17} and measurements from spatially resolved studies of QSOs \citep{bruno21, kakkad20, vayner21} and Seyfert galaxies  \citep{M1066KIN, rogemarN7582, schnorr-muller14, schnorr-muller16,rogemar_m1157, rogemar_n5929, muller-sanchez11, Shimizu19, Diniz19, barbosa14,rogemar_cygnus, couto20, rogemar_n5643, bianchin22}. These values are represented as grey dots in Fig.\ref{fig:rates_lbol}. Our comparison includes the mass outflow rates and kinetic powers calculated in \citet{rogemar_agnifskin} as these authors use a method similar to ours for analysing the outflows for a sample of nearby active galaxies  selected with hard X-rays (grey triangles in the plots). We also include the results obtained by \citet{bessiere24} for a sample of $z<0.14$ type 2 QSOs observed by SDSS where the densities were measured using the TR method (grey diamonds). We obtain the bolometric luminosity ($L_{\rm bol}$) for our data using the $L_{\rm [O III]}$ values SDSS measurements (Tab. \ref{tab:obs}) and the correction from \citep{heckman04}, which tends to result in higher $L_{\rm bol}$ than if estimated from other methods \citep[see][]{jarvis21}.

The mass outflow rates and kinetic powers are in agreement with the values obtained in the literature for galaxies at similar luminosities (see Sect.\ref{sec:out_context}), but with the values of $\dot{M}_{\rm out}$ from \citet{bessiere24} consistently higher than ours. However, our measurements and the scaled values from the literature are below the correlation found in \citet[dashed line in the top panel of Fig.~\ref{fig:rates_lbol}]{fiore17}. Our measurements show a greater agreement with the relation obtained by \citet{cezar26}, where they fit $\dot{M}_{\rm out}$ versus $L_{\rm bol}$ in a sample of type 2 QSOs.

Fig.~\ref{fig:rates_lbol} also shows the  $\dot{M}_{\rm out}$ and $\dot{E}_{\rm kin}$ values obtained using the [Ar\:{\sc iv}] densities. Our measured densities tend to produce mass outflow rates and kinetic powers $\sim 1$dex lower than with $n_e=10^3$~\cmm. The densities derived using different observational tracers, or assumed uniform values, might produce overestimated outflow masses and therefore $\dot{M}_{\rm out}$ and $\dot{E}_{\rm kin}$ by a few orders of magnitude \citep[see][and references therein]{almeida26}. However, high-density gas (which we trace here) might be a small fraction of the total mass that is outflowing (dominated by lower density and slower material), as pointed out in \citet{ward24, almeida26}.

We also estimate the fraction of gas in the outflows given by $M_{\rm out}/M_{\rm ion}$, where $M_{\rm out}$ is the sum of the mass in the spaxels with $W_{80}$ above the threshold present in Table \ref{tab:dens}, and $M_{\rm ion}$ corresponds to the total mass of the gas emitting \oiii, calculated according to Eq.\:\ref{eq:moiii}.
We obtain fractions of gas in the outflow in the range of 60-100\%. Such high values can be explained by the fact that we adopt the total \oiiir~line flux to estimate $M_{\rm out}$, rather than the perhaps more realistic scenario where only the broad component is considered as outflow. Thus, our results are much larger than the inferences from previous works. \citet{ulivi24} measured 25 to 60\% of the gas in the outflows for J1000+1242,  J1010+1413, J1010+0612, J1100+0846. For local Seyfert galaxies in which ionised outflows, traced by  Br$\gamma$ emission, are observed, fractions of 42--84\% \citep{bianchin22}, and up to 60\% are observed \citep{rogemar_agnifskin}. However, if instead of adopting the thresholds mentioned above, we use the fixed limit of $W_{80}>600$\,\kms, we obtain 36-100\% of the gas in the outflows, and two galaxies are considered to have no outflowing gas: J1055+1102 and J1518+1403. These two sources have the lowest outflow velocities (Table \ref{tab:out_props}); radio emission from the first source was recently associated with an AGN origin \citep{njeri26} while for the second, radio emission is not known to be associated with an AGN \citep{jarvis21}. Thus, the entire mass of the \oiii\ being carried in the outflows for these two galaxies might be an oversimplification and our method might not be applicable in such cases.

The blue lines in the bottom panel of Figure~\ref{fig:rates_lbol} represent the different fractions of $\dot{E}_{\rm kin}$ with respect to $L_{\rm bol}$. The dot-dashed line corresponds to $\dot{E}_{\rm kin}=0.01\% L_{\rm bol}$ and is the most consistent with our measurements. This result is a direct account of the coupling efficiencies of the outflow ($\varepsilon_f=\dot{E}_{\rm kin}/L_{\rm bol}$) presented in Table~\ref{tab:fracs}. In theoretical models, coupling efficiency, the fraction of energy released by the AGN that couples with the interstellar medium (ISM), is predicted to be as small as $0.5$\,\% of the AGN bolometric luminosity \citep{hopkins_elvis10} or requires higher values of 5\%~\citep[e.g.][]{dimatteo05,zubovas18}. 
The coupling efficiency encompasses all forms of energy released by the AGN, from the kinetic power of gas outflows in its multiple phases to the gas heating by the accretion disk radiation field. As noted in \citet[see also \citet{ward24}]{harrison18}, a direct comparison between the kinetic coupling efficiency and the coupling efficiencies derived from the models is not straightforward. 
We are not estimating the total contribution to the energy injected by the AGN into the ISM, only the kinetic part of it. Thus, the low kinetic coupling efficiencies 
presented in Table\,\ref{tab:fracs} are not unexpected. In addition, the equation adopted for estimating the kinetic power (Eq.\,\ref{eq:ekin}) does not include the velocity dispersion term of $\frac{3}{2}\sigma^2$. Even using $W_{80}$ as an estimator of the outflow velocity, some turbulent contribution to the energy input into the ISM might not be accounted for. 
Another caveat of our study is the limitation to one gas phase, when, in fact, the molecular gas accounts for most of the gas mass in galaxies \citep[e.g.][]{fiore17, veilleux20, fleutsch21, harrison24} and the outflows in this colder phase, as traced by the CO transitions observed with ALMA, can have higher powers and therefore greater efficiencies than the ionised phase \citep[e.g.][]{feruglio10,ramos-almeida22, audibert25}. Despite these results from observations, simulation work shows that although the cold phase might carry most of the mass, it is actually not likely to dominate the energy content \citep{ward24}. Thus, all kinetic powers and efficiencies obtained in this work are merely lower limits to the total energy injected by the AGN into the ISM. Also, the \oiii~gas we trace here is in a relatively high-density environment and, therefore, can be cold, and carry a limited fraction of the total energy according to simulation predictions \citep{ward24, almeida26}.

\begin{table}
    \centering
    \caption{Outflow gas fraction and kinetic efficiencies.}
    \begin{tabular}{cccc}
    \hline\hline
    (1) & (2) & (3) &(4)\\
    Galaxy & $M_{\rm out}/M_{\rm ion}$& $\log{\varepsilon_{f{\rm [Ar\,IV]}}}$ & $\log{\varepsilon_{f10^3}}$\\
    \hline
J0945+1737 & 0.94 & -5.61 $\pm$ 0.67 & -5.02 $\pm$ 0.67\\
J0958+1439 & 0.99 & -4.15 $\pm$ 0.02 & -3.82 $\pm$ 0.01\\
J1000+1242 & 0.96 & -4.39 $\pm$ 0.12 & -4.72 $\pm$ 0.03\\
J1010+0612 & 1.0 & -4.43 $\pm$ 0.10 & -3.10 $\pm$ 0.09\\
J1010+1413 & 0.84 & -5.24 $\pm$ 0.48 & -3.80 $\pm$ 0.48\\
J1016+0028 & 0.88 & -6.05 $\pm$ 0.02 & -5.17 $\pm$ 0.01\\
J1055+1102 & 1.0 & -6.81 $\pm$ 0.01 & -5.40 $\pm$ 0.01\\
J1100+0846 & 1.0 & -4.95 $\pm$ 0.18 & -4.17 $\pm$ 0.16\\
J1114+1939 & 0.96 & $-$  & -5.40 $\pm$ 0.21\\
J1116+2200 & 1.0 & -5.94 $\pm$ 0.03 & -5.23 $\pm$ 0.01\\
J1316+1753 & 0.97 & -5.16 $\pm$ 0.08 & -4.38 $\pm$ 0.08\\
J1347+1217 & 1.0 & $-$ & -3.15 $\pm$ 2.01\\
J1356+1026 & 0.87 & -5.52 $\pm$ 0.09 & -5.20 $\pm$ 0.09\\
J1430+1339 & 0.63 & -4.55 $\pm$ 0.08 & -4.09 $\pm$ 0.07\\
J1518+1403 & 0.98 & -5.87 $\pm$ 0.04 & -5.01 $\pm$ 0.03\\
\hline
    \end{tabular}
\tablefoot{(1) Galaxy name; (2) Fraction of outflowing ionised gas; (3)-(4) Kinetic efficiencies considering the [Ar\:{\sc iv}] and $10^3$\cmm~densities and adopting the {\em global} measurements in Tables \ref{tab:out_props} and \ref{tab:out_103}.}
    \label{tab:fracs}
\end{table}

\subsection{The relation between electron density and outflow velocity}

Figure \ref{fig:dens_vout} shows $n_e$ based on the [Ar\:{\sc iv}] lines that we measure in this work versus $V_{\rm out}$ for the global and peak\textit{} methods. The shaded grey area corresponds to galaxies with $V_{\rm out}<630$\:\kms, where there is a clear scatter of the electron density values. When considering only the values with velocities higher than this threshold, there is a clear positive correlation %
between these two quantities. We fit a linear correlation, indicated by the dashed line in Fig.~\ref{fig:dens_vout} to the combined {global} and peak} data points with $V_{\rm out}>630$\:\kms. 
We excluded the four galaxies at the lower $V_{\rm out}$ end from the linear fit, as indicated by the grey region in Fig.\:\ref{fig:dens_vout}. This cut is motivated mainly because two of the four are not outflow-dominated according to the $W_{80}>600$~\kms~ criterion (J1055+1102 and J1518+1403, see Sect.~\ref{sec:origin_outs}) or most of the disturbed kinematics might be associated with the merger (J1016+0028) or have one of the lowest outflow velocity and no AGN origin to its radio emission (see Appendix \ref{app:maps}). If we consider only the {peak} or {global} velocities above the $V_{\rm out}$ threshold, the correlation is still present but weaker than the one with the combined data points.

\begin{figure}[h!]
    \centering
\includegraphics[width=0.87\linewidth]{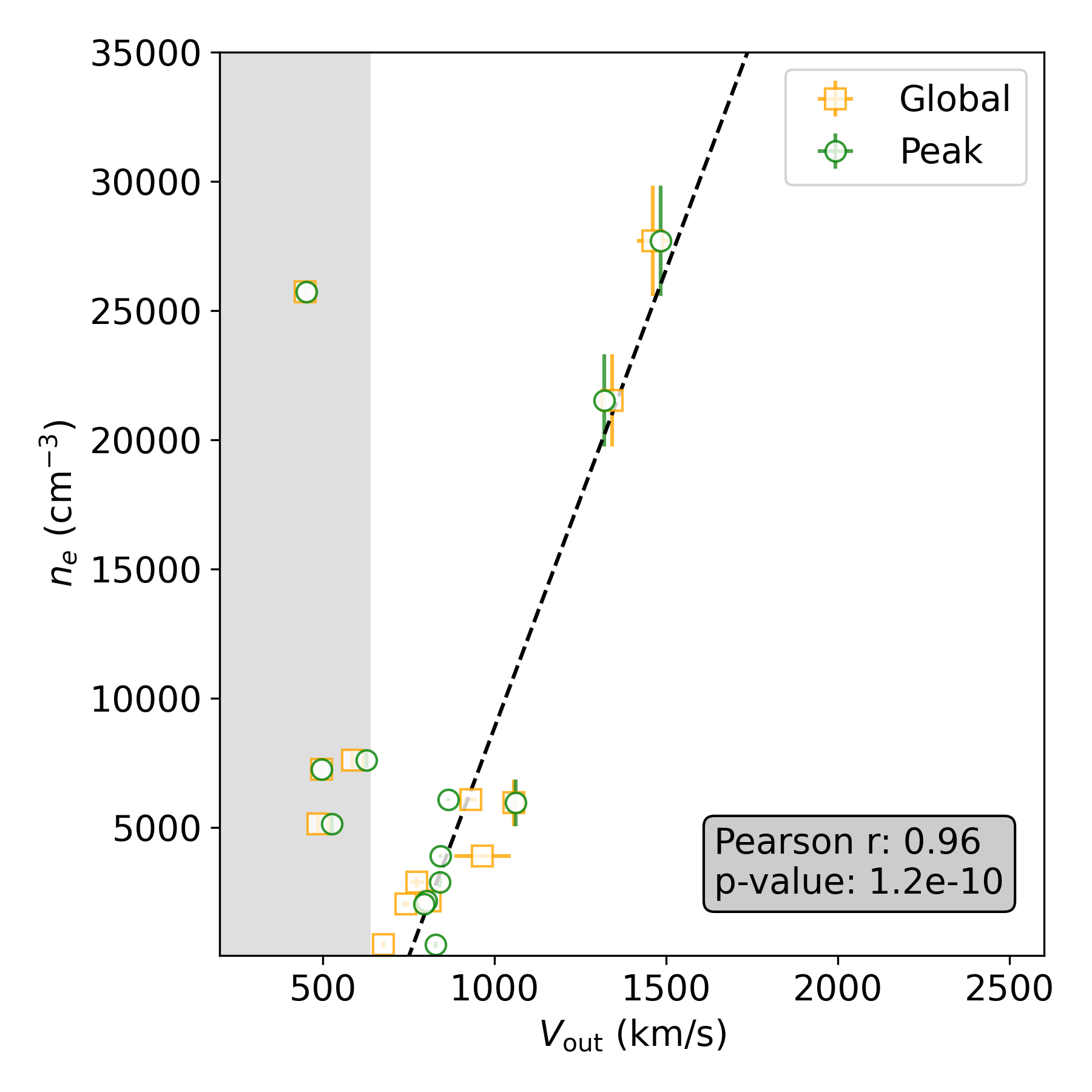}
    \caption{Electron density derived from the [Ar\:{\sc iv}] line doublet versus outflow velocity derived using the two different methods. There is a clear positive correlation between the two quantities. 
    The dashed line corresponds to the linear fit to the data, and the grey region represents the data points excluded from it, which have $V_{\rm out}<630$\:\kms.}
    \label{fig:dens_vout}
\end{figure}

In recent simulation work, \citet{almeida26} found that the gas density ($n_{\rm H}$) shows a positive correlation with $L_{\rm AGN}$ for simulated galaxies with $43<\log L_{\rm AGN}/{\rm erg~s^{-1}}<47$. Since $n_e$ is often a direct proxy for $n_H$ and $L_{\rm bol}\approx L_{\rm AGN}$, we tested whether this correlation is also present in our sample. 
Probably due to the small range of luminosities covered ($45.8<\log L_{\rm bol} ({\rm erg~s^{-1}})<46.7$), we do not find a correlation between $n_e$ and $L_{\rm bol}$. Within this range of luminosities, \citet{almeida26} observe the largest scatter in their relations when considering all the outflowing cool gas clouds ($T\sim10^4$\:K)  in their simulations.

The underlying assumption in \citet{almeida26} simulations is that wind power scales with $L_{\rm AGN}$, which we also see in our data and in the overall observational results (see Fig.~\ref{fig:rates_lbol} and Sec. \ref{sec:origin_outs}). \citet{costa20, ward24, almeida26} showed that the outflow velocity, for example  traced by gas at $T\sim10^4$K (which is the typical \oiii~temperature), scales with wind power and therefore with $L_{\rm AGN}$. Therefore, although indirect, the correlation of $n_e$ and $V_{\rm out}$ we find here is consistent with hydrodynamical simulations of AGN-driven outflows propagating through the ISM.

A possible physical interpretation of the correlation we observe may be related to a stronger compression of the ionised gas clouds (higher $n_e$ should imply higher $n_{\rm H}$) induced by stronger AGN-driven outflows. A similar effect was observed in the simulations, but at much smaller outflow velocities \citep[$V_{\rm out}<400$~\kms]{lauzikas24}. A deeper investigation of the implications and physical origins of this relation, as well as the expansion of the sample to check if it is still present in other $L_{\rm AGN}$ is beyond the scope of this paper.

\section{Conclusions}
\label{sec:conclusions}
We studied a sample of 18 type 2 QSOs observed with VLT/MUSE at spatial resolutions $1-3$\,kpc with a FoV of tens of kiloparsec that are part of the QFeedS project. All galaxies where the \oiiir~line can be observed (15 galaxies) present disturbed kinematics based on our criteria of spaxels with $W_{80}$ greater than the value of the narrow component measured at the nucleus (Sec.\:\ref{sec:outflowcalc}). We assume these motions are due to AGN-driven outflows and measure their related properties. We also obtain integrated ionised gas properties, whenever possible, for the entire sample. The main conclusions from our work are summarised below. 

\begin{itemize}

\item The highest $W_{80}$ values and the radio peaks and/or hot spots are co-spatial in most galaxies (Fig.\:\ref{fig:w80+radio}). This is also the case even when the $W_{80}$ and the radio jet are displaced from the peak of the \oiiir~emission. We note that this is simply a qualitative assessment, but informs how the radio jet and the disturbed motions might be connected.

\item The stellar continuum and emission-line flux distributions are distinct (see Figs.\:\ref{fig:maps_J0945}-\ref{fig:maps_J1518}). In some cases, the gas component shows signatures of {disruptions}
and/or of gas outflows, whereas the continuum traces the 
stellar bulges and disks.

\item Electron densities are derived from the [Ar\,{\sc iv}]$\lambda\lambda4711,4740$ emission lines, detected only in the integrated spectra ($r=1.5$\arcsec), which produce  values in the range of $\approx 500$--$27\,700\,\mathrm{cm^{-3}}$, with a median of $\approx 5900\,\mathrm{cm^{-3}}$ (see Table~\ref{tab:dens}). These densities lead to a reduction of approximately $\sim 1$ dex in both mass outflow rates and kinetic powers relative to the commonly assumed value of $10^3\,\mathrm{cm^{-3}}$ for all galaxies (see Fig.~\ref{fig:rates_lbol}).

\item The peak of the mass outflow is concentrated in the inner 2\,kpc with a general trend of decreasing radial profiles (Fig.\ref{fig:vel_mass_prof}). 
The velocity and mass profiles indicate that disturbed motions extend up to tens of kiloparsec, although at such distances the presence of \oiii-emitting gas might not be connected to the outflow. The general decreasing trend on the profiles might come from the destruction of the clouds at larger radii or simply because the radiation from AGN is not capable of ionising the gas at those distances.

\item A tight positive correlation is observed between $n_e$, obtained from the [Ar{\sc iv}] doublet, and $V_{\rm out}$ for $V_{\rm out} > 600\,\mathrm{km\,s^{-1}}$ (see Fig.~\ref{fig:dens_vout}). Although an exact prediction is not available, this trend is likely consistent with recent simulations of quasar outflows, which show correlations between $L_{\rm bol}$ and both the density and $V_{\rm out}$. This behaviour may be linked to the compression of \oiii-emitting clouds at higher outflow velocities. A deeper analysis is needed to understand whether such a scenario is physically plausible and if this correlation still holds in other samples of galaxies with AGN-driven outflows.

The results presented in this paper highlight how 
{density measurements obtained with different methods and/or tracers} impact the derived outflow-related quantities by at least an order of magnitude. The relation between $n_e$ and $V_{\rm out}$, obtained here for the first time, should be validated in samples that encompass galaxies with a wider range of $L_{\rm bol}$. Such tests would also make it possible to verify the relation between $n_e$ and $L_{\rm bol}$ predicted by simulations \citep{almeida26}, and to assess whether there are also empirical correlations between other quantities. Moreover, a refinement of these correlations may provide valuable observational constraints for realistic simulations.

\end{itemize}

\begin{acknowledgements}
Based on observations collected at the European Southern Observatory under ESO programmes O103B-0071, O104.B-0476 and O102.B-0107. This research made use of Astropy (\url{http://www.astropy.org}) a community-developed core Python package for Astronomy \citep{astropy:2013, astropy:2018}. 
M.B. thanks the financial support from Coordena\c c\~ao de Aperfei\c coamento de Pessoal de N\'ivel Superior - Brasil (CAPES) - Finance Code 001, from IAU - Gruber Foundation Fellowship program and from the Juan de La Cierva scholarship with reference JDC2023-052684-I, funded by MICIU/AEI/10.13039/501100011033 and FSE+. MB, CRA and PHC thank the Agencia Estatal de Investigaci\'on of the Ministerio de Ciencia, Innovaci\'on y Universidades (MCIU/AEI) under the grant ``Tracking active galactic nuclei feedback from parsec to kiloparsec scales'', with reference PID2022$-$141105NB$-$I00 and the European Regional Development Fund (ERDF). They also thank Agencia Estatal de Investigación (AEI) through the Severo Ochoa Centre of Excellence accreditation awarded to the Instituto de Astrofísica de Canarias, grant CEX2025-001609-S, funded by MICIU/AEI/10.13039/501100011033.
CMH acknowledge funding from an United Kingdom Research and Innovation grant (code: UKRI2730). 
RAR acknowledges the support from the Conselho Nacional de Desenvolvimento Científico e Tecnológico (CNPq; Projects 303450/2022-3, and 403398/2023-1), the Coordenação de Aperfeiçoamento de Pessoal de Nível Superior (CAPES; Project 88887.894973/2023-00), and Fundação de Amparo à Pesquisa do Estado do Rio Grande do Sul (FAPERGS; Project 25/2551-0002765-9). 
GV acknowledges financial support from the Italian National Institute for Astrophysics (INAF) under the IAF - Astrophysics Fellowships in Italy grant CUP C59J21034720001 (AD MAJORA) and from European Union’s HE ERC Starting Grant No. 101040227 - WINGS.
EPF is supported by the international Gemini Observatory, a program of NSF NOIRLab, which is managed by the Association of Universities for Research in Astronomy (AURA) under a cooperative agreement with the U.S. National Science Foundation, on behalf of the Gemini partnership of Argentina, Brazil, Canada, Chile, the Republic of Korea, and the United States of America.
This work was performed in part at the Aspen Center for Physics, which is supported by National Science Foundation grant PHY-2210452.
\end{acknowledgements}

\bibliographystyle{bibtex/aa.bst}
\bibliography{refs.bib}

@ARTICLE{xu99,
       author = {{Xu}, Chun and {Livio}, Mario and {Baum}, Stefi},
        title = "{Radio-loud and Radio-quiet Active Galactic Nuclei}",
      journal = {\aj},
         year = 1999,
        month = sep,
       volume = {118},
       number = {3},
        pages = {1169-1176},
          doi = {10.1086/301007},
archivePrefix = {arXiv},
       eprint = {astro-ph/9905322},
 primaryClass = {astro-ph},
       adsurl = {https://ui.adsabs.harvard.edu/abs/1999AJ....118.1169X}
}

@ARTICLE{almeida26,
       author = {{Almeida}, Ivan and {Costa}, Tiago and {Harrison}, Chris M. and {Ward}, Samuel R.},
        title = "{Tracing AGN feedback power with cool/warm outflow densities: predictions and observational implications}",
      journal = {\mnras},
         year = 2026,
        month = mar,
       volume = {546},
       number = {4},
          eid = {stag231},
        pages = {stag231},
          doi = {10.1093/mnras/stag231},
archivePrefix = {arXiv},
       eprint = {2602.05954},
 primaryClass = {astro-ph.GA},
       adsurl = {https://ui.adsabs.harvard.edu/abs/2026MNRAS.546ag231A}
}

@ARTICLE{venturi23,
       author = {{Venturi}, G. and {Treister}, E. and {Finlez}, C. and {D'Ago}, G. and {Bauer}, F. and {Harrison}, C.~M. and {Ramos Almeida}, C. and {Revalski}, M. and {Ricci}, F. and {Sartori}, L.~F. and {Girdhar}, A. and {Keel}, W.~C. and {Tub{\'\i}n}, D.},
        title = "{Complex AGN feedback in the Teacup galaxy. A powerful ionised galactic outflow, jet-ISM interaction, and evidence for AGN-triggered star formation in a giant bubble}",
      journal = {\aap},
         year = 2023,
        month = oct,
       volume = {678},
          eid = {A127},
        pages = {A127},
          doi = {10.1051/0004-6361/202347375},
archivePrefix = {arXiv},
       eprint = {2309.02498},
 primaryClass = {astro-ph.GA},
       adsurl = {https://ui.adsabs.harvard.edu/abs/2023A&A...678A.127V}
}

@ARTICLE{revalski18,
       author = {{Revalski}, M. and {Crenshaw}, D.~M. and {Kraemer}, S.~B. and {Fischer}, T.~C. and {Schmitt}, H.~R. and {Machuca}, C.},
        title = "{Quantifying Feedback from Narrow Line Region Outflows in Nearby Active Galaxies. I. Spatially Resolved Mass Outflow Rates for the Seyfert 2 Galaxy Markarian 573}",
      journal = {\apj},
         year = 2018,
        month = mar,
       volume = {856},
       number = {1},
          eid = {46},
        pages = {46},
          doi = {10.3847/1538-4357/aab107},
archivePrefix = {arXiv},
       eprint = {1802.07734},
 primaryClass = {astro-ph.GA},
       adsurl = {https://ui.adsabs.harvard.edu/abs/2018ApJ...856...46R}
}

@ARTICLE{ramos-almeida22,
       author = {{Ramos Almeida}, C. and {Bischetti}, M. and {Garc{\'\i}a-Burillo}, S. and {Alonso-Herrero}, A. and {Audibert}, A. and {Cicone}, C. and {Feruglio}, C. and {Tadhunter}, C.~N. and {Pierce}, J.~C.~S. and {Pereira-Santaella}, M. and {Bessiere}, P.~S.},
        title = "{The diverse cold molecular gas contents, morphologies, and kinematics of type-2 quasars as seen by ALMA}",
      journal = {\aap},
         year = 2022,
        month = feb,
       volume = {658},
          eid = {A155},
        pages = {A155},
          doi = {10.1051/0004-6361/202141906},
archivePrefix = {arXiv},
       eprint = {2111.13578},
 primaryClass = {astro-ph.GA},
       adsurl = {https://ui.adsabs.harvard.edu/abs/2022A&A...658A.155R}
}

@ARTICLE{villar-martin23,
       author = {{Villar Mart{\'\i}n}, M. and {Castro-Rodr{\'\i}guez}, N. and {Pereira Santaella}, M. and {Lamperti}, I. and {Tadhunter}, C. and {Emonts}, B. and {Colina}, L. and {Alonso Herrero}, A. and {Cabrera-Lavers}, A. and {Bellocchi}, E.},
        title = "{Limited impact of jet-induced feedback in the multi-phase nuclear interstellar medium of 4C12.50}",
      journal = {\aap},
         year = 2023,
        month = may,
       volume = {673},
          eid = {A25},
        pages = {A25},
          doi = {10.1051/0004-6361/202245418},
archivePrefix = {arXiv},
       eprint = {2303.00291},
 primaryClass = {astro-ph.GA},
       adsurl = {https://ui.adsabs.harvard.edu/abs/2023A&A...673A..25V}
}

@ARTICLE{couto13,
       author = {{Couto}, Guilherme S. and {Storchi-Bergmann}, Thaisa and {Axon}, David J. and {Robinson}, Andrew and {Kharb}, Preeti and {Riffel}, Rogemar A.},
        title = "{Kinematics and excitation of the nuclear spiral in the active galaxy Arp 102B}",
      journal = {\mnras},
         year = 2013,
        month = nov,
       volume = {435},
       number = {4},
        pages = {2982-3000},
          doi = {10.1093/mnras/stt1491},
archivePrefix = {arXiv},
       eprint = {1308.1891},
 primaryClass = {astro-ph.CO},
       adsurl = {https://ui.adsabs.harvard.edu/abs/2013MNRAS.435.2982C}
}

@ARTICLE{lena15,
       author = {{Lena}, D. and {Robinson}, A. and {Storchi-Bergman}, T. and {Schnorr-M{\"u}ller}, A. and {Seelig}, T. and {Riffel}, R.~A. and {Nagar}, N.~M. and {Couto}, G.~S. and {Shadler}, L.},
        title = "{The Complex Gas Kinematics in the Nucleus of the Seyfert 2 Galaxy NGC 1386: Rotation, Outflows, and Inflows}",
      journal = {\apj},
         year = 2015,
        month = jun,
       volume = {806},
       number = {1},
          eid = {84},
        pages = {84},
          doi = {10.1088/0004-637X/806/1/84},
archivePrefix = {arXiv},
       eprint = {1504.05089},
 primaryClass = {astro-ph.GA},
       adsurl = {https://ui.adsabs.harvard.edu/abs/2015ApJ...806...84L}
}

@ARTICLE{finlez18,
       author = {{Finlez}, Carolina and {Nagar}, Neil M. and {Storchi-Bergmann}, Thaisa and {Schnorr-M{\"u}ller}, Allan and {Riffel}, Rogemar A. and {Lena}, Davide and {Mundell}, C.~G. and {Elvis}, Martin S.},
        title = "{The complex jet- and bar-perturbed kinematics in NGC 3393 as revealed with ALMA and GEMINI-GMOS/IFU}",
      journal = {\mnras},
         year = 2018,
        month = sep,
       volume = {479},
       number = {3},
        pages = {3892-3908},
          doi = {10.1093/mnras/sty1555},
archivePrefix = {arXiv},
       eprint = {1806.02756},
 primaryClass = {astro-ph.GA},
       adsurl = {https://ui.adsabs.harvard.edu/abs/2018MNRAS.479.3892F}
}

@ARTICLE{vayner24,
       author = {{Vayner}, Andrey and {Zakamska}, Nadia L. and {Ishikawa}, Yuzo and {Sankar}, Swetha and {Wylezalek}, Dominika and {Rupke}, David S.~N. and {Veilleux}, Sylvain and {Bertemes}, Caroline and {Barrera-Ballesteros}, Jorge K. and {Chen}, Hsiao-Wen and {Diachenko}, Nadiia and {Goulding}, Andy D. and {Greene}, Jenny E. and {Hainline}, Kevin N. and {Hamann}, Fred and {Heckman}, Timothy and {Johnson}, Sean D. and {Grace Lim}, Hui Xian and {Liu}, Weizhe and {Lutz}, Dieter and {L{\"u}tzgendorf}, Nora and {Mainieri}, Vincenzo and {McCrory}, Ryan and {Murphree}, Grey and {Nesvadba}, Nicole P.~H. and {Ogle}, Patrick and {Sturm}, Eckhard and {Whitesell}, Lillian},
        title = "{First Results from the JWST Early Release Science Program Q3D: Powerful Quasar-driven Galactic Scale Outflow at z = 3}",
      journal = {\apj},
         year = 2024,
        month = jan,
       volume = {960},
       number = {2},
          eid = {126},
        pages = {126},
          doi = {10.3847/1538-4357/ad0be9},
archivePrefix = {arXiv},
       eprint = {2307.13751},
 primaryClass = {astro-ph.GA},
       adsurl = {https://ui.adsabs.harvard.edu/abs/2024ApJ...960..126V}
}

@ARTICLE{mukherjee25,
       author = {{Mukherjee}, Dipanjan},
        title = "{Jet Feedback on kpc Scales: A Review}",
      journal = {Galaxies},
         year = 2025,
        month = sep,
       volume = {13},
       number = {5},
          eid = {102},
        pages = {102},
          doi = {10.3390/galaxies13050102},
archivePrefix = {arXiv},
       eprint = {2506.03888},
 primaryClass = {astro-ph.GA},
       adsurl = {https://ui.adsabs.harvard.edu/abs/2025Galax..13..102M}
}

@ARTICLE{riffel26,
       author = {{Riffel}, Rogemar A. and {Colina}, Luis and {Costa-Souza}, Jos{\'e} Henrique and {Mainieri}, Vincenzo and {Pereira Santaella}, Miguel and {Dors}, Oli L. and {Garc{\'\i}a-Bernete}, Ismael and {Alonso-Herrero}, Almudena and {Audibert}, Anelise and {Bellocchi}, Enrica and {Bunker}, Andrew J. and {Campbell}, Steph and {Combes}, Fran{\c{c}}oise and {Davies}, Richard I. and {D{\'\i}az-Santos}, Tanio and {Donnan}, Fergus R. and {Esposito}, Federico and {Garc{\'\i}a-Burillo}, Santiago and {Garc{\'\i}a-Lorenzo}, Bego{\~n}a and {Gonz{\'a}lez Mart{\'\i}n}, Omaira and {Haidar}, Houda and {Hicks}, Erin K.~S. and {Hoenig}, Sebastian F. and {Imanishi}, Masatoshi and {Labiano}, Alvaro and {Lopez-Rodriguez}, Enrique and {Packham}, Christopher and {Ramos Almeida}, Cristina and {Rigopoulou}, Dimitra and {Rosario}, David and {Souza-Oliveira}, Gabriel Luan and {Villar Mart{\'\i}n}, Montserrat and {Veenema}, Oscar and {Zhang}, Lulu},
        title = "{Impact of active galactic nuclei and nuclear star formation on the ISM turbulence of galaxies: Insights from JWST/MIRI spectroscopy}",
      journal = {\aap},
         year = 2026,
        month = jan,
       volume = {705},
          eid = {A59},
        pages = {A59},
          doi = {10.1051/0004-6361/202556775},
archivePrefix = {arXiv},
       eprint = {2510.02517},
 primaryClass = {astro-ph.GA},
       adsurl = {https://ui.adsabs.harvard.edu/abs/2026A&A...705A..59R}
}

@ARTICLE{njeri26,
       author = {{Njeri}, Ann and {Harrison}, Chris M. and {Kharb}, Preeti and {Alexander}, David M. and {Mainieri}, Vincenzo and {Circosta}, Chiara and {Fawcett}, Victoria A. and {Kakkad}, Darshan and {Mukherjee}, Dipanjan and {Molyneux}, Stephen and {Sasikumar}, Silpa},
        title = "{The Quasar Feedback Survey: revealing the importance of sensitive radio imaging for AGN identification deeper into the radio-quiet regime}",
      journal = {\mnras},
         year = 2026,
        month = mar,
       volume = {546},
       number = {4},
          eid = {stag097},
        pages = {stag097},
          doi = {10.1093/mnras/stag097},
archivePrefix = {arXiv},
       eprint = {2601.09218},
 primaryClass = {astro-ph.GA},
       adsurl = {https://ui.adsabs.harvard.edu/abs/2026MNRAS.546ag097N}
}

@ARTICLE{begelman06,
       author = {{Begelman}, Mitchell C. and {Volonteri}, Marta and {Rees}, Martin J.},
        title = "{Formation of supermassive black holes by direct collapse in pre-galactic haloes}",
      journal = {\mnras},
         year = 2006,
        month = jul,
       volume = {370},
       number = {1},
        pages = {289-298},
          doi = {10.1111/j.1365-2966.2006.10467.x},
archivePrefix = {arXiv},
       eprint = {astro-ph/0602363},
 primaryClass = {astro-ph},
       adsurl = {https://ui.adsabs.harvard.edu/abs/2006MNRAS.370..289B}
}

@ARTICLE{zubovas18,
       author = {{Zubovas}, Kastytis},
        title = "{AGN must be very efficient at powering outflows}",
      journal = {\mnras},
         year = 2018,
        month = sep,
       volume = {479},
       number = {3},
        pages = {3189-3196},
          doi = {10.1093/mnras/sty1679},
archivePrefix = {arXiv},
       eprint = {1806.08914},
 primaryClass = {astro-ph.GA},
       adsurl = {https://ui.adsabs.harvard.edu/abs/2018MNRAS.479.3189Z}
}

@ARTICLE{costa18,
       author = {{Costa}, Tiago and {Rosdahl}, Joakim and {Sijacki}, Debora and {Haehnelt}, Martin G.},
        title = "{Quenching star formation with quasar outflows launched by trapped IR radiation}",
      journal = {\mnras},
         year = 2018,
        month = sep,
       volume = {479},
       number = {2},
        pages = {2079-2111},
          doi = {10.1093/mnras/sty1514},
archivePrefix = {arXiv},
       eprint = {1709.08638},
 primaryClass = {astro-ph.GA},
       adsurl = {https://ui.adsabs.harvard.edu/abs/2018MNRAS.479.2079C}
}

@INPROCEEDINGS{bacon10,
       author = {{Bacon}, R. and {Accardo}, M. and {Adjali}, L. and {Anwand}, H. and {Bauer}, S. and {Biswas}, I. and {Blaizot}, J. and {Boudon}, D. and {Brau-Nogue}, S. and {Brinchmann}, J. and {Caillier}, P. and {Capoani}, L. and {Carollo}, C.~M. and {Contini}, T. and {Couderc}, P. and {Daguis{\'e}}, E. and {Deiries}, S. and {Delabre}, B. and {Dreizler}, S. and {Dubois}, J. and {Dupieux}, M. and {Dupuy}, C. and {Emsellem}, E. and {Fechner}, T. and {Fleischmann}, A. and {Fran{\c{c}}ois}, M. and {Gallou}, G. and {Gharsa}, T. and {Glindemann}, A. and {Gojak}, D. and {Guiderdoni}, B. and {Hansali}, G. and {Hahn}, T. and {Jarno}, A. and {Kelz}, A. and {Koehler}, C. and {Kosmalski}, J. and {Laurent}, F. and {Le Floch}, M. and {Lilly}, S.~J. and {Lizon}, J. -L. and {Loupias}, M. and {Manescau}, A. and {Monstein}, C. and {Nicklas}, H. and {Olaya}, J. -C. and {Pares}, L. and {Pasquini}, L. and {P{\'e}contal-Rousset}, A. and {Pell{\'o}}, R. and {Petit}, C. and {Popow}, E. and {Reiss}, R. and {Remillieux}, A. and {Renault}, E. and {Roth}, M. and {Rupprecht}, G. and {Serre}, D. and {Schaye}, J. and {Soucail}, G. and {Steinmetz}, M. and {Streicher}, O. and {Stuik}, R. and {Valentin}, H. and {Vernet}, J. and {Weilbacher}, P. and {Wisotzki}, L. and {Yerle}, N.},
        title = "{The MUSE second-generation VLT instrument}",
    booktitle = {Ground-based and Airborne Instrumentation for Astronomy III},
         year = 2010,
       editor = {{McLean}, Ian S. and {Ramsay}, Suzanne K. and {Takami}, Hideki},
       series = {Society of Photo-Optical Instrumentation Engineers (SPIE) Conference Series},
       volume = {7735},
        month = jul,
          eid = {773508},
        pages = {773508},
          doi = {10.1117/12.856027},
       adsurl = {https://ui.adsabs.harvard.edu/abs/2010SPIE.7735E..08B}
}

@MISC{qfitsview,
       author = {{Ott}, Thomas},
        title = "{QFitsView: FITS file viewer}",
 howpublished = {Astrophysics Source Code Library, record ascl:1210.019},
         year = 2012,
        month = oct,
          eid = {ascl:1210.019},
        pages = {ascl:1210.019},
archivePrefix = {ascl},
       eprint = {1210.019},
       adsurl = {https://ui.adsabs.harvard.edu/abs/2012ascl.soft10019O}
}

@ARTICLE{heckman04,
       author = {{Heckman}, Timothy M. and {Kauffmann}, Guinevere and {Brinchmann}, Jarle and {Charlot}, St{\'e}phane and {Tremonti}, Christy and {White}, Simon D.~M.},
        title = "{Present-Day Growth of Black Holes and Bulges: The Sloan Digital Sky Survey Perspective}",
      journal = {\apj},
         year = 2004,
        month = sep,
       volume = {613},
       number = {1},
        pages = {109-118},
          doi = {10.1086/422872},
archivePrefix = {arXiv},
       eprint = {astro-ph/0406218},
 primaryClass = {astro-ph},
       adsurl = {https://ui.adsabs.harvard.edu/abs/2004ApJ...613..109H}
}

@ARTICLE{bianchin22,
       author = {{Bianchin}, M. and {Riffel}, R.~A. and {Storchi-Bergmann}, T. and {Riffel}, R. and {Ruschel-Dutra}, D. and {Harrison}, C.~M. and {Dahmer-Hahn}, L.~G. and {Mainieri}, V. and {Sch{\"o}nell}, A.~J. and {Dametto}, N.~Z.},
        title = "{Gemini NIFS survey of feeding and feedback in nearby active galaxies - V. Molecular and ionized gas kinematics}",
      journal = {\mnras},
         year = 2022,
        month = feb,
       volume = {510},
       number = {1},
        pages = {639-657},
          doi = {10.1093/mnras/stab3468},
archivePrefix = {arXiv},
       eprint = {2111.09130},
 primaryClass = {astro-ph.GA},
       adsurl = {https://ui.adsabs.harvard.edu/abs/2022MNRAS.510..639B}
}

@ARTICLE{vayner21,
       author = {{Vayner}, Andrey and {Zakamska}, Nadia L. and {Riffel}, Rogemar A. and {Alexandroff}, Rachael and {Cosens}, Maren and {Hamann}, Fred and {Perrotta}, Serena and {Rupke}, David S.~N. and {Bergmann}, Thaisa Storchi and {Veilleux}, Sylvain and {Walth}, Greg and {Wright}, Shelley and {Wylezalek}, Dominika},
        title = "{Powerful winds in high-redshift obscured and red quasars}",
      journal = {\mnras},
         year = 2021,
        month = jul,
       volume = {504},
       number = {3},
        pages = {4445-4459},
          doi = {10.1093/mnras/stab1176},
archivePrefix = {arXiv},
       eprint = {2101.04688},
 primaryClass = {astro-ph.GA},
       adsurl = {https://ui.adsabs.harvard.edu/abs/2021MNRAS.504.4445V}
}

@ARTICLE{kakkad22,
       author = {{Kakkad}, D. and {Sani}, E. and {Rojas}, A.~F. and {Mallmann}, Nicolas D. and {Veilleux}, S. and {Bauer}, Franz E. and {Ricci}, F. and {Mushotzky}, R. and {Koss}, M. and {Ricci}, C. and {Treister}, E. and {Privon}, George C. and {Nguyen}, N. and {B{\"a}r}, R. and {Harrison}, F. and {Oh}, K. and {Powell}, M. and {Riffel}, R. and {Stern}, D. and {Trakhtenbrot}, B. and {Urry}, C.~M.},
        title = "{BASS XXXI: Outflow scaling relations in low redshift X-ray AGN host galaxies with MUSE}",
      journal = {\mnras},
         year = 2022,
        month = apr,
       volume = {511},
       number = {2},
        pages = {2105-2124},
          doi = {10.1093/mnras/stac103},
archivePrefix = {arXiv},
       eprint = {2201.04149},
 primaryClass = {astro-ph.GA},
       adsurl = {https://ui.adsabs.harvard.edu/abs/2022MNRAS.511.2105K}
}

@ARTICLE{storchi-bergmann18,
       author = {{Storchi-Bergmann}, T. and {Dall'Agnol de Oliveira}, B. and {Longo Micchi}, L.~F. and {Schmitt}, H.~R. and {Fischer}, T.~C. and {Kraemer}, S. and {Crenshaw}, M. and {Maksym}, P. and {Elvis}, M. and {Fabbiano}, G. and {Colina}, L.},
        title = "{Bipolar Ionization Cones in the Extended Narrow-line Region of Nearby QSO2s}",
      journal = {\apj},
         year = 2018,
        month = nov,
       volume = {868},
       number = {1},
          eid = {14},
        pages = {14},
          doi = {10.3847/1538-4357/aae7cd},
archivePrefix = {arXiv},
       eprint = {1810.06246},
 primaryClass = {astro-ph.GA},
       adsurl = {https://ui.adsabs.harvard.edu/abs/2018ApJ...868...14S}
}

@ARTICLE{fischer13,
       author = {{Fischer}, T.~C. and {Crenshaw}, D.~M. and {Kraemer}, S.~B. and {Schmitt}, H.~R.},
        title = "{Determining Inclinations of Active Galactic Nuclei via their Narrow-line Region Kinematics. I. Observational Results}",
      journal = {\apjs},
         year = 2013,
        month = nov,
       volume = {209},
       number = {1},
          eid = {1},
        pages = {1},
          doi = {10.1088/0067-0049/209/1/1},
archivePrefix = {arXiv},
       eprint = {1308.4129},
 primaryClass = {astro-ph.CO},
       adsurl = {https://ui.adsabs.harvard.edu/abs/2013ApJS..209....1F}
}

@ARTICLE{rogemar_n5643,
       author = {{Riffel}, Rogemar A. and {Hekatelyne}, C. and {Freitas}, Izabel C.},
        title = "{Outflows in the Seyfert 2 galaxy NGC 5643 traced by the [S III] emission}",
      journal = {\pasa},
         year = 2018,
        month = nov,
       volume = {35},
          eid = {e040},
        pages = {e040},
          doi = {10.1017/pasa.2018.31},
archivePrefix = {arXiv},
       eprint = {1807.02743},
 primaryClass = {astro-ph.GA},
       adsurl = {https://ui.adsabs.harvard.edu/abs/2018PASA...35...40R}
}

@ARTICLE{lutz20,
       author = {{Lutz}, D. and {Sturm}, E. and {Janssen}, A. and {Veilleux}, S. and {Aalto}, S. and {Cicone}, C. and {Contursi}, A. and {Davies}, R.~I. and {Feruglio}, C. and {Fischer}, J. and {Fluetsch}, A. and {Garcia-Burillo}, S. and {Genzel}, R. and {Gonz{\'a}lez-Alfonso}, E. and {Graci{\'a}-Carpio}, J. and {Herrera-Camus}, R. and {Maiolino}, R. and {Schruba}, A. and {Shimizu}, T. and {Sternberg}, A. and {Tacconi}, L.~J. and {Wei{\ss}}, A.},
        title = "{Molecular outflows in local galaxies: Method comparison and a role of intermittent AGN driving}",
      journal = {\aap},
         year = 2020,
        month = jan,
       volume = {633},
          eid = {A134},
        pages = {A134},
          doi = {10.1051/0004-6361/201936803},
archivePrefix = {arXiv},
       eprint = {1911.05608},
 primaryClass = {astro-ph.GA},
       adsurl = {https://ui.adsabs.harvard.edu/abs/2020A&A...633A.134L}
}

@ARTICLE{rogemar_cygnus,
       author = {{Riffel}, R.~A.},
        title = "{Powerful multiphase outflows in the central region of Cygnus A}",
      journal = {\mnras},
         year = 2021,
        month = sep,
       volume = {506},
       number = {2},
        pages = {2950-2962},
          doi = {10.1093/mnras/stab1877},
archivePrefix = {arXiv},
       eprint = {2106.15279},
 primaryClass = {astro-ph.GA},
       adsurl = {https://ui.adsabs.harvard.edu/abs/2021MNRAS.506.2950R}
}

@ARTICLE{harrison16,
       author = {{Harrison}, C.~M. and {Alexander}, D.~M. and {Mullaney}, J.~R. and {Stott}, J.~P. and {Swinbank}, A.~M. and {Arumugam}, V. and {Bauer}, F.~E. and {Bower}, R.~G. and {Bunker}, A.~J. and {Sharples}, R.~M.},
        title = "{The KMOS AGN Survey at High redshift (KASHz): the prevalence and drivers of ionized outflows in the host galaxies of X-ray AGN}",
      journal = {\mnras},
         year = 2016,
        month = feb,
       volume = {456},
       number = {2},
        pages = {1195-1220},
          doi = {10.1093/mnras/stv2727},
archivePrefix = {arXiv},
       eprint = {1511.00008},
 primaryClass = {astro-ph.GA},
       adsurl = {https://ui.adsabs.harvard.edu/abs/2016MNRAS.456.1195H}
}

@ARTICLE{rupke05,
       author = {{Rupke}, David S. and {Veilleux}, Sylvain and {Sanders}, D.~B.},
        title = "{Outflows in Active Galactic Nucleus/Starburst-Composite Ultraluminous Infrared Galaxies1,}",
      journal = {\apj},
         year = 2005,
        month = oct,
       volume = {632},
       number = {2},
        pages = {751-780},
          doi = {10.1086/444451},
archivePrefix = {arXiv},
       eprint = {astro-ph/0507037},
 primaryClass = {astro-ph},
       adsurl = {https://ui.adsabs.harvard.edu/abs/2005ApJ...632..751R}
}

@ARTICLE{genzel11,
       author = {{Genzel}, R. and {Newman}, S. and {Jones}, T. and {F{\"o}rster Schreiber}, N.~M. and {Shapiro}, K. and {Genel}, S. and {Lilly}, S.~J. and {Renzini}, A. and {Tacconi}, L.~J. and {Bouch{\'e}}, N. and {Burkert}, A. and {Cresci}, G. and {Buschkamp}, P. and {Carollo}, C.~M. and {Ceverino}, D. and {Davies}, R. and {Dekel}, A. and {Eisenhauer}, F. and {Hicks}, E. and {Kurk}, J. and {Lutz}, D. and {Mancini}, C. and {Naab}, T. and {Peng}, Y. and {Sternberg}, A. and {Vergani}, D. and {Zamorani}, G.},
        title = "{The Sins Survey of z \raisebox{-0.5ex}\textasciitilde 2 Galaxy Kinematics: Properties of the Giant Star-forming Clumps}",
      journal = {\apj},
         year = 2011,
        month = jun,
       volume = {733},
       number = {2},
          eid = {101},
        pages = {101},
          doi = {10.1088/0004-637X/733/2/101},
archivePrefix = {arXiv},
       eprint = {1011.5360},
 primaryClass = {astro-ph.CO},
       adsurl = {https://ui.adsabs.harvard.edu/abs/2011ApJ...733..101G}
}

@ARTICLE{carniani15,
       author = {{Carniani}, S. and {Marconi}, A. and {Maiolino}, R. and {Balmaverde}, B. and {Brusa}, M. and {Cano-D{\'\i}az}, M. and {Cicone}, C. and {Comastri}, A. and {Cresci}, G. and {Fiore}, F. and {Feruglio}, C. and {La Franca}, F. and {Mainieri}, V. and {Mannucci}, F. and {Nagao}, T. and {Netzer}, H. and {Piconcelli}, E. and {Risaliti}, G. and {Schneider}, R. and {Shemmer}, O.},
        title = "{Ionised outflows in z \raisebox{-0.5ex}\textasciitilde 2.4 quasar host galaxies}",
      journal = {\aap},
         year = 2015,
        month = aug,
       volume = {580},
          eid = {A102},
        pages = {A102},
          doi = {10.1051/0004-6361/201526557},
archivePrefix = {arXiv},
       eprint = {1506.03096},
 primaryClass = {astro-ph.GA},
       adsurl = {https://ui.adsabs.harvard.edu/abs/2015A&A...580A.102C}
}

@ARTICLE{rogemar_agnifskin,
       author = {{Riffel}, R.~A. and {Storchi-Bergmann}, T. and {Riffel}, R. and {Bianchin}, M. and {Zakamska}, N.~L. and {Ruschel-Dutra}, D. and {Bentz}, M.~C. and {Burtscher}, L. and {Crenshaw}, D.~M. and {Dahmer-Hahn}, L.~G. and {Dametto}, N.~Z. and {Davies}, R.~I. and {Diniz}, M.~R. and {Fischer}, T.~C. and {Harrison}, C.~M. and {Mainieri}, V. and {Revalski}, M. and {Rodriguez-Ardila}, A. and {Rosario}, D.~J. and {Sch{\"o}nell}, A.~J.},
        title = "{The AGNIFS survey: spatially resolved observations of hot molecular and ionized outflows in nearby active galaxies}",
      journal = {\mnras},
         year = 2023,
        month = may,
       volume = {521},
       number = {2},
        pages = {1832-1848},
          doi = {10.1093/mnras/stad599},
archivePrefix = {arXiv},
       eprint = {2302.11324},
 primaryClass = {astro-ph.GA},
       adsurl = {https://ui.adsabs.harvard.edu/abs/2023MNRAS.521.1832R}
}

@ARTICLE{greene12,
       author = {{Greene}, Jenny E. and {Zakamska}, Nadia L. and {Smith}, Paul S.},
        title = "{A Spectacular Outflow in an Obscured Quasar}",
      journal = {\apj},
         year = 2012,
        month = feb,
       volume = {746},
       number = {1},
          eid = {86},
        pages = {86},
          doi = {10.1088/0004-637X/746/1/86},
archivePrefix = {arXiv},
       eprint = {1112.3358},
 primaryClass = {astro-ph.CO},
       adsurl = {https://ui.adsabs.harvard.edu/abs/2012ApJ...746...86G}
}

@ARTICLE{somalwar20,
       author = {{Somalwar}, Jean and {Johnson}, Sean D. and {Stern}, Jonathan and {Goulding}, Andy D. and {Greene}, Jenny E. and {Zakamska}, Nadia L. and {Alexandroff}, Rachael M. and {Chen}, Hsiao-Wen},
        title = "{Spatially Resolved UV Diagnostics of AGN Feedback: Radiation Pressure Dominates in a Prototypical Quasar-driven Superwind}",
      journal = {\apjl},
         year = 2020,
        month = feb,
       volume = {890},
       number = {2},
          eid = {L28},
        pages = {L28},
          doi = {10.3847/2041-8213/ab733d},
archivePrefix = {arXiv},
       eprint = {2002.02454},
 primaryClass = {astro-ph.GA},
       adsurl = {https://ui.adsabs.harvard.edu/abs/2020ApJ...890L..28S}
}

@ARTICLE{lintott09,
       author = {{Lintott}, Chris J. and {Schawinski}, Kevin and {Keel}, William and {van Arkel}, Hanny and {Bennert}, Nicola and {Edmondson}, Edward and {Thomas}, Daniel and {Smith}, Daniel J.~B. and {Herbert}, Peter D. and {Jarvis}, Matt J. and {Virani}, Shanil and {Andreescu}, Dan and {Bamford}, Steven P. and {Land}, Kate and {Murray}, Phil and {Nichol}, Robert C. and {Raddick}, M. Jordan and {Slosar}, An{\v{z}}e and {Szalay}, Alex and {Vandenberg}, Jan},
        title = "{Galaxy Zoo: `Hanny's Voorwerp', a quasar light echo?}",
      journal = {\mnras},
         year = 2009,
        month = oct,
       volume = {399},
       number = {1},
        pages = {129-140},
          doi = {10.1111/j.1365-2966.2009.15299.x},
archivePrefix = {arXiv},
       eprint = {0906.5304},
 primaryClass = {astro-ph.CO},
       adsurl = {https://ui.adsabs.harvard.edu/abs/2009MNRAS.399..129L}
}

@ARTICLE{keel12,
       author = {{Keel}, William C. and {Lintott}, Chris J. and {Schawinski}, Kevin and {Bennert}, Vardha N. and {Thomas}, Daniel and {Manning}, Anna and {Chojnowski}, S. Drew and {van Arkel}, Hanny and {Lynn}, Stuart},
        title = "{The History and Environment of a Faded Quasar: Hubble Space Telescope Observations of Hanny's Voorwerp and IC 2497}",
      journal = {\aj},
         year = 2012,
        month = aug,
       volume = {144},
       number = {2},
          eid = {66},
        pages = {66},
          doi = {10.1088/0004-6256/144/2/66},
archivePrefix = {arXiv},
       eprint = {1206.3797},
 primaryClass = {astro-ph.CO},
       adsurl = {https://ui.adsabs.harvard.edu/abs/2012AJ....144...66K}
}

@ARTICLE{keel19,
       author = {{Keel}, William C. and {Bennert}, Vardha N. and {Pancoast}, Anna and {Harris}, Chelsea E. and {Nierenberg}, Anna and {Chojnowski}, S. Drew and {Moiseev}, Alexei V. and {Oparin}, Dmitry V. and {Lintott}, Chris J. and {Schawinski}, Kevin and {Mitchell}, Graham and {Cornen}, Claude},
        title = "{AGN photoionization of gas in companion galaxies as a probe of AGN radiation in time and direction}",
      journal = {\mnras},
         year = 2019,
        month = mar,
       volume = {483},
       number = {4},
        pages = {4847-4865},
          doi = {10.1093/mnras/sty3332},
archivePrefix = {arXiv},
       eprint = {1711.09936},
 primaryClass = {astro-ph.GA},
       adsurl = {https://ui.adsabs.harvard.edu/abs/2019MNRAS.483.4847K}
}

@ARTICLE{rose18,
       author = {{Rose}, Marvin and {Tadhunter}, Clive and {Ramos Almeida}, Cristina and {Rodr{\'\i}guez Zaur{\'\i}n}, Javier and {Santoro}, Francesco and {Spence}, Robert},
        title = "{Quantifying the AGN-driven outflows in ULIRGs (QUADROS) - I: VLT/Xshooter observations of nine nearby objects}",
      journal = {\mnras},
         year = 2018,
        month = feb,
       volume = {474},
       number = {1},
        pages = {128-156},
          doi = {10.1093/mnras/stx2590},
archivePrefix = {arXiv},
       eprint = {1710.06600},
 primaryClass = {astro-ph.GA},
       adsurl = {https://ui.adsabs.harvard.edu/abs/2018MNRAS.474..128R}
}

@ARTICLE{holt11,
       author = {{Holt}, J. and {Tadhunter}, C.~N. and {Morganti}, R. and {Emonts}, B.~H.~C.},
        title = "{The impact of the warm outflow in the young (GPS) radio source and ULIRG PKS 1345+12 (4C 12.50)}",
      journal = {\mnras},
         year = 2011,
        month = jan,
       volume = {410},
       number = {3},
        pages = {1527-1536},
          doi = {10.1111/j.1365-2966.2010.17535.x},
archivePrefix = {arXiv},
       eprint = {1008.2846},
 primaryClass = {astro-ph.CO},
       adsurl = {https://ui.adsabs.harvard.edu/abs/2011MNRAS.410.1527H}
}

@ARTICLE{jarvis20,
       author = {{Jarvis}, M.~E. and {Harrison}, C.~M. and {Mainieri}, V. and {Calistro Rivera}, G. and {Jethwa}, P. and {Zhang}, Z. -Y. and {Alexander}, D.~M. and {Circosta}, C. and {Costa}, T. and {De Breuck}, C. and {Kakkad}, D. and {Kharb}, P. and {Lansbury}, G.~B. and {Thomson}, A.~P.},
        title = "{High molecular gas content and star formation rates in local galaxies that host quasars, outflows, and jets}",
      journal = {\mnras},
         year = 2020,
        month = oct,
       volume = {498},
       number = {2},
        pages = {1560-1575},
          doi = {10.1093/mnras/staa2196},
archivePrefix = {arXiv},
       eprint = {2007.10351},
 primaryClass = {astro-ph.GA},
       adsurl = {https://ui.adsabs.harvard.edu/abs/2020MNRAS.498.1560J}
}

@ARTICLE{jarvis21,
       author = {{Jarvis}, M.~E. and {Harrison}, C.~M. and {Mainieri}, V. and {Alexander}, D.~M. and {Arrigoni Battaia}, F. and {Calistro Rivera}, G. and {Circosta}, C. and {Costa}, T. and {De Breuck}, C. and {Edge}, A.~C. and {Girdhar}, A. and {Kakkad}, D. and {Kharb}, P. and {Lansbury}, G.~B. and {Molyneux}, S.~J. and {Mukherjee}, D. and {Mullaney}, J.~R. and {Farina}, E.~P. and {Silpa}, S. and {Thomson}, A.~P. and {Ward}, S.~R.},
        title = "{The quasar feedback survey: discovering hidden Radio-AGN and their connection to the host galaxy ionized gas}",
      journal = {\mnras},
         year = 2021,
        month = may,
       volume = {503},
       number = {2},
        pages = {1780-1797},
          doi = {10.1093/mnras/stab549},
archivePrefix = {arXiv},
       eprint = {2103.00014},
 primaryClass = {astro-ph.GA},
       adsurl = {https://ui.adsabs.harvard.edu/abs/2021MNRAS.503.1780J}
}

@ARTICLE{kakkad20,
       author = {{Kakkad}, D. and {Mainieri}, V. and {Vietri}, G. and {Carniani}, S. and {Harrison}, C.~M. and {Perna}, M. and {Scholtz}, J. and {Circosta}, C. and {Cresci}, G. and {Husemann}, B. and {Bischetti}, M. and {Feruglio}, C. and {Fiore}, F. and {Marconi}, A. and {Padovani}, P. and {Brusa}, M. and {Cicone}, C. and {Comastri}, A. and {Lanzuisi}, G. and {Mannucci}, F. and {Menci}, N. and {Netzer}, H. and {Piconcelli}, E. and {Puglisi}, A. and {Salvato}, M. and {Schramm}, M. and {Silverman}, J. and {Vignali}, C. and {Zamorani}, G. and {Zappacosta}, L.},
        title = "{SUPER. II. Spatially resolved ionised gas kinematics and scaling relations in z {\ensuremath{\sim}} 2 AGN host galaxies}",
      journal = {\aap},
         year = 2020,
        month = oct,
       volume = {642},
          eid = {A147},
        pages = {A147},
          doi = {10.1051/0004-6361/202038551},
archivePrefix = {arXiv},
       eprint = {2008.01728},
 primaryClass = {astro-ph.GA},
       adsurl = {https://ui.adsabs.harvard.edu/abs/2020A&A...642A.147K}
}

@ARTICLE{wylezalek20,
       author = {{Wylezalek}, Dominika and {Flores}, Anthony M. and {Zakamska}, Nadia L. and
         {Greene}, Jenny E. and {Riffel}, Rogemar A.},
        title = "{Ionized gas outflow signatures in SDSS-IV MaNGA active galactic nuclei}",
      journal = {\mnras},
         year = 2020,
        month = mar,
       volume = {492},
       number = {4},
        pages = {4680-4696},
          doi = {10.1093/mnras/staa062},
archivePrefix = {arXiv},
       eprint = {1911.10212},
 primaryClass = {astro-ph.GA},
       adsurl = {https://ui.adsabs.harvard.edu/abs/2020MNRAS.492.4680W}
}

@ARTICLE{weilbacher20,
       author = {{Weilbacher}, Peter M. and {Palsa}, Ralf and {Streicher}, Ole and {Bacon}, Roland and {Urrutia}, Tanya and {Wisotzki}, Lutz and {Conseil}, Simon and {Husemann}, Bernd and {Jarno}, Aur{\'e}lien and {Kelz}, Andreas and {P{\'e}contal-Rousset}, Arlette and {Richard}, Johan and {Roth}, Martin M. and {Selman}, Fernando and {Vernet}, Jo{\"e}l},
        title = "{The data processing pipeline for the MUSE instrument}",
      journal = {\aap},
         year = 2020,
        month = sep,
       volume = {641},
          eid = {A28},
        pages = {A28},
          doi = {10.1051/0004-6361/202037855},
archivePrefix = {arXiv},
       eprint = {2006.08638},
 primaryClass = {astro-ph.IM},
       adsurl = {https://ui.adsabs.harvard.edu/abs/2020A&A...641A..28W}
}

@ARTICLE{astropy:2013,
       author = {{Astropy Collaboration} and {Robitaille}, Thomas P. and {Tollerud}, Erik J. and {Greenfield}, Perry and {Droettboom}, Michael and {Bray}, Erik and {Aldcroft}, Tom and {Davis}, Matt and {Ginsburg}, Adam and {Price-Whelan}, Adrian M. and {Kerzendorf}, Wolfgang E. and {Conley}, Alexander and {Crighton}, Neil and {Barbary}, Kyle and {Muna}, Demitri and {Ferguson}, Henry and {Grollier}, Fr{\'e}d{\'e}ric and {Parikh}, Madhura M. and {Nair}, Prasanth H. and {Unther}, Hans M. and {Deil}, Christoph and {Woillez}, Julien and {Conseil}, Simon and {Kramer}, Roban and {Turner}, James E.~H. and {Singer}, Leo and {Fox}, Ryan and {Weaver}, Benjamin A. and {Zabalza}, Victor and {Edwards}, Zachary I. and {Azalee Bostroem}, K. and {Burke}, D.~J. and {Casey}, Andrew R. and {Crawford}, Steven M. and {Dencheva}, Nadia and {Ely}, Justin and {Jenness}, Tim and {Labrie}, Kathleen and {Lim}, Pey Lian and {Pierfederici}, Francesco and {Pontzen}, Andrew and {Ptak}, Andy and {Refsdal}, Brian and {Servillat}, Mathieu and {Streicher}, Ole},
        title = "{Astropy: A community Python package for astronomy}",
      journal = {\aap},
         year = 2013,
        month = oct,
       volume = {558},
          eid = {A33},
        pages = {A33},
          doi = {10.1051/0004-6361/201322068},
archivePrefix = {arXiv},
       eprint = {1307.6212},
 primaryClass = {astro-ph.IM},
       adsurl = {https://ui.adsabs.harvard.edu/abs/2013A&A...558A..33A}
}

@ARTICLE{astropy:2018,
       author = {{Astropy Collaboration} and {Price-Whelan}, A.~M. and {Sip{\H{o}}cz}, B.~M. and {G{\"u}nther}, H.~M. and {Lim}, P.~L. and {Crawford}, S.~M. and {Conseil}, S. and {Shupe}, D.~L. and {Craig}, M.~W. and {Dencheva}, N. and {Ginsburg}, A. and {VanderPlas}, J.~T. and {Bradley}, L.~D. and {P{\'e}rez-Su{\'a}rez}, D. and {de Val-Borro}, M. and {Aldcroft}, T.~L. and {Cruz}, K.~L. and {Robitaille}, T.~P. and {Tollerud}, E.~J. and {Ardelean}, C. and {Babej}, T. and {Bach}, Y.~P. and {Bachetti}, M. and {Bakanov}, A.~V. and {Bamford}, S.~P. and {Barentsen}, G. and {Barmby}, P. and {Baumbach}, A. and {Berry}, K.~L. and {Biscani}, F. and {Boquien}, M. and {Bostroem}, K.~A. and {Bouma}, L.~G. and {Brammer}, G.~B. and {Bray}, E.~M. and {Breytenbach}, H. and {Buddelmeijer}, H. and {Burke}, D.~J. and {Calderone}, G. and {Cano Rodr{\'\i}guez}, J.~L. and {Cara}, M. and {Cardoso}, J.~V.~M. and {Cheedella}, S. and {Copin}, Y. and {Corrales}, L. and {Crichton}, D. and {D'Avella}, D. and {Deil}, C. and {Depagne}, {\'E}. and {Dietrich}, J.~P. and {Donath}, A. and {Droettboom}, M. and {Earl}, N. and {Erben}, T. and {Fabbro}, S. and {Ferreira}, L.~A. and {Finethy}, T. and {Fox}, R.~T. and {Garrison}, L.~H. and {Gibbons}, S.~L.~J. and {Goldstein}, D.~A. and {Gommers}, R. and {Greco}, J.~P. and {Greenfield}, P. and {Groener}, A.~M. and {Grollier}, F. and {Hagen}, A. and {Hirst}, P. and {Homeier}, D. and {Horton}, A.~J. and {Hosseinzadeh}, G. and {Hu}, L. and {Hunkeler}, J.~S. and {Ivezi{\'c}}, {\v{Z}}. and {Jain}, A. and {Jenness}, T. and {Kanarek}, G. and {Kendrew}, S. and {Kern}, N.~S. and {Kerzendorf}, W.~E. and {Khvalko}, A. and {King}, J. and {Kirkby}, D. and {Kulkarni}, A.~M. and {Kumar}, A. and {Lee}, A. and {Lenz}, D. and {Littlefair}, S.~P. and {Ma}, Z. and {Macleod}, D.~M. and {Mastropietro}, M. and {McCully}, C. and {Montagnac}, S. and {Morris}, B.~M. and {Mueller}, M. and {Mumford}, S.~J. and {Muna}, D. and {Murphy}, N.~A. and {Nelson}, S. and {Nguyen}, G.~H. and {Ninan}, J.~P. and {N{\"o}the}, M. and {Ogaz}, S. and {Oh}, S. and {Parejko}, J.~K. and {Parley}, N. and {Pascual}, S. and {Patil}, R. and {Patil}, A.~A. and {Plunkett}, A.~L. and {Prochaska}, J.~X. and {Rastogi}, T. and {Reddy Janga}, V. and {Sabater}, J. and {Sakurikar}, P. and {Seifert}, M. and {Sherbert}, L.~E. and {Sherwood-Taylor}, H. and {Shih}, A.~Y. and {Sick}, J. and {Silbiger}, M.~T. and {Singanamalla}, S. and {Singer}, L.~P. and {Sladen}, P.~H. and {Sooley}, K.~A. and {Sornarajah}, S. and {Streicher}, O. and {Teuben}, P. and {Thomas}, S.~W. and {Tremblay}, G.~R. and {Turner}, J.~E.~H. and {Terr{\'o}n}, V. and {van Kerkwijk}, M.~H. and {de la Vega}, A. and {Watkins}, L.~L. and {Weaver}, B.~A. and {Whitmore}, J.~B. and {Woillez}, J. and {Zabalza}, V. and {Astropy Contributors}},
        title = "{The Astropy Project: Building an Open-science Project and Status of the v2.0 Core Package}",
      journal = {\aj},
         year = 2018,
        month = sep,
       volume = {156},
       number = {3},
          eid = {123},
        pages = {123},
          doi = {10.3847/1538-3881/aabc4f},
archivePrefix = {arXiv},
       eprint = {1801.02634},
 primaryClass = {astro-ph.IM},
       adsurl = {https://ui.adsabs.harvard.edu/abs/2018AJ....156..123A}
}

@ARTICLE{pyneb,
       author = {{Luridiana}, V. and {Morisset}, C. and {Shaw}, R.~A.},
        title = "{PyNeb: a new tool for analyzing emission lines. I. Code description and validation of results}",
      journal = {\aap},
         year = 2015,
        month = jan,
       volume = {573},
          eid = {A42},
        pages = {A42},
          doi = {10.1051/0004-6361/201323152},
archivePrefix = {arXiv},
       eprint = {1410.6662},
 primaryClass = {astro-ph.IM},
       adsurl = {https://ui.adsabs.harvard.edu/abs/2015A&A...573A..42L}
}

@ARTICLE{zakamska14,
       author = {{Zakamska}, Nadia L. and {Greene}, Jenny E.},
        title = "{Quasar feedback and the origin of radio emission in radio-quiet quasars}",
      journal = {\mnras},
         year = "2014",
        month = "Jul",
       volume = {442},
       number = {1},
        pages = {784-804},
          doi = {10.1093/mnras/stu842},
archivePrefix = {arXiv},
       eprint = {1402.6736},
 primaryClass = {astro-ph.GA},
       adsurl = {https://ui.adsabs.harvard.edu/abs/2014MNRAS.442..784Z}
}

@ARTICLE{harrison15,
       author = {{Harrison}, C.~M. and {Thomson}, A.~P. and {Alexander}, D.~M. and {Bauer}, F.~E. and {Edge}, A.~C. and {Hogan}, M.~T. and {Mullaney}, J.~R. and {Swinbank}, A.~M.},
        title = "{Storm in a ``Teacup'': A Radio-quiet Quasar with {\ensuremath{\approx}}10 kpc Radio-emitting Bubbles and Extreme Gas Kinematics}",
      journal = {\apj},
         year = 2015,
        month = feb,
       volume = {800},
       number = {1},
          eid = {45},
        pages = {45},
          doi = {10.1088/0004-637X/800/1/45},
archivePrefix = {arXiv},
       eprint = {1410.4198},
 primaryClass = {astro-ph.GA},
       adsurl = {https://ui.adsabs.harvard.edu/abs/2015ApJ...800...45H}
}

@ARTICLE{mullaney13,
       author = {{Mullaney}, J.~R. and {Alexander}, D.~M. and {Fine}, S. and {Goulding}, A.~D. and {Harrison}, C.~M. and {Hickox}, R.~C.},
        title = "{Narrow-line region gas kinematics of 24 264 optically selected AGN: the radio connection}",
      journal = {\mnras},
         year = 2013,
        month = jul,
       volume = {433},
       number = {1},
        pages = {622-638},
          doi = {10.1093/mnras/stt751},
archivePrefix = {arXiv},
       eprint = {1305.0263},
 primaryClass = {astro-ph.CO},
       adsurl = {https://ui.adsabs.harvard.edu/abs/2013MNRAS.433..622M}
}

@ARTICLE{reyes08,
       author = {{Reyes}, Reinabelle and {Zakamska}, Nadia L. and {Strauss}, Michael A. and {Green}, Joshua and {Krolik}, Julian H. and {Shen}, Yue and {Richards}, Gordon T. and {Anderson}, Scott F. and {Schneider}, Donald P.},
        title = "{Space Density of Optically Selected Type 2 Quasars}",
      journal = {\aj},
         year = 2008,
        month = dec,
       volume = {136},
       number = {6},
        pages = {2373-2390},
          doi = {10.1088/0004-6256/136/6/2373},
archivePrefix = {arXiv},
       eprint = {0801.1115},
 primaryClass = {astro-ph},
       adsurl = {https://ui.adsabs.harvard.edu/abs/2008AJ....136.2373R}
}

@ARTICLE{girdhar22,
       author = {{Girdhar}, A. and {Harrison}, C.~M. and {Mainieri}, V. and {Bittner}, A. and {Costa}, T. and {Kharb}, P. and {Mukherjee}, D. and {Arrigoni Battaia}, F. and {Alexander}, D.~M. and {Calistro Rivera}, G. and {Circosta}, C. and {De Breuck}, C. and {Edge}, A.~C. and {Farina}, E.~P. and {Kakkad}, D. and {Lansbury}, G.~B. and {Molyneux}, S.~J. and {Mullaney}, J.~R. and {S}, Silpa and {Thomson}, A.~P. and {Ward}, S.~R.},
        title = "{Quasar feedback survey: multiphase outflows, turbulence, and evidence for feedback caused by low power radio jets inclined into the galaxy disc}",
      journal = {\mnras},
         year = 2022,
        month = may,
       volume = {512},
       number = {2},
        pages = {1608-1628},
          doi = {10.1093/mnras/stac073},
archivePrefix = {arXiv},
       eprint = {2201.02208},
 primaryClass = {astro-ph.GA},
       adsurl = {https://ui.adsabs.harvard.edu/abs/2022MNRAS.512.1608G}
}

@ARTICLE{lansbury18,
       author = {{Lansbury}, George B. and {Jarvis}, Miranda E. and {Harrison}, Chris M. and {Alexander}, David M. and {Del Moro}, Agnese and {Edge}, Alastair C. and {Mullaney}, James R. and {Thomson}, Alasdair P.},
        title = "{Storm in a Teacup: X-Ray View of an Obscured Quasar and Superbubble}",
      journal = {\apjl},
         year = 2018,
        month = mar,
       volume = {856},
       number = {1},
          eid = {L1},
        pages = {L1},
          doi = {10.3847/2041-8213/aab357},
archivePrefix = {arXiv},
       eprint = {1803.00009},
 primaryClass = {astro-ph.GA},
       adsurl = {https://ui.adsabs.harvard.edu/abs/2018ApJ...856L...1L}
}

@ARTICLE{bower06,
       author = {{Bower}, R.~G. and {Benson}, A.~J. and {Malbon}, R. and {Helly}, J.~C. and {Frenk}, C.~S. and {Baugh}, C.~M. and {Cole}, S. and {Lacey}, C.~G.},
        title = "{Breaking the hierarchy of galaxy formation}",
      journal = {\mnras},
         year = 2006,
        month = aug,
       volume = {370},
       number = {2},
        pages = {645-655},
          doi = {10.1111/j.1365-2966.2006.10519.x},
archivePrefix = {arXiv},
       eprint = {astro-ph/0511338},
 primaryClass = {astro-ph},
       adsurl = {https://ui.adsabs.harvard.edu/abs/2006MNRAS.370..645B}
}

@ARTICLE{mukherjee18,
       author = {{Mukherjee}, Dipanjan and {Bicknell}, Geoffrey V. and {Wagner}, Alexander Y. and {Sutherland}, Ralph S. and {Silk}, Joseph},
        title = "{Relativistic jet feedback - III. Feedback on gas discs}",
      journal = {\mnras},
         year = 2018,
        month = oct,
       volume = {479},
       number = {4},
        pages = {5544-5566},
          doi = {10.1093/mnras/sty1776},
archivePrefix = {arXiv},
       eprint = {1803.08305},
 primaryClass = {astro-ph.HE},
       adsurl = {https://ui.adsabs.harvard.edu/abs/2018MNRAS.479.5544M}
}

@ARTICLE{granato04,
       author = {{Granato}, Gian Luigi and {De Zotti}, Gianfranco and {Silva}, Laura and {Bressan}, Alessandro and {Danese}, Luigi},
        title = "{A Physical Model for the Coevolution of QSOs and Their Spheroidal Hosts}",
      journal = {\apj},
         year = 2004,
        month = jan,
       volume = {600},
       number = {2},
        pages = {580-594},
          doi = {10.1086/379875},
archivePrefix = {arXiv},
       eprint = {astro-ph/0307202},
 primaryClass = {astro-ph},
       adsurl = {https://ui.adsabs.harvard.edu/abs/2004ApJ...600..580G}
}

@ARTICLE{king15,
       author = {{King}, Andrew and {Pounds}, Ken},
        title = "{Powerful Outflows and Feedback from Active Galactic Nuclei}",
      journal = {\araa},
         year = 2015,
        month = aug,
       volume = {53},
        pages = {115-154},
          doi = {10.1146/annurev-astro-082214-122316},
archivePrefix = {arXiv},
       eprint = {1503.05206},
 primaryClass = {astro-ph.GA},
       adsurl = {https://ui.adsabs.harvard.edu/abs/2015ARA&A..53..115K}
}

@ARTICLE{zubovas17,
       author = {{Zubovas}, Kastytis and {Bourne}, Martin A.},
        title = "{Do AGN outflows quench or enhance star formation?}",
      journal = {\mnras},
         year = 2017,
        month = jul,
       volume = {468},
       number = {4},
        pages = {4956-4967},
          doi = {10.1093/mnras/stx787},
archivePrefix = {arXiv},
       eprint = {1703.10782},
 primaryClass = {astro-ph.GA},
       adsurl = {https://ui.adsabs.harvard.edu/abs/2017MNRAS.468.4956Z}
}

@ARTICLE{ruschel-dutra21,
       author = {{Ruschel-Dutra}, D. and {Storchi-Bergmann}, T. and {Schnorr-M{\"u}ller}, A. and {Riffel}, R.~A. and {de Oliveira}, B. Dall'Agnol and {Lena}, D. and {Robinson}, A. and {Nagar}, N. and {Elvis}, M.},
        title = "{AGNIFS survey of local AGN: GMOS-IFU data and outflows in 30 sources}",
      journal = {\mnras},
         year = 2021,
        month = jul,
       volume = {507},
       number = {1},
        pages = {74–89},
          doi = {10.1093/mnras/stab2058},
       adsurl = {https://ui.adsabs.harvard.edu/abs/2021MNRAS.tmp.1841R}
}

@ARTICLE{M1066KIN,
       author = {{Riffel}, Rogemar A. and {Storchi-Bergmann}, Thaisa},
        title = "{Compact molecular disc and ionized gas outflows within 350 pc of the active nucleus of Mrk 1066}",
      journal = {\mnras},
         year = 2011,
        month = feb,
       volume = {411},
       number = {1},
        pages = {469-486},
          doi = {10.1111/j.1365-2966.2010.17721.x},
archivePrefix = {arXiv},
       eprint = {1009.4832},
 primaryClass = {astro-ph.CO},
       adsurl = {https://ui.adsabs.harvard.edu/abs/2011MNRAS.411..469R}
}

@ARTICLE{rogemar_m1157,
       author = {{Riffel}, Rogemar A. and {Storchi-Bergmann}, Thaisa},
        title = "{Feeding and feedback in the active nucleus of Mrk 1157 probed with the Gemini Near-Infrared Integral-Field Spectrograph}",
      journal = {\mnras},
         year = 2011,
        month = nov,
       volume = {417},
       number = {4},
        pages = {2752-2769},
          doi = {10.1111/j.1365-2966.2011.19441.x},
archivePrefix = {arXiv},
       eprint = {1107.2564},
 primaryClass = {astro-ph.CO},
       adsurl = {https://ui.adsabs.harvard.edu/abs/2011MNRAS.417.2752R}
}

@ARTICLE{rogemar_N1275,
       author = {{Riffel}, Rogemar A. and {Storchi-Bergmann}, Thaisa and
         {Zakamska}, Nadia L. and {Riffel}, Rog{\'e}rio},
        title = "{Ionized and hot molecular outflows in the inner 500 pc of NGC 1275}",
      journal = {\mnras},
         year = 2020,
        month = jul,
       volume = {496},
       number = {4},
        pages = {4857-4873},
           doi = {10.1093/mnras/staa1922},
archivePrefix = {arXiv},
       eprint = {2006.15198},
 primaryClass = {astro-ph.GA},
       adsurl = {https://ui.adsabs.harvard.edu/abs/2020MNRAS.tmp.2053R}
}

@ARTICLE{Shimizu19,
       author = {{Shimizu}, T. Taro and {Davies}, R.~I. and {Lutz}, D. and
         {Burtscher}, L. and {Lin}, M. and {Baron}, D. and {Davies}, R.~L. and
         {Genzel}, R. and {Hicks}, E.~K.~S. and {Koss}, M. and
         {Maciejewski}, W. and {M{\"u}ller-S{\'a}nchez}, F. and
         {Orban de Xivry}, G. and {Price}, S.~H. and {Ricci}, C. and
         {Riffel}, R. and {Riffel}, R.~A. and {Rosario}, D. and
         {Schartmann}, M. and {Schnorr-M{\"u}ller}, A. and {Sternberg}, A. and
         {Sturm}, E. and {Storchi-Bergmann}, T. and {Tacconi}, L. and
         {Veilleux}, S.},
        title = "{The multiphase gas structure and kinematics in the circumnuclear region of NGC 5728}",
      journal = {\mnras},
         year = 2019,
        month = dec,
       volume = {490},
       number = {4},
        pages = {5860-5887},
          doi = {10.1093/mnras/stz2802},
archivePrefix = {arXiv},
       eprint = {1907.03801},
 primaryClass = {astro-ph.GA},
       adsurl = {https://ui.adsabs.harvard.edu/abs/2019MNRAS.490.5860S}
}

@ARTICLE{kakkad18,
       author = {{Kakkad}, D. and {Groves}, B. and {Dopita}, M. and {Thomas}, A.~D. and
         {Davies}, R.~L. and {Mainieri}, V. and {Kharb}, P. and
         {Scharw{\"a}chter}, J. and {Hampton}, E.~J. and {Ho}, I. -T.},
        title = "{Spatially resolved electron density in the narrow line region of z \&lt; 0.02 radio AGNs}",
      journal = {\aap},
         year = 2018,
        month = oct,
       volume = {618},
          eid = {A6},
        pages = {A6},
          doi = {10.1051/0004-6361/201832790},
archivePrefix = {arXiv},
       eprint = {1806.02839},
 primaryClass = {astro-ph.GA},
       adsurl = {https://ui.adsabs.harvard.edu/abs/2018A&A...618A...6K}
}

@ARTICLE{trindade-falcao21,
       author = {{Trindade Falc{\~a}o}, Anna and {Kraemer}, S.~B. and {Fischer}, T.~C. and {Crenshaw}, D.~M. and {Revalski}, M. and {Schmitt}, H.~R. and {Vestergaard}, M. and {Elvis}, M. and {Gaskell}, C.~M. and {Hamann}, F. and {Ho}, L.~C. and {Hutchings}, J. and {Mushotzky}, R. and {Netzer}, H. and {Storchi-Bergmann}, T. and {Turner}, T.~J. and {Ward}, M.~J.},
        title = "{Hubble Space Telescope observations of [O III] emission in nearby QSO2s: physical properties of the ionized outflows}",
      journal = {\mnras},
         year = 2021,
        month = jan,
       volume = {500},
       number = {1},
        pages = {1491-1504},
          doi = {10.1093/mnras/staa3239},
archivePrefix = {arXiv},
       eprint = {2010.08050},
 primaryClass = {astro-ph.GA},
       adsurl = {https://ui.adsabs.harvard.edu/abs/2021MNRAS.500.1491T}
}

@ARTICLE{heckman81,
       author = {{Heckman}, T.~M. and {Miley}, G.~K. and {van Breugel}, W.~J.~M. and {Butcher}, H.~R.},
        title = "{Emission-line profiles and kinematics of the narrow-line region in Seyfert and radio galaxies.}",
      journal = {\apj},
         year = 1981,
        month = jul,
       volume = {247},
        pages = {403-418},
          doi = {10.1086/159050},
       adsurl = {https://ui.adsabs.harvard.edu/abs/1981ApJ...247..403H}
}

@ARTICLE{zakamska16,
       author = {{Zakamska}, Nadia L. and {Lampayan}, Kelly and {Petric}, Andreea and {Dicken}, Daniel and {Greene}, Jenny E. and {Heckman}, Timothy M. and {Hickox}, Ryan C. and {Ho}, Luis C. and {Krolik}, Julian H. and {Nesvadba}, Nicole P.~H. and {Strauss}, Michael A. and {Geach}, James E. and {Oguri}, Masamune and {Strateva}, Iskra V.},
        title = "{Star formation in quasar hosts and the origin of radio emission in radio-quiet quasars}",
      journal = {\mnras},
         year = 2016,
        month = feb,
       volume = {455},
       number = {4},
        pages = {4191-4211},
          doi = {10.1093/mnras/stv2571},
archivePrefix = {arXiv},
       eprint = {1511.00013},
 primaryClass = {astro-ph.GA},
       adsurl = {https://ui.adsabs.harvard.edu/abs/2016MNRAS.455.4191Z}
}

@ARTICLE{fleutsch21,
       author = {{Fluetsch}, A. and {Maiolino}, R. and {Carniani}, S. and {Arribas}, S. and {Belfiore}, F. and {Bellocchi}, E. and {Cazzoli}, S. and {Cicone}, C. and {Cresci}, G. and {Fabian}, A.~C. and {Gallagher}, R. and {Ishibashi}, W. and {Mannucci}, F. and {Marconi}, A. and {Perna}, M. and {Sturm}, E. and {Venturi}, G.},
        title = "{Properties of the multiphase outflows in local (ultra)luminous infrared galaxies}",
      journal = {\mnras},
         year = 2021,
        month = aug,
       volume = {505},
       number = {4},
        pages = {5753-5783},
          doi = {10.1093/mnras/stab1666},
archivePrefix = {arXiv},
       eprint = {2006.13232},
 primaryClass = {astro-ph.GA},
       adsurl = {https://ui.adsabs.harvard.edu/abs/2021MNRAS.505.5753F}
}

@ARTICLE{revalski21,
       author = {{Revalski}, Mitchell and {Meena}, Beena and {Martinez}, Francisco and {Polack}, Garrett E. and {Crenshaw}, D. Michael and {Kraemer}, Steven B. and {Collins}, Nicholas R. and {Fischer}, Travis C. and {Schmitt}, Henrique R. and {Schmidt}, Judy and {Maksym}, W. Peter and {Rafelski}, Marc},
        title = "{Quantifying Feedback from Narrow Line Region Outflows in Nearby Active Galaxies. III. Results for the Seyfert 2 Galaxies Markarian 3, Markarian 78, and NGC 1068}",
      journal = {\apj},
         year = 2021,
        month = apr,
       volume = {910},
       number = {2},
          eid = {139},
        pages = {139},
          doi = {10.3847/1538-4357/abdcad},
archivePrefix = {arXiv},
       eprint = {2101.06270},
 primaryClass = {astro-ph.GA},
       adsurl = {https://ui.adsabs.harvard.edu/abs/2021ApJ...910..139R}
}

@ARTICLE{fischer18,
       author = {{Fischer}, Travis C. and {Kraemer}, S.~B. and {Schmitt}, H.~R. and {Longo Micchi}, L.~F. and {Crenshaw}, D.~M. and {Revalski}, M. and {Vestergaard}, M. and {Elvis}, M. and {Gaskell}, C.~M. and {Hamann}, F. and et al.},
        title = "{Hubble Space Telescope Observations of Extended [O III]{\ensuremath{\lambda}} 5007 Emission in Nearby QSO2s: New Constraints on AGN Host Galaxy Interaction}",
      journal = {\apj},
         year = 2018,
        month = apr,
       volume = {856},
       number = {2},
          eid = {102},
        pages = {102},
          doi = {10.3847/1538-4357/aab03e},
archivePrefix = {arXiv},
       eprint = {1802.06184},
 primaryClass = {astro-ph.GA},
       adsurl = {https://ui.adsabs.harvard.edu/abs/2018ApJ...856..102F}
}

@ARTICLE{revalski22,
       author = {{Revalski}, Mitchell and {Crenshaw}, D. Michael and {Rafelski}, Marc and {Kraemer}, Steven B. and {Polack}, Garrett E. and {Falc{\~a}o}, Anna Trindade and {Fischer}, Travis C. and {Meena}, Beena and {Martinez}, Francisco and {Schmitt}, Henrique R. and {Collins}, Nicholas R. and {Falcone}, Julia},
        title = "{Quantifying Feedback from Narrow Line Region Outflows in Nearby Active Galaxies. IV. The Effects of Different Density Estimates on the Ionized Gas Masses and Outflow Rates}",
      journal = {\apj},
         year = 2022,
        month = may,
       volume = {930},
       number = {1},
          eid = {14},
        pages = {14},
          doi = {10.3847/1538-4357/ac5f3d},
archivePrefix = {arXiv},
       eprint = {2203.07387},
 primaryClass = {astro-ph.GA},
       adsurl = {https://ui.adsabs.harvard.edu/abs/2022ApJ...930...14R}
}

@ARTICLE{baron19,
       author = {{Baron}, Dalya and {Netzer}, Hagai},
        title = "{Discovering AGN-driven winds through their infrared emission - II. Mass outflow rate and energetics}",
      journal = {\mnras},
         year = 2019,
        month = jul,
       volume = {486},
       number = {3},
        pages = {4290-4303},
          doi = {10.1093/mnras/stz1070},
archivePrefix = {arXiv},
       eprint = {1903.11076},
 primaryClass = {astro-ph.GA},
       adsurl = {https://ui.adsabs.harvard.edu/abs/2019MNRAS.486.4290B}
}

@ARTICLE{davies20,
       author = {{Davies}, R. and {Baron}, D. and {Shimizu}, T. and {Netzer}, H. and {Burtscher}, L. and {de Zeeuw}, P.~T. and {Genzel}, R. and {Hicks}, E.~K.~S. and {Koss}, M. and {Lin}, M. -Y. and {Lutz}, D. and {Maciejewski}, W. and {M{\"u}ller-S{\'a}nchez}, F. and {Orban de Xivry}, G. and {Ricci}, C. and {Riffel}, R. and {Riffel}, R.~A. and {Rosario}, D. and {Schartmann}, M. and {Schnorr-M{\"u}ller}, A. and {Shangguan}, J. and {Sternberg}, A. and {Sturm}, E. and {Storchi-Bergmann}, T. and {Tacconi}, L. and {Veilleux}, S.},
        title = "{Ionized outflows in local luminous AGN: what are the real densities and outflow rates?}",
      journal = {\mnras},
         year = 2020,
        month = nov,
       volume = {498},
       number = {3},
        pages = {4150-4177},
          doi = {10.1093/mnras/staa2413},
archivePrefix = {arXiv},
       eprint = {2003.06153},
 primaryClass = {astro-ph.GA},
       adsurl = {https://ui.adsabs.harvard.edu/abs/2020MNRAS.498.4150D}
}

@ARTICLE{rogemar_abundances,
       author = {{Riffel}, R.~A. and {Dors}, O.~L. and {Armah}, M. and {Storchi-Bergmann}, T. and {Feltre}, A. and {H{\"a}gele}, G.~F. and {Cardaci}, M.~V. and {Ruschel-Dutra}, D. and {Krabbe}, A.~C. and {P{\'e}rez-Montero}, E. and {Zakamska}, N.~L. and {Freitas}, I.~C.},
        title = "{Chemical abundances in Seyfert galaxies - V. The discovery of shocked emission outside the AGN ionization axis}",
      journal = {\mnras},
         year = 2021,
        month = feb,
       volume = {501},
       number = {1},
        pages = {L54-L59},
          doi = {10.1093/mnrasl/slaa194},
archivePrefix = {arXiv},
       eprint = {2012.02013},
 primaryClass = {astro-ph.GA},
       adsurl = {https://ui.adsabs.harvard.edu/abs/2021MNRAS.501L..54R}
}

@ARTICLE{fiore17,
       author = {{Fiore}, F. and {Feruglio}, C. and {Shankar}, F. and {Bischetti}, M. and {Bongiorno}, A. and {Brusa}, M. and {Carniani}, S. and {Cicone}, C. and {Duras}, F. and {Lamastra}, A. and {Mainieri}, V. and {Marconi}, A. and {Menci}, N. and {Maiolino}, R. and {Piconcelli}, E. and {Vietri}, G. and {Zappacosta}, L.},
        title = "{AGN wind scaling relations and the co-evolution of black holes and galaxies}",
      journal = {\aap},
         year = 2017,
        month = may,
       volume = {601},
          eid = {A143},
        pages = {A143},
          doi = {10.1051/0004-6361/201629478},
archivePrefix = {arXiv},
       eprint = {1702.04507},
 primaryClass = {astro-ph.GA},
       adsurl = {https://ui.adsabs.harvard.edu/abs/2017A&A...601A.143F}
}

@ARTICLE{couto20,
       author = {{Couto}, Guilherme S. and {Storchi-Bergmann}, Thaisa and {Siemiginowska}, Aneta and {Riffel}, Rogemar A. and {Morganti}, Raffaella},
        title = "{Powerful ionized gas outflows in the interacting radio galaxy 4C+29.30}",
      journal = {\mnras},
         year = 2020,
        month = oct,
       volume = {497},
       number = {4},
        pages = {5103-5117},
          doi = {10.1093/mnras/staa2268},
archivePrefix = {arXiv},
       eprint = {2007.14977},
 primaryClass = {astro-ph.GA},
       adsurl = {https://ui.adsabs.harvard.edu/abs/2020MNRAS.497.5103C}
}

@ARTICLE{emonts16,
       author = {{Emonts}, B.~H.~C. and {Morganti}, R. and {Villar-Mart{\'\i}n}, M. and {Hodgson}, J. and {Brogt}, E. and {Tadhunter}, C.~N. and {Mahony}, E. and {Oosterloo}, T.~A.},
        title = "{From galaxy-scale fueling to nuclear-scale feedback. The merger-state of radio galaxies 3C 293, 3C 305, and 4C 12.50}",
      journal = {\aap},
         year = 2016,
        month = nov,
       volume = {596},
          eid = {A19},
        pages = {A19},
          doi = {10.1051/0004-6361/201628592},
archivePrefix = {arXiv},
       eprint = {1609.06539},
 primaryClass = {astro-ph.GA},
       adsurl = {https://ui.adsabs.harvard.edu/abs/2016A&A...596A..19E}
}

@ARTICLE{bessiere24,
       author = {{Bessiere}, P.~S. and {Ramos Almeida}, C. and {Holden}, L.~R. and {Tadhunter}, C.~N. and {Canalizo}, G.},
        title = "{QSOFEED: Relationship between star formation and active galactic nuclei feedback}",
      journal = {\aap},
         year = 2024,
        month = sep,
       volume = {689},
          eid = {A271},
        pages = {A271},
          doi = {10.1051/0004-6361/202348795},
archivePrefix = {arXiv},
       eprint = {2405.06421},
 primaryClass = {astro-ph.GA},
       adsurl = {https://ui.adsabs.harvard.edu/abs/2024A&A...689A.271B}
}

@ARTICLE{holden24,
       author = {{Holden}, Luke R. and {Tadhunter}, Clive and {Audibert}, Anelise and {Oosterloo}, Tom and {Ramos Almeida}, Cristina and {Morganti}, Raffaella and {Pereira-Santaella}, Miguel and {Lamperti}, Isabella},
        title = "{ALMA reveals a compact and massive molecular outflow driven by the young AGN in a nearby ULIRG}",
      journal = {\mnras},
         year = 2024,
        month = may,
       volume = {530},
       number = {1},
        pages = {446-456},
          doi = {10.1093/mnras/stae810},
archivePrefix = {arXiv},
       eprint = {2403.08869},
 primaryClass = {astro-ph.GA},
       adsurl = {https://ui.adsabs.harvard.edu/abs/2024MNRAS.530..446H}
}

@ARTICLE{evans99,
       author = {{Evans}, A.~S. and {Kim}, D.~C. and {Mazzarella}, J.~M. and {Scoville}, N.~Z. and {Sanders}, D.~B.},
        title = "{Molecular Gas in the Powerful Radio Nucleus of the Ultraluminous Infrared Galaxy PKS 1345+12}",
      journal = {\apjl},
         year = 1999,
        month = aug,
       volume = {521},
       number = {2},
        pages = {L107-L110},
          doi = {10.1086/312198},
archivePrefix = {arXiv},
       eprint = {astro-ph/9907397},
 primaryClass = {astro-ph},
       adsurl = {https://ui.adsabs.harvard.edu/abs/1999ApJ...521L.107E}
}

@ARTICLE{holt03,
       author = {{Holt}, J. and {Tadhunter}, C.~N. and {Morganti}, R.},
        title = "{Highly extinguished emission line outflows in the young radio source PKS 1345+12}",
      journal = {\mnras},
         year = 2003,
        month = jun,
       volume = {342},
       number = {1},
        pages = {227-238},
          doi = {10.1046/j.1365-8711.2003.06532.x},
archivePrefix = {arXiv},
       eprint = {astro-ph/0302311},
 primaryClass = {astro-ph},
       adsurl = {https://ui.adsabs.harvard.edu/abs/2003MNRAS.342..227H}
}

@ARTICLE{tadhunter18,
       author = {{Tadhunter}, C. and {Rodr{\'\i}guez Zaur{\'\i}n}, J. and {Rose}, M. and {Spence}, R.~A.~W. and {Batcheldor}, D. and {Berg}, M.~A. and {Ramos Almeida}, C. and {Spoon}, H.~W.~W. and {Sparks}, W. and {Chiaberge}, M.},
        title = "{Quantifying the AGN-driven outflows in ULIRGs (QUADROS) - II. Evidence for compact outflow regions from HST [O III] imaging observations}",
      journal = {\mnras},
         year = 2018,
        month = aug,
       volume = {478},
       number = {2},
        pages = {1558-1569},
          doi = {10.1093/mnras/sty1064},
archivePrefix = {arXiv},
       eprint = {1805.00514},
 primaryClass = {astro-ph.GA},
       adsurl = {https://ui.adsabs.harvard.edu/abs/2018MNRAS.478.1558T}
}

@ARTICLE{wylezalek18,
       author = {{Wylezalek}, Dominika and {Morganti}, Raffaella},
        title = "{Questions and challenges of what powers galactic outflows in active galactic nuclei}",
      journal = {Nature Astronomy},
         year = 2018,
        month = feb,
       volume = {2},
        pages = {181-182},
          doi = {10.1038/s41550-018-0409-0},
archivePrefix = {arXiv},
       eprint = {1802.10307},
 primaryClass = {astro-ph.GA},
       adsurl = {https://ui.adsabs.harvard.edu/abs/2018NatAs...2..181W}
}

@ARTICLE{bruno21,
       author = {{Dall'Agnol de Oliveira}, B. and {Storchi-Bergmann}, T. and {Kraemer}, S.~B. and {Villar Mart{\'\i}n}, M. and {Schnorr-M{\"u}ller}, A. and {Schmitt}, H.~R. and {Ruschel-Dutra}, D. and {Crenshaw}, D.~M. and {Fischer}, T.~C.},
        title = "{Gauging the effect of supermassive black holes feedback on quasar host galaxies}",
      journal = {\mnras},
         year = 2021,
        month = jun,
       volume = {504},
       number = {3},
        pages = {3890-3908},
          doi = {10.1093/mnras/stab1067},
archivePrefix = {arXiv},
       eprint = {2104.06223},
 primaryClass = {astro-ph.GA},
       adsurl = {https://ui.adsabs.harvard.edu/abs/2021MNRAS.504.3890D}
}

@ARTICLE{rogemar_n5929,
       author = {{Riffel}, Rogemar A. and {Storchi-Bergmann}, Thaisa and
         {Riffel}, Rog{\'e}rio},
        title = "{Feeding versus feedback in active galactic nuclei from near-infrared integral field spectroscopy - X. NGC 5929}",
      journal = {\mnras},
         year = 2015,
        month = aug,
       volume = {451},
       number = {4},
        pages = {3587-3605},
          doi = {10.1093/mnras/stv1129},
archivePrefix = {arXiv},
       eprint = {1505.04052},
 primaryClass = {astro-ph.GA},
       adsurl = {https://ui.adsabs.harvard.edu/abs/2015MNRAS.451.3587R}
}

@ARTICLE{rogemar_n5929letter,
       author = {{Riffel}, Rogemar A. and {Storchi-Bergmann}, Thaisa and {Riffel}, Rog{\'e}rio},
        title = "{An Outflow Perpendicular to the Radio Jet in the Seyfert Nucleus of NGC 5929}",
      journal = {\apjl},
         year = 2014,
        month = jan,
       volume = {780},
       number = {2},
          eid = {L24},
        pages = {L24},
          doi = {10.1088/2041-8205/780/2/L24},
archivePrefix = {arXiv},
       eprint = {1311.6142},
 primaryClass = {astro-ph.GA},
       adsurl = {https://ui.adsabs.harvard.edu/abs/2014ApJ...780L..24R}
}

@ARTICLE{muller-sanchez11,
       author = {{M{\"u}ller-S{\'a}nchez}, F. and {Prieto}, M.~A. and {Hicks}, E.~K.~S. and {Vives-Arias}, H. and {Davies}, R.~I. and {Malkan}, M. and {Tacconi}, L.~J. and {Genzel}, R.},
        title = "{Outflows from Active Galactic Nuclei: Kinematics of the Narrow-line and Coronal-line Regions in Seyfert Galaxies}",
      journal = {\apj},
         year = 2011,
        month = oct,
       volume = {739},
       number = {2},
          eid = {69},
        pages = {69},
          doi = {10.1088/0004-637X/739/2/69},
archivePrefix = {arXiv},
       eprint = {1107.3140},
 primaryClass = {astro-ph.CO},
       adsurl = {https://ui.adsabs.harvard.edu/abs/2011ApJ...739...69M}
}

@ARTICLE{ramos-almeida17,
       author = {{Ramos Almeida}, C. and {Piqueras L{\'o}pez}, J. and {Villar-Mart{\'\i}n}, M. and {Bessiere}, P.~S.},
        title = "{An infrared view of AGN feedback in a type-2 quasar: the case of the Teacup galaxy}",
      journal = {\mnras},
         year = 2017,
        month = sep,
       volume = {470},
       number = {1},
        pages = {964-976},
          doi = {10.1093/mnras/stx1287},
archivePrefix = {arXiv},
       eprint = {1705.07631},
 primaryClass = {astro-ph.GA},
       adsurl = {https://ui.adsabs.harvard.edu/abs/2017MNRAS.470..964R}
}

@ARTICLE{costa20,
       author = {{Costa}, Tiago and {Pakmor}, R{\"u}diger and {Springel}, Volker},
        title = "{Powering galactic superwinds with small-scale AGN winds}",
      journal = {\mnras},
         year = 2020,
        month = oct,
       volume = {497},
       number = {4},
        pages = {5229-5255},
          doi = {10.1093/mnras/staa2321},
archivePrefix = {arXiv},
       eprint = {2006.05997},
 primaryClass = {astro-ph.GA},
       adsurl = {https://ui.adsabs.harvard.edu/abs/2020MNRAS.497.5229C}
}

@ARTICLE{fabian12,
       author = {{Fabian}, A.~C.},
        title = "{Observational Evidence of Active Galactic Nuclei Feedback}",
      journal = {\araa},
         year = 2012,
        month = sep,
       volume = {50},
        pages = {455-489},
          doi = {10.1146/annurev-astro-081811-125521},
archivePrefix = {arXiv},
       eprint = {1204.4114},
 primaryClass = {astro-ph.CO},
       adsurl = {https://ui.adsabs.harvard.edu/abs/2012ARA&A..50..455F}
}

@ARTICLE{harrison14,
       author = {{Harrison}, C.~M. and {Alexander}, D.~M. and {Mullaney}, J.~R. and {Swinbank}, A.~M.},
        title = "{Kiloparsec-scale outflows are prevalent among luminous AGN: outflows and feedback in the context of the overall AGN population}",
      journal = {\mnras},
         year = 2014,
        month = jul,
       volume = {441},
       number = {4},
        pages = {3306-3347},
          doi = {10.1093/mnras/stu515},
archivePrefix = {arXiv},
       eprint = {1403.3086},
 primaryClass = {astro-ph.GA},
       adsurl = {https://ui.adsabs.harvard.edu/abs/2014MNRAS.441.3306H}
}

@ARTICLE{liu13,
       author = {{Liu}, Guilin and {Zakamska}, Nadia L. and {Greene}, Jenny E. and
         {Nesvadba}, Nicole P.~H. and {Liu}, Xin},
        title = "{Observations of feedback from radio-quiet quasars - II. Kinematics of ionized gas nebulae}",
      journal = {\mnras},
         year = 2013,
        month = dec,
       volume = {436},
       number = {3},
        pages = {2576-2597},
          doi = {10.1093/mnras/stt1755},
archivePrefix = {arXiv},
       eprint = {1305.6922},
 primaryClass = {astro-ph.CO},
       adsurl = {https://ui.adsabs.harvard.edu/abs/2013MNRAS.436.2576L}
}

@ARTICLE{Diniz19,
       author = {{Diniz}, Marlon R. and {Riffel}, Rogemar A. and {Storchi-Bergmann}, Thaisa and {Riffel}, Rog{\'e}rio},
        title = "{Outflows, inflows, and young stars in the inner 200 pc of the Seyfert galaxy NGC 2110}",
      journal = {\mnras},
         year = 2019,
        month = aug,
       volume = {487},
       number = {3},
        pages = {3958-3970},
          doi = {10.1093/mnras/stz1329},
archivePrefix = {arXiv},
       eprint = {1905.04168},
 primaryClass = {astro-ph.GA},
       adsurl = {https://ui.adsabs.harvard.edu/abs/2019MNRAS.487.3958D}
}

@ARTICLE{schnorr-muller14,
       author = {{Schnorr-M{\"u}ller}, Allan and {Storchi-Bergmann}, Thaisa and {Nagar}, Neil M. and {Robinson}, Andrew and {Lena}, Davide and {Riffel}, Rogemar A. and {Couto}, Guilherme S.},
        title = "{Feeding and feedback in the inner kiloparsec of the active galaxy NGC 2110}",
      journal = {\mnras},
         year = 2014,
        month = jan,
       volume = {437},
       number = {2},
        pages = {1708-1724},
          doi = {10.1093/mnras/stt2001},
archivePrefix = {arXiv},
       eprint = {1310.7916},
 primaryClass = {astro-ph.CO},
       adsurl = {https://ui.adsabs.harvard.edu/abs/2014MNRAS.437.1708S}
}

@ARTICLE{schnorr-muller16,
       author = {{Schnorr-M{\"u}ller}, Allan and {Storchi-Bergmann}, Thaisa and {Robinson}, Andrew and {Lena}, Davide and {Nagar}, Neil M.},
        title = "{Feeding and feedback in NGC 3081}",
      journal = {\mnras},
         year = 2016,
        month = mar,
       volume = {457},
       number = {1},
        pages = {972-985},
          doi = {10.1093/mnras/stw037},
archivePrefix = {arXiv},
       eprint = {1601.05432},
 primaryClass = {astro-ph.GA},
       adsurl = {https://ui.adsabs.harvard.edu/abs/2016MNRAS.457..972S}
}

@ARTICLE{harrison18,
       author = {{Harrison}, C.~M. and {Costa}, T. and {Tadhunter}, C.~N. and
         {Fl{\"u}tsch}, A. and {Kakkad}, D. and {Perna}, M. and {Vietri}, G.},
        title = "{AGN outflows and feedback twenty years on}",
      journal = {Nature Astronomy},
         year = 2018,
        month = feb,
       volume = {2},
        pages = {198-205},
          doi = {10.1038/s41550-018-0403-6},
archivePrefix = {arXiv},
       eprint = {1802.10306},
 primaryClass = {astro-ph.GA},
       adsurl = {https://ui.adsabs.harvard.edu/abs/2018NatAs...2..198H}
}

@ARTICLE{dimatteo05,
       author = {{Di Matteo}, Tiziana and {Springel}, Volker and {Hernquist}, Lars},
        title = "{Energy input from quasars regulates the growth and activity of black holes and their host galaxies}",
      journal = {\nat},
         year = 2005,
        month = feb,
       volume = {433},
       number = {7026},
        pages = {604-607},
          doi = {10.1038/nature03335},
archivePrefix = {arXiv},
       eprint = {astro-ph/0502199},
 primaryClass = {astro-ph},
       adsurl = {https://ui.adsabs.harvard.edu/abs/2005Natur.433..604D}
}

@BOOK{Osterbrock06,
       author = {{Osterbrock}, Donald E. and {Ferland}, Gary J.},
        title = "{Astrophysics of gaseous nebulae and active galactic nuclei}",
         year = "2006",
       adsurl = {https://ui.adsabs.harvard.edu/abs/2006agna.book.....O},
       publisher = {University Science Books}
}

@ARTICLE{rogemar_agnifs,
       author = {{Riffel}, R.~A. and {Storchi-Bergmann}, T. and {Riffel}, R. and {Bianchin}, M. and {Zakamska}, N.~L. and {Ruschel-Dutra}, D. and {Sch{\"o}nell}, A.~J. and {Rosario}, D.~J. and {Rodriguez-Ardila}, A. and {Fischer}, T.~C. and {Davies}, R.~I. and {Dametto}, N.~Z. and {Dahmer-Hahn}, L.~G. and {Crenshaw}, D.~M. and {Burtscher}, L. and {Bentz}, M.~C.},
        title = "{The AGNIFS survey: distribution and excitation of the hot molecular and ionized gas in the inner kpc of nearby AGN hosts}",
      journal = {\mnras},
         year = 2021,
        month = jul,
       volume = {504},
       number = {3},
        pages = {3265-3283},
          doi = {10.1093/mnras/stab998},
archivePrefix = {arXiv},
       eprint = {2104.03105},
 primaryClass = {astro-ph.GA},
       adsurl = {https://ui.adsabs.harvard.edu/abs/2021MNRAS.504.3265R}
}

@ARTICLE{rogemarN7582,
       author = {{Riffel}, Rogemar A. and {Storchi-Bergmann}, Thaisa and {Dors}, Oli L. and
         {Winge}, Cl{\'a}udia},
        title = "{AGN-starburst connection in NGC7582: Gemini near-infrared spectrograph integral field unit observations}",
      journal = {\mnras},
         year = 2009,
        month = mar,
       volume = {393},
       number = {3},
        pages = {783-797},
          doi = {10.1111/j.1365-2966.2008.14250.x},
archivePrefix = {arXiv},
       eprint = {0811.2327},
 primaryClass = {astro-ph},
       adsurl = {https://ui.adsabs.harvard.edu/abs/2009MNRAS.393..783R}
}

@ARTICLE{barbosa14,
       author = {{Barbosa}, F.~K.~B. and {Storchi-Bergmann}, T. and {McGregor}, P. and
         {Vale}, T.~B. and {Rogemar Riffel}, A.},
        title = "{Modelling the [Fe II] {\ensuremath{\lambda}}1.644 {\ensuremath{\mu}}m outflow and comparison with H$_{2}$ and H$^{+}$ kinematics in the inner 200 pc of NGC 1068}",
      journal = {\mnras},
         year = 2014,
        month = dec,
       volume = {445},
       number = {3},
        pages = {2353-2370},
          doi = {10.1093/mnras/stu1637},
archivePrefix = {arXiv},
       eprint = {1408.4750},
 primaryClass = {astro-ph.GA},
       adsurl = {https://ui.adsabs.harvard.edu/abs/2014MNRAS.445.2353B}
}

@ARTICLE{hopkins_elvis10,
       author = {{Hopkins}, Philip F. and {Elvis}, Martin},
        title = "{Quasar feedback: more bang for your buck}",
      journal = {\mnras},
         year = 2010,
        month = jan,
       volume = {401},
       number = {1},
        pages = {7-14},
          doi = {10.1111/j.1365-2966.2009.15643.x},
archivePrefix = {arXiv},
       eprint = {0904.0649},
 primaryClass = {astro-ph.CO},
       adsurl = {https://ui.adsabs.harvard.edu/abs/2010MNRAS.401....7H}
}

@ARTICLE{sb19,
       author = {{Storchi-Bergmann}, Thaisa and {Schnorr-M{\"u}ller}, Allan},
        title = "{Observational constraints on the feeding of supermassive black holes}",
      journal = {Nature Astronomy},
         year = 2019,
        month = jan,
       volume = {3},
        pages = {48-61},
          doi = {10.1038/s41550-018-0611-0},
archivePrefix = {arXiv},
       eprint = {1904.03338},
 primaryClass = {astro-ph.GA},
       adsurl = {https://ui.adsabs.harvard.edu/abs/2019NatAs...3...48S}
}

@ARTICLE{kormendy13,
       author = {{Kormendy}, John and {Ho}, Luis C.},
        title = "{Coevolution (Or Not) of Supermassive Black Holes and Host Galaxies}",
      journal = {\araa},
         year = 2013,
        month = aug,
       volume = {51},
       number = {1},
        pages = {511-653},
          doi = {10.1146/annurev-astro-082708-101811},
archivePrefix = {arXiv},
       eprint = {1304.7762},
 primaryClass = {astro-ph.CO},
       adsurl = {https://ui.adsabs.harvard.edu/abs/2013ARA&A..51..511K}
}

@ARTICLE{veilleux20,
       author = {{Veilleux}, Sylvain and {Maiolino}, Roberto and {Bolatto}, Alberto D. and {Aalto}, Susanne},
        title = "{Cool outflows in galaxies and their implications}",
      journal = {\aapr},
         year = 2020,
        month = apr,
       volume = {28},
       number = {1},
          eid = {2},
        pages = {2},
          doi = {10.1007/s00159-019-0121-9},
archivePrefix = {arXiv},
       eprint = {2002.07765},
 primaryClass = {astro-ph.GA},
       adsurl = {https://ui.adsabs.harvard.edu/abs/2020A&ARv..28....2V}
}

@ARTICLE{jarvis19,
       author = {{Jarvis}, M.~E. and {Harrison}, C.~M. and {Thomson}, A.~P. and {Circosta}, C. and {Mainieri}, V. and {Alexander}, D.~M. and {Edge}, A.~C. and {Lansbury}, G.~B. and {Molyneux}, S.~J. and {Mullaney}, J.~R.},
        title = "{Prevalence of radio jets associated with galactic outflows and feedback from quasars}",
      journal = {\mnras},
         year = 2019,
        month = may,
       volume = {485},
       number = {2},
        pages = {2710-2730},
          doi = {10.1093/mnras/stz556},
archivePrefix = {arXiv},
       eprint = {1902.07727},
 primaryClass = {astro-ph.GA},
       adsurl = {https://ui.adsabs.harvard.edu/abs/2019MNRAS.485.2710J}
}

@software{ifscube,
  author       = {Daniel Ruschel-Dutra},
  title        = {danielrd6/ifscube v1.0},
  month        = jul,
  year         = 2020,
  publisher    = {Zenodo},
  version      = {v1.0},
  doi          = {10.5281/zenodo.3945237},
  url          = {}
}

@ARTICLE{binette24,
       author = {{Binette}, Luc and {Zovaro}, Henry R.~M. and {Villar Mart{\'\i}n}, Montserrat and {Dors}, Oli L. and {Krongold}, Yair and {Morisset}, Christophe and {Revalski}, Mitchell and {Alarie}, Alexandre and {Riffel}, Rogemar A. and {Dopita}, Michael A.},
        title = "{Constraints on the densities and temperature of the Seyfert 2 narrow line region}",
      journal = {\aap},
         year = 2024,
        month = apr,
       volume = {684},
          eid = {A53},
        pages = {A53},
          doi = {10.1051/0004-6361/202245754},
archivePrefix = {arXiv},
       eprint = {2401.06972},
 primaryClass = {astro-ph.GA},
       adsurl = {https://ui.adsabs.harvard.edu/abs/2024A&A...684A..53B}
}

@ARTICLE{ulivi24,
       author = {{Ulivi}, L. and {Venturi}, G. and {Cresci}, G. and {Marconi}, A. and {Marconcini}, C. and {Amiri}, A. and {Belfiore}, F. and {Bertola}, E. and {Carniani}, S. and {D'Amato}, Q. and {Di Teodoro}, E. and {Ginolfi}, M. and {Girdhar}, A. and {Harrison}, C. and {Maiolino}, R. and {Mannucci}, F. and {Mingozzi}, M. and {Perna}, M. and {Scialpi}, M. and {Tomicic}, N. and {Tozzi}, G. and {Treister}, E.},
        title = "{Feedback and ionized gas outflows in four low-radio power AGN at z {\ensuremath{\sim}} 0.15}",
      journal = {\aap},
         year = 2024,
        month = may,
       volume = {685},
          eid = {A122},
        pages = {A122},
          doi = {10.1051/0004-6361/202347436},
archivePrefix = {arXiv},
       eprint = {2403.01258},
 primaryClass = {astro-ph.GA},
       adsurl = {https://ui.adsabs.harvard.edu/abs/2024A&A...685A.122U}
}

@ARTICLE{silpa22,
       author = {{Silpa}, S. and {Kharb}, P. and {Harrison}, C.~M. and {Girdhar}, A. and {Mukherjee}, D. and {Mainieri}, V. and {Jarvis}, M.~E.},
        title = "{The Quasar Feedback Survey: revealing the interplay of jets, winds, and emission-line gas in type 2 quasars with radio polarization}",
      journal = {\mnras},
         year = 2022,
        month = jul,
       volume = {513},
       number = {3},
        pages = {4208-4223},
          doi = {10.1093/mnras/stac1044},
archivePrefix = {arXiv},
       eprint = {2204.05613},
 primaryClass = {astro-ph.GA},
       adsurl = {https://ui.adsabs.harvard.edu/abs/2022MNRAS.513.4208S}
}

@ARTICLE{molyneux24,
       author = {{Molyneux}, S.~J. and {Calistro Rivera}, G. and {De Breuck}, C. and {Harrison}, C.~M. and {Mainieri}, V. and {Lundgren}, A. and {Kakkad}, D. and {Circosta}, C. and {Girdhar}, A. and {Costa}, T. and {Mullaney}, J.~R. and {Kharb}, P. and {Arrigoni Battaia}, F. and {Farina}, E.~P. and {Alexander}, D.~M. and {Ward}, S.~R. and {Silpa}, S. and {Smit}, R.},
        title = "{The Quasar Feedback Survey: characterizing CO excitation in quasar host galaxies}",
      journal = {\mnras},
         year = 2024,
        month = jan,
       volume = {527},
       number = {3},
        pages = {4420-4439},
          doi = {10.1093/mnras/stad3133},
archivePrefix = {arXiv},
       eprint = {2310.10235},
 primaryClass = {astro-ph.GA},
       adsurl = {https://ui.adsabs.harvard.edu/abs/2024MNRAS.527.4420M}
}

@ARTICLE{ramos-almeida19,
       author = {{Ramos Almeida}, C. and {Acosta-Pulido}, J.~A. and {Tadhunter}, C.~N. and {Gonz{\'a}lez-Fern{\'a}ndez}, C. and {Cicone}, C. and {Fern{\'a}ndez-Torreiro}, M.},
        title = "{A near-infrared study of the multiphase outflow in the type-2 quasar J1509+0434}",
      journal = {\mnras},
         year = 2019,
        month = jul,
       volume = {487},
       number = {1},
        pages = {L18-L23},
          doi = {10.1093/mnrasl/slz072},
archivePrefix = {arXiv},
       eprint = {1905.06288},
 primaryClass = {astro-ph.GA},
       adsurl = {https://ui.adsabs.harvard.edu/abs/2019MNRAS.487L..18R}
}

@ARTICLE{gatto24,
       author = {{Gatto}, Lara and {Storchi-Bergmann}, T. and {Riffel}, Rogemar A. and {Riffel}, Rog{\'e}rio and {Rembold}, Sandro B. and {Schimoia}, Jaderson S. and {Mallmann}, Nicolas D. and {Ilha}, Gabriele S.},
        title = "{The extent and power of 'maintenance mode' feedback in MaNGA AGN}",
      journal = {\mnras},
         year = 2024,
        month = may,
       volume = {530},
       number = {3},
        pages = {3059-3074},
          doi = {10.1093/mnras/stae989},
archivePrefix = {arXiv},
       eprint = {2404.14502},
 primaryClass = {astro-ph.GA},
       adsurl = {https://ui.adsabs.harvard.edu/abs/2024MNRAS.530.3059G}
}

@ARTICLE{riffeln3884,
       author = {{Riffel}, Rogemar A. and {Riffel}, Rog{\'e}rio and {Storchi-Bergmann}, Thaisa and {Costa-Souza}, Jos{\'e} Henrique and {Souza-Oliveira}, Gabriel Luan and {Bianchin}, Marina},
        title = "{Revealing the kinematic puzzle of the AGN host NGC 3884: optical integral field spectroscopy unravels stellar and gas motions}",
      journal = {\mnras},
         year = 2024,
        month = feb,
       volume = {528},
       number = {2},
        pages = {1476-1486},
          doi = {10.1093/mnras/stae055},
       adsurl = {https://ui.adsabs.harvard.edu/abs/2024MNRAS.528.1476R}
}

@ARTICLE{pierce23,
       author = {{Pierce}, J.~C.~S. and {Tadhunter}, C. and {Ramos Almeida}, C. and {Bessiere}, P. and {Heaton}, J.~V. and {Ellison}, S.~L. and {Speranza}, G. and {Gordon}, Y. and {O'Dea}, C. and {Grimmett}, L. and {Makrygianni}, L.},
        title = "{Galaxy interactions are the dominant trigger for local type 2 quasars}",
      journal = {\mnras},
         year = 2023,
        month = jun,
       volume = {522},
       number = {2},
        pages = {1736-1751},
          doi = {10.1093/mnras/stad455},
archivePrefix = {arXiv},
       eprint = {2303.15506},
 primaryClass = {astro-ph.GA},
       adsurl = {https://ui.adsabs.harvard.edu/abs/2023MNRAS.522.1736P}
}

@ARTICLE{audibert25,
       author = {{Audibert}, A. and {Ramos Almeida}, C. and {Garc{\'\i}a-Burillo}, S. and {Speranza}, G. and {Lamperti}, I. and {Pereira-Santaella}, M. and {Panessa}, F.},
        title = "{Molecular gas excitation and outflow properties of obscured quasars at z {\ensuremath{\sim}} 0.1}",
      journal = {\aap},
         year = 2025,
        month = jul,
       volume = {699},
          eid = {A83},
        pages = {A83},
          doi = {10.1051/0004-6361/202453291},
archivePrefix = {arXiv},
       eprint = {2505.02759},
 primaryClass = {astro-ph.GA},
       adsurl = {https://ui.adsabs.harvard.edu/abs/2025A&A...699A..83A}
}

@ARTICLE{audibert23,
       author = {{Audibert}, A. and {Ramos Almeida}, C. and {Garc{\'\i}a-Burillo}, S. and {Combes}, F. and {Bischetti}, M. and {Meenakshi}, M. and {Mukherjee}, D. and {Bicknell}, G. and {Wagner}, A.~Y.},
        title = "{Jet-induced molecular gas excitation and turbulence in the Teacup}",
      journal = {\aap},
         year = 2023,
        month = mar,
       volume = {671},
          eid = {L12},
        pages = {L12},
          doi = {10.1051/0004-6361/202345964},
archivePrefix = {arXiv},
       eprint = {2302.13884},
 primaryClass = {astro-ph.GA},
       adsurl = {https://ui.adsabs.harvard.edu/abs/2023A&A...671L..12A}
}

@ARTICLE{holden26,
       author = {{Holden}, Luke R. and {Smith}, Daniel J.~B. and {Arnaudova}, Marina I. and {Tadhunter}, Clive N. and {Ramos Ameida}, Cristina and {Shenoy}, Shravya and {Cezar}, Pedro H. and {Das}, Soumyadeep and {Binu}, Akshara},
        title = "{Electron densities from [S II] lines significantly overestimate the impact of ionized AGN outflows}",
      journal = {\mnras},
         year = 2026,
        month = jan,
       volume = {545},
       number = {3},
          eid = {staf2075},
        pages = {staf2075},
          doi = {10.1093/mnras/staf2075},
archivePrefix = {arXiv},
       eprint = {2511.15791},
 primaryClass = {astro-ph.GA},
       adsurl = {https://ui.adsabs.harvard.edu/abs/2026MNRAS.545f2075H}
}

@ARTICLE{cezar26,
       author = {{Cezar}, P.~H. and {Coloma Puga}, M. and {Ramos Almeida}, C. and {Acosta-Pulido}, J.~A. and {Speranza}, G. and {Holden}, L.~R. and {Tadhunter}, C.~N. and {Zanchettin}, M.~V. and {Audibert}, A.},
        title = "{QSOFEED: Investigating warm molecular, low- and high-ionization atomic gas in six type-2 quasars with GTC/EMIR}",
      journal = {\aap},
         year = 2026,
        month = apr,
       volume = {708},
          eid = {A12},
        pages = {A12},
          doi = {10.1051/0004-6361/202557799},
archivePrefix = {arXiv},
       eprint = {2601.22906},
 primaryClass = {astro-ph.GA},
       adsurl = {https://ui.adsabs.harvard.edu/abs/2026A&A...708A..12C}
}

@ARTICLE{revalski25,
       author = {{Revalski}, Mitchell and {Crenshaw}, D. Michael and {Polack}, Garrett E. and {Rafelski}, Marc and {Kraemer}, Steven B. and {Fischer}, Travis C. and {Meena}, Beena and {Schmitt}, Henrique R. and {Trindade Falc{\~a}o}, Anna and {Falcone}, Julia and {Shea}, Maura Kathleen},
        title = "{Quantifying Feedback from Narrow Line Region Outflows in Nearby Active Galaxies. V. The Expanded Sample}",
      journal = {\apj},
         year = 2025,
        month = may,
       volume = {984},
       number = {1},
          eid = {32},
        pages = {32},
          doi = {10.3847/1538-4357/adc131},
archivePrefix = {arXiv},
       eprint = {2503.17444},
 primaryClass = {astro-ph.GA},
       adsurl = {https://ui.adsabs.harvard.edu/abs/2025ApJ...984...32R}
}

@ARTICLE{rupke11,
       author = {{Rupke}, David S.~N. and {Veilleux}, Sylvain},
        title = "{Integral Field Spectroscopy of Massive, Kiloparsec-scale Outflows in the Infrared-luminous QSO Mrk 231}",
      journal = {\apjl},
         year = 2011,
        month = mar,
       volume = {729},
       number = {2},
          eid = {L27},
        pages = {L27},
          doi = {10.1088/2041-8205/729/2/L27},
archivePrefix = {arXiv},
       eprint = {1102.4349},
 primaryClass = {astro-ph.GA},
       adsurl = {https://ui.adsabs.harvard.edu/abs/2011ApJ...729L..27R}
}

@ARTICLE{ramos-almeida25,
       author = {{Ramos Almeida}, C. and {Garc{\'\i}a-Bernete}, I. and {Pereira-Santaella}, M. and {Speranza}, G. and {Maiolino}, R. and {Ji}, X. and {Audibert}, A. and {Cezar}, P.~H. and {Acosta-Pulido}, J.~A. and {Alonso-Herrero}, A. and {Garc{\'\i}a-Burillo}, S. and {Gonz{\'a}lez-Mart{\'\i}n}, O. and {Rigopoulou}, D. and {Tadhunter}, C.~N. and {Labiano}, A. and {Levenson}, N.~A. and {Donnan}, F.~R.},
        title = "{JWST MIRI reveals the diversity of nuclear mid-infrared spectra of nearby type 2 quasars}",
      journal = {\aap},
         year = 2025,
        month = jun,
       volume = {698},
          eid = {A194},
        pages = {A194},
          doi = {10.1051/0004-6361/202453549},
archivePrefix = {arXiv},
       eprint = {2504.01595},
 primaryClass = {astro-ph.GA},
       adsurl = {https://ui.adsabs.harvard.edu/abs/2025A&A...698A.194R}
}

@ARTICLE{marconcini25,
       author = {{Marconcini}, Cosimo and {Marconi}, Alessandro and {Cresci}, Giovanni and {Mannucci}, Filippo and {Ulivi}, Lorenzo and {Venturi}, Giacomo and {Scialpi}, Martina and {Tozzi}, Giulia and {Belfiore}, Francesco and {Bertola}, Elena and {Carniani}, Stefano and {Cataldi}, Elisa and {Chakraborty}, Avinanda and {D'Amato}, Quirino and {Di Teodoro}, Enrico and {Feltre}, Anna and {Ginolfi}, Michele and {Moreschini}, Bianca and {Orientale}, Nicole and {Trefoloni}, Bartolomeo and {King}, Andrew},
        title = "{Evidence of the fast acceleration of AGN-driven winds at kiloparsec scales}",
      journal = {Nature Astronomy},
         year = 2025,
        month = jun,
       volume = {9},
        pages = {907-915},
          doi = {10.1038/s41550-025-02518-6},
archivePrefix = {arXiv},
       eprint = {2503.24359},
 primaryClass = {astro-ph.GA},
       adsurl = {https://ui.adsabs.harvard.edu/abs/2025NatAs...9..907M}
}

@ARTICLE{lauzikas24,
       author = {{Lau{\v{z}}ikas}, M. and {Zubovas}, K.},
        title = "{Slow and steady does the trick: Slow outflows enhance the fragmentation of molecular clouds}",
      journal = {\aap},
         year = 2024,
        month = oct,
       volume = {690},
          eid = {A396},
        pages = {A396},
          doi = {10.1051/0004-6361/202450286},
archivePrefix = {arXiv},
       eprint = {2409.13234},
 primaryClass = {astro-ph.GA},
       adsurl = {https://ui.adsabs.harvard.edu/abs/2024A&A...690A.396L}
}

@ARTICLE{hervella23,
       author = {{Hervella Seoane}, K. and {Ramos Almeida}, C. and {Acosta-Pulido}, J.~A. and {Speranza}, G. and {Tadhunter}, C.~N. and {Bessiere}, P.~S.},
        title = "{Investigating the impact of quasar-driven outflows on galaxies at z {\ensuremath{\sim}} 0.3-0.4}",
      journal = {\aap},
         year = 2023,
        month = dec,
       volume = {680},
          eid = {A71},
        pages = {A71},
          doi = {10.1051/0004-6361/202347756},
archivePrefix = {arXiv},
       eprint = {2309.10572},
 primaryClass = {astro-ph.GA},
       adsurl = {https://ui.adsabs.harvard.edu/abs/2023A&A...680A..71H}
}

@ARTICLE{njeri25,
       author = {{Njeri}, Ann and {Harrison}, Chris M. and {Kharb}, Preeti and {Beswick}, Robert and {Calistro-Rivera}, Gabriela and {Circosta}, Chiara and {Mainieri}, Vincenzo and {Molyneux}, Stephen and {Mullaney}, James and {Silpa}, Sasikumar},
        title = "{The Quasar Feedback Survey: zooming into the origin of radio emission with e-MERLIN}",
      journal = {\mnras},
         year = 2025,
        month = feb,
       volume = {537},
       number = {2},
        pages = {705-722},
          doi = {10.1093/mnras/staf020},
archivePrefix = {arXiv},
       eprint = {2501.03433},
 primaryClass = {astro-ph.GA},
       adsurl = {https://ui.adsabs.harvard.edu/abs/2025MNRAS.537..705N}
}

@ARTICLE{girdhar24,
       author = {{Girdhar}, A. and {Harrison}, C.~M. and {Mainieri}, V. and {Fern{\'a}ndez Aranda}, R. and {Alexander}, D.~M. and {Arrigoni Battaia}, F. and {Bianchin}, M. and {Calistro Rivera}, G. and {Circosta}, C. and {Costa}, T. and {Edge}, A.~C. and {Farina}, E.~P. and {Kakkad}, D. and {Kharb}, P. and {Molyneux}, S.~J. and {Mukherjee}, D. and {Njeri}, A. and {Silpa}, S. and {Venturi}, G. and {Ward}, S.~R.},
        title = "{Quasar feedback survey: molecular gas affected by central outflows and by  10-kpc radio lobes reveal dual feedback effects in 'radio quiet' quasars}",
      journal = {\mnras},
         year = 2024,
        month = jan,
       volume = {527},
       number = {3},
        pages = {9322-9342},
          doi = {10.1093/mnras/stad3453},
archivePrefix = {arXiv},
       eprint = {2311.03453},
 primaryClass = {astro-ph.GA},
       adsurl = {https://ui.adsabs.harvard.edu/abs/2024MNRAS.527.9322G}
}

@ARTICLE{speranza22,
       author = {{Speranza}, G. and {Ramos Almeida}, C. and {Acosta-Pulido}, J.~A. and {Riffel}, R.~A. and {Tadhunter}, C. and {Pierce}, J.~C.~S. and {Rodr{\'\i}guez-Ardila}, A. and {Coloma Puga}, M. and {Brusa}, M. and {Musiimenta}, B. and {Alexander}, D.~M. and {Lapi}, A. and {Shankar}, F. and {Villforth}, C.},
        title = "{Warm molecular and ionized gas kinematics in the type-2 quasar J0945+1737}",
      journal = {\aap},
         year = 2022,
        month = sep,
       volume = {665},
          eid = {A55},
        pages = {A55},
          doi = {10.1051/0004-6361/202243585},
archivePrefix = {arXiv},
       eprint = {2206.15347},
 primaryClass = {astro-ph.GA},
       adsurl = {https://ui.adsabs.harvard.edu/abs/2022A&A...665A..55S}
}

@ARTICLE{speranza24,
       author = {{Speranza}, G. and {Ramos Almeida}, C. and {Acosta-Pulido}, J.~A. and {Audibert}, A. and {Holden}, L.~R. and {Tadhunter}, C.~N. and {Lapi}, A. and {Gonz{\'a}lez-Mart{\'\i}n}, O. and {Brusa}, M. and {L{\'o}pez}, I.~E. and {Musiimenta}, B. and {Shankar}, F.},
        title = "{Multiphase characterization of AGN winds in five local type-2 quasars}",
      journal = {\aap},
         year = 2024,
        month = jan,
       volume = {681},
          eid = {A63},
        pages = {A63},
          doi = {10.1051/0004-6361/202347715},
archivePrefix = {arXiv},
       eprint = {2311.10132},
 primaryClass = {astro-ph.GA},
       adsurl = {https://ui.adsabs.harvard.edu/abs/2024A&A...681A..63S}
}

@ARTICLE{harrison24,
       author = {{Harrison}, Chris M. and {Ramos Almeida}, Cristina},
        title = "{Observational Tests of Active Galactic Nuclei Feedback: An Overview of Approaches and Interpretation}",
      journal = {Galaxies},
         year = 2024,
        month = apr,
       volume = {12},
       number = {2},
          eid = {17},
        pages = {17},
          doi = {10.3390/galaxies12020017},
archivePrefix = {arXiv},
       eprint = {2404.08050},
 primaryClass = {astro-ph.GA},
       adsurl = {https://ui.adsabs.harvard.edu/abs/2024Galax..12...17H}
}

@ARTICLE{bianchin26,
       author = {{Bianchin}, M. and {Ramos Almeida}, C. and {Gonz{\'a}lez-Mart{\'\i}n}, O. and {Zanchettin}, M.~V. and {Carneiro}, M. and {Pereira-Santaella}, M. and {Tadhunter}, C. and {Speranza}, G. and {Garc{\'\i}a-Bernete}, I. and {Audibert}, A. and {Alonso-Herrero}, A. and {Rigopoulou}, D. and {Labiano}, A. and {Acosta-Pulido}, J.~A. and {Garc{\'\i}a-Burillo}, S.},
        title = "{Extended coronal line emission and new clues to a possible dual AGN in the merger J1356+1026}",
      journal = {\aap},
         year = 2026,
        month = may,
       volume = {709},
          eid = {L9},
        pages = {L9},
          doi = {10.1051/0004-6361/202659271},
archivePrefix = {arXiv},
       eprint = {2604.08239},
 primaryClass = {astro-ph.GA},
       adsurl = {https://ui.adsabs.harvard.edu/abs/2026A&A...709L...9B}
}

@ARTICLE{alexander25,
       author = {{Alexander}, D.~M. and {Hickox}, R.~C. and {Aird}, J. and {Combes}, F. and {Costa}, T. and {Habouzit}, M. and {Harrison}, C.~M. and {Leng}, R.~I. and {Morabito}, L.~K. and {Uckelman}, S.~L. and {Vickers}, P.},
        title = "{What drives the growth of black holes: A decade of progress}",
      journal = {\nar},
         year = 2025,
        month = dec,
       volume = {101},
          eid = {101733},
        pages = {101733},
          doi = {10.1016/j.newar.2025.101733},
archivePrefix = {arXiv},
       eprint = {2506.19166},
 primaryClass = {astro-ph.GA},
       adsurl = {https://ui.adsabs.harvard.edu/abs/2025NewAR.10101733A}
}

@ARTICLE{bertola25,
       author = {{Bertola}, E. and {Cresci}, G. and {Venturi}, G. and {Perna}, M. and {Circosta}, C. and {Tozzi}, G. and {Lamperti}, I. and {Vignali}, C. and {Arribas}, S. and {Bunker}, A.~J. and {Charlot}, S. and {Carniani}, S. and {Maiolino}, R. and {Rodr{\'\i}guez Del Pino}, B. and {{\"U}bler}, H. and {Willott}, C.~J. and {B{\"o}ker}, T. and {Marshall}, M.~A. and {Parlanti}, E. and {Scholtz}, J.},
        title = "{GA-NIFS: Mapping z ≃ 3.5 AGN-driven ionized outflows in the COSMOS field}",
      journal = {\aap},
         year = 2025,
        month = jul,
       volume = {699},
          eid = {A220},
        pages = {A220},
          doi = {10.1051/0004-6361/202554281},
archivePrefix = {arXiv},
       eprint = {2505.08867},
 primaryClass = {astro-ph.GA},
       adsurl = {https://ui.adsabs.harvard.edu/abs/2025A&A...699A.220B}
}

@ARTICLE{zubovas25,
       author = {{Zubovas}, K. and {Tart{\.{e}}nas}, M.},
        title = "{Active galactic nucleus outflows accelerate when they escape the bulge}",
      journal = {\aap},
         year = 2025,
        month = nov,
       volume = {703},
          eid = {A230},
        pages = {A230},
          doi = {10.1051/0004-6361/202555667},
archivePrefix = {arXiv},
       eprint = {2510.14667},
 primaryClass = {astro-ph.GA},
       adsurl = {https://ui.adsabs.harvard.edu/abs/2025A&A...703A.230Z}
}

@ARTICLE{villar-martin24,
       author = {{Villar Mart{\'\i}n}, M. and {L{\'o}pez Cob{\'a}}, C. and {Cazzoli}, S. and {P{\'e}rez Montero}, E. and {Cabrera Lavers}, A.},
        title = "{AGN feedback can produce metal enrichment on galaxy scales}",
      journal = {\aap},
         year = 2024,
        month = oct,
       volume = {690},
          eid = {A397},
        pages = {A397},
          doi = {10.1051/0004-6361/202449621},
archivePrefix = {arXiv},
       eprint = {2407.02115},
 primaryClass = {astro-ph.GA},
       adsurl = {https://ui.adsabs.harvard.edu/abs/2024A&A...690A.397V}
}

@ARTICLE{tanner22,
       author = {{Tanner}, Ryan and {Weaver}, Kimberly A.},
        title = "{Simulations of AGN-driven Galactic Outflow Morphology and Content}",
      journal = {\aj},
         year = 2022,
        month = mar,
       volume = {163},
       number = {3},
          eid = {134},
        pages = {134},
          doi = {10.3847/1538-3881/ac4d23},
archivePrefix = {arXiv},
       eprint = {2201.08360},
 primaryClass = {astro-ph.GA},
       adsurl = {https://ui.adsabs.harvard.edu/abs/2022AJ....163..134T}
}

\begin{appendix}

\section{Supplementary material}
\subsection{Comments on flux and kinematic maps}
\label{app:maps}
\begin{figure}[h!]
\centering
\includegraphics[width=0.97\textwidth]{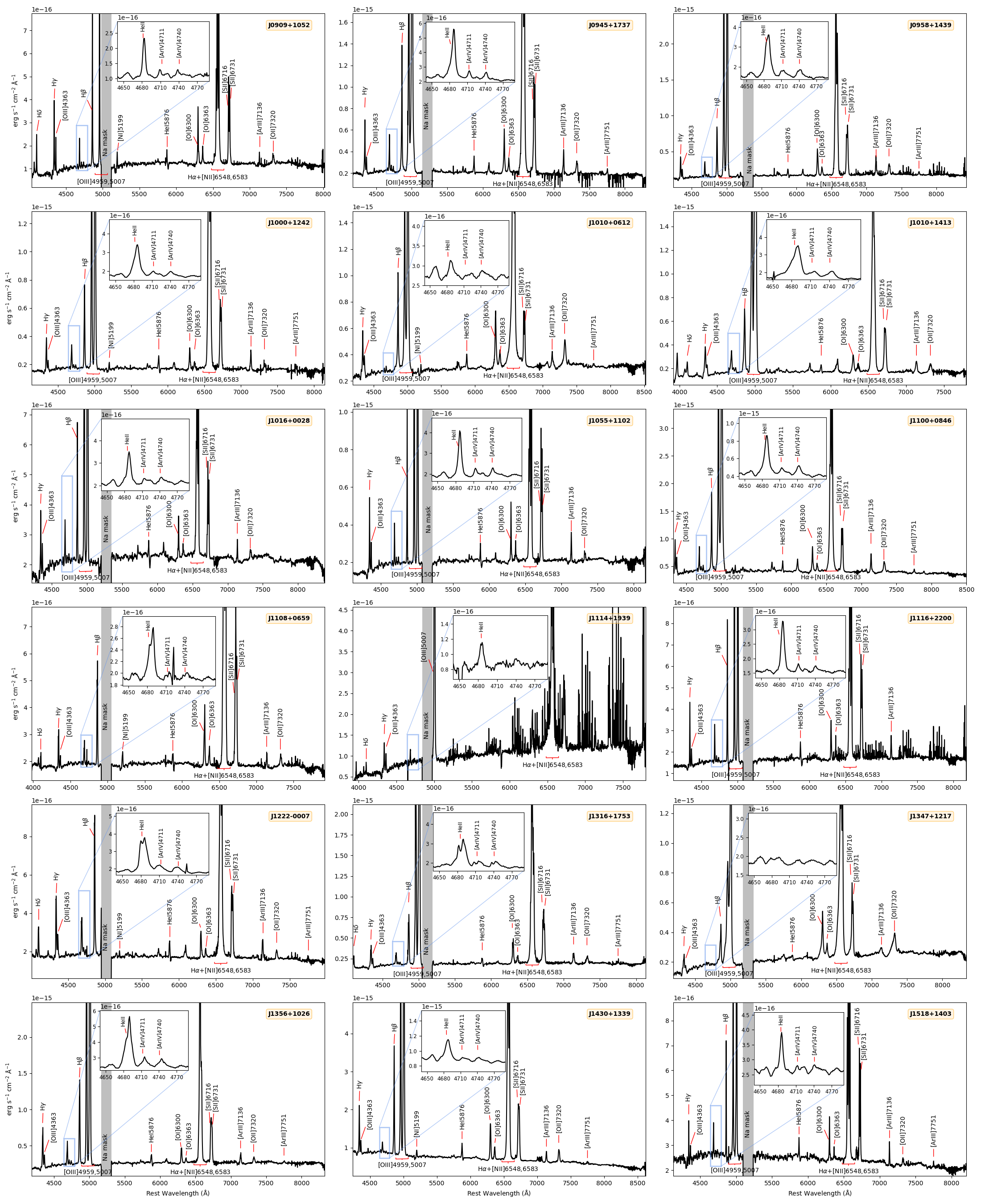}
    \caption{Extracted spectra at $r=1.5\arcsec$ centred on the nucleus of each galaxy. The most prominent emission lines are labelled, and the inset zoom shows the spectral region over which the [Ar~{\sc iv}] doublet is detected. The grey vertical area corresponds to the region masked out due to the sodium emission of the AO system.}
    \label{fig:spec_sample}
\end{figure}

\begin{figure*}[ht!]
    \centering
    \includegraphics[width=0.91\textwidth]{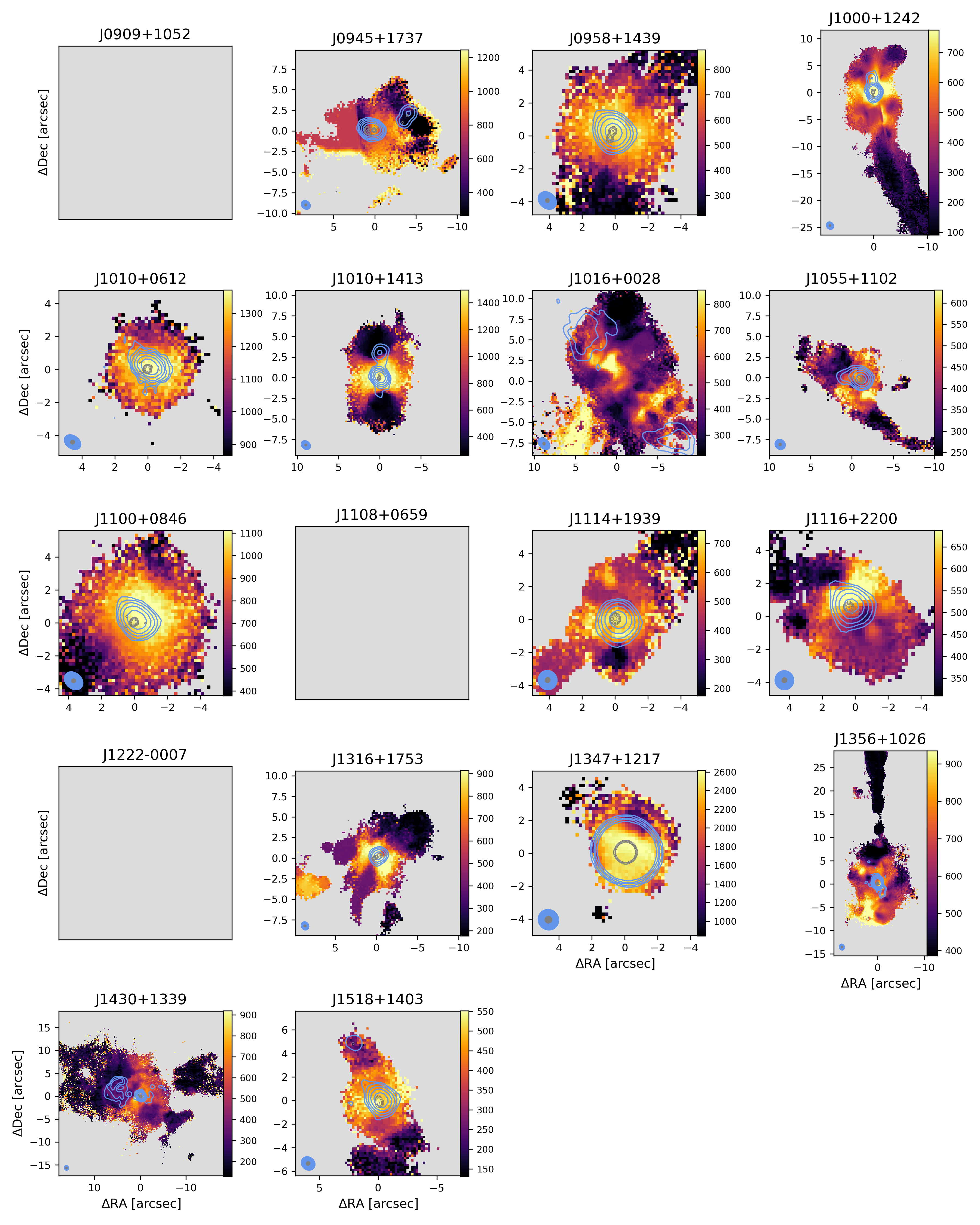}
    \caption{Same as Fig. \ref{fig:flux+radio} but for the $W_{80}$ plots. The blank panels are the galaxies where the \oiiir~is not observed due to the sodium mask used as part of the optical AO. The colorbars are in units of \kms.}
    \label{fig:w80+radio}
\end{figure*}

\begin{figure}
    \centering
    \includegraphics[width=0.95\textwidth]{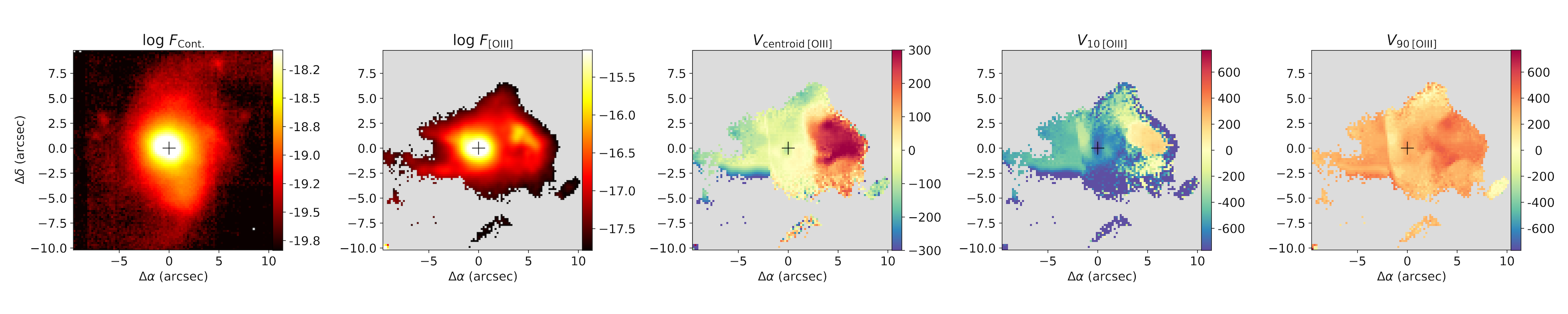}
    \caption{Continuum, integrated flux of the \oiii~emission line, $V_{\rm cen}$, $V_{10}$, and $V_{90}$ for the galaxy J0945+1737. The colour bars are in units of erg~s$^{-1}$~cm$^{-2}$ for the first two panels and in \kms~for the other three.}
    \label{fig:maps_J0945}
\end{figure}

\begin{figure}
    \centering
    \includegraphics[width=0.95\textwidth]{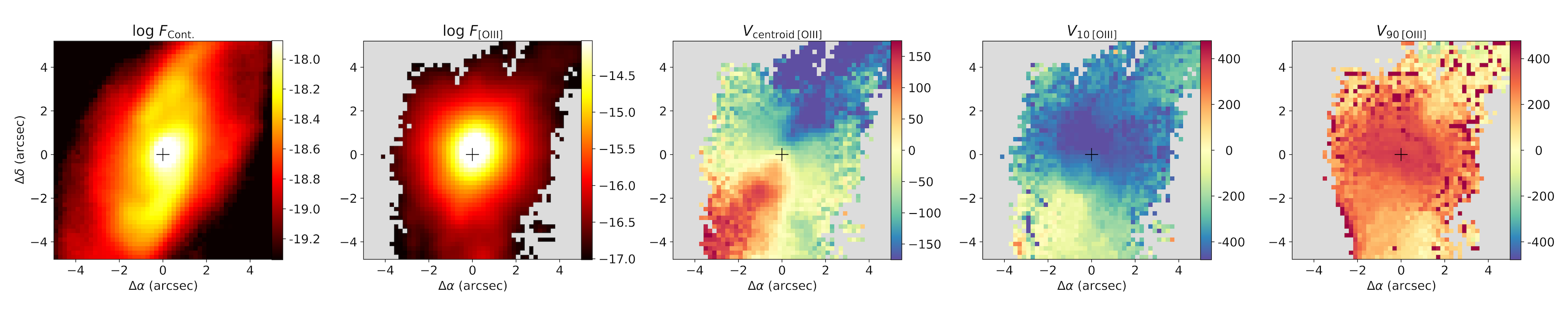}
    \caption{Same as Fig.\ref{fig:maps_J0945} for the galaxy J0958+1439.}
    \label{fig:maps_J0958}
\end{figure}

\begin{figure}
    \centering
    \includegraphics[width=0.95\textwidth]{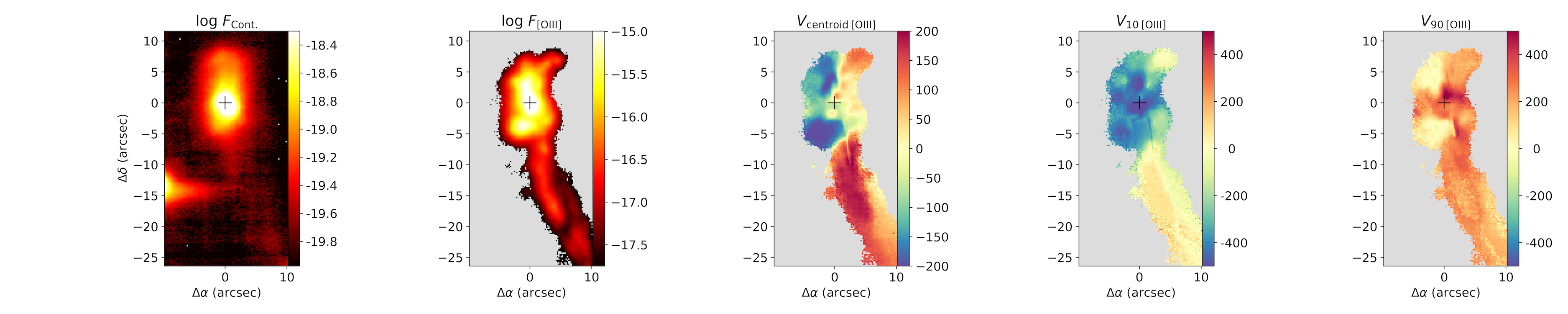}
    \caption{Same as Fig.\ref{fig:maps_J0945} for the galaxy J1000+1242.}
    \label{fig:maps_J1000}
\end{figure}

\begin{figure}
    \centering
    \includegraphics[width=0.95\textwidth]{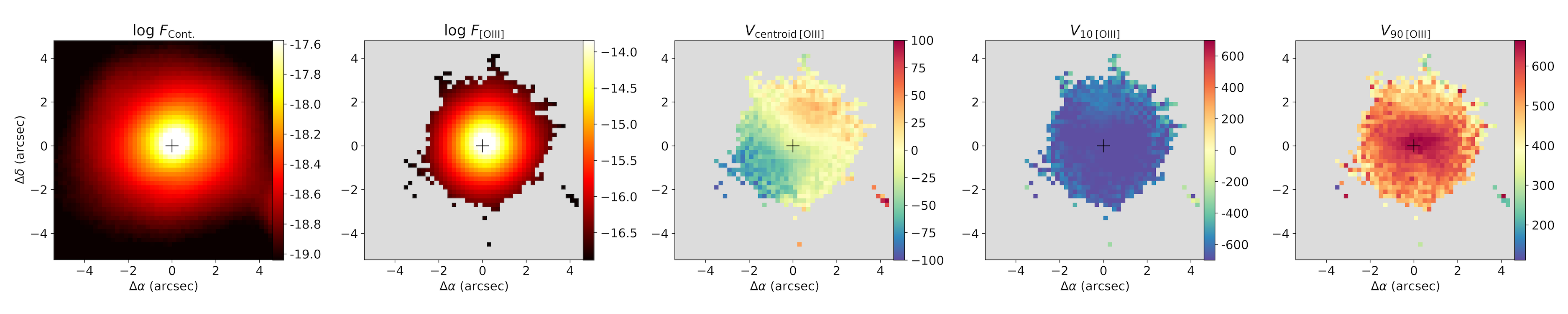}
    \caption{Same as Fig.\ref{fig:maps_J0945} for the galaxy J1010+0612.}
    \label{fig:maps_J1010+06}
\end{figure}

\begin{figure}
    \centering
    \includegraphics[width=0.95\textwidth]{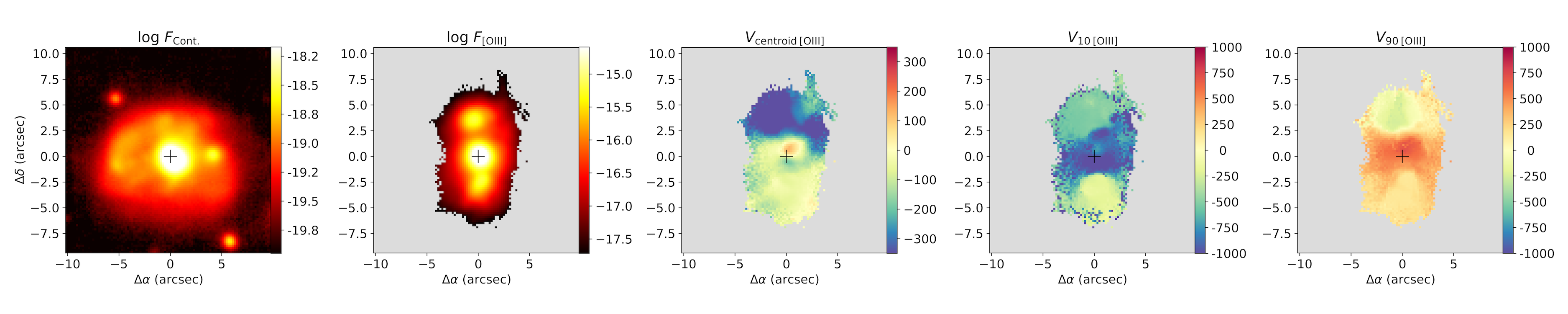}
    \caption{Same as Fig.\ref{fig:maps_J0945} for the galaxy J1010+1413.}
    \label{fig:maps_J1010+14}
\end{figure}

\begin{figure}
    \centering
    \includegraphics[width=0.95\textwidth]{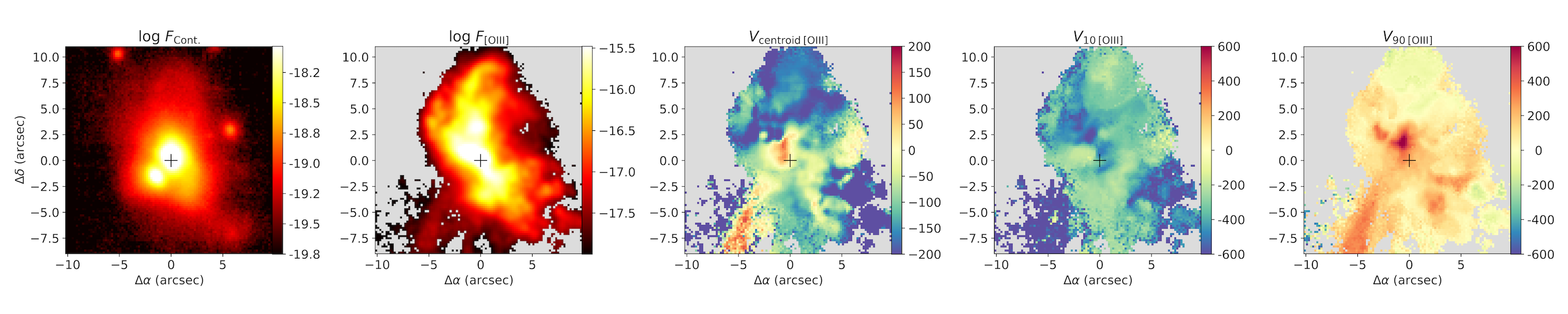}
    \caption{Same as Fig.\ref{fig:maps_J0945} for the galaxy J1016+0028.}
    \label{fig:maps_J1016+0028}
\end{figure}

\begin{figure}
    \centering
    \includegraphics[width=0.95\textwidth]{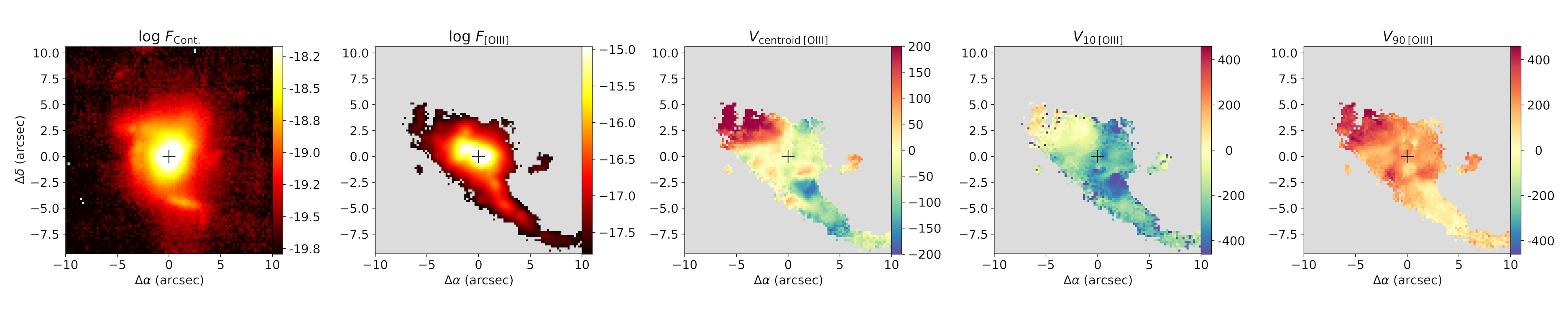}
    \caption{Same as Fig.\ref{fig:maps_J0945} for the galaxy J1055+1102.}
    \label{fig:maps_J1055}
\end{figure}

\begin{figure}
    \centering
    \includegraphics[width=0.95\textwidth]{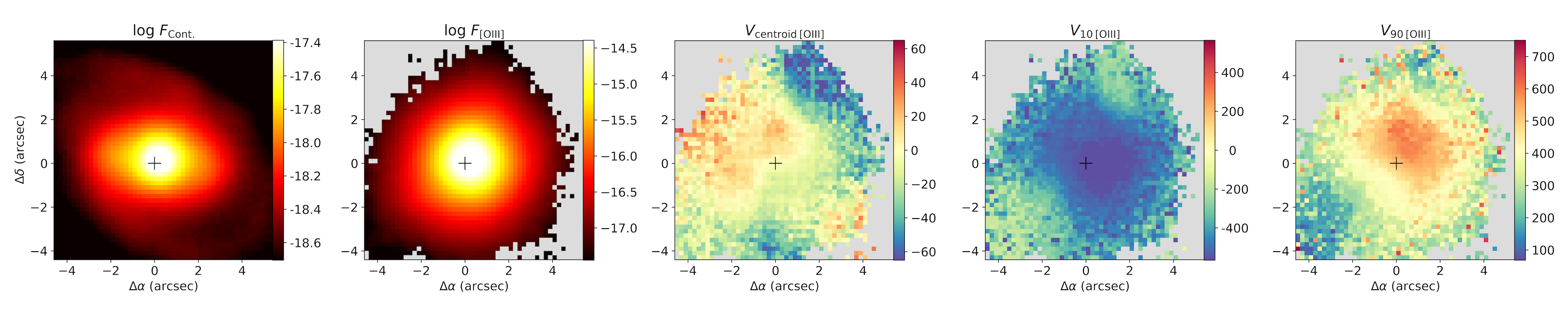}
    \caption{Same as Fig.\ref{fig:maps_J0945} for the galaxy J1100+0846.}
    \label{fig:maps_J1100}
\end{figure}

\begin{figure}
    \centering
    \includegraphics[width=0.95\textwidth]{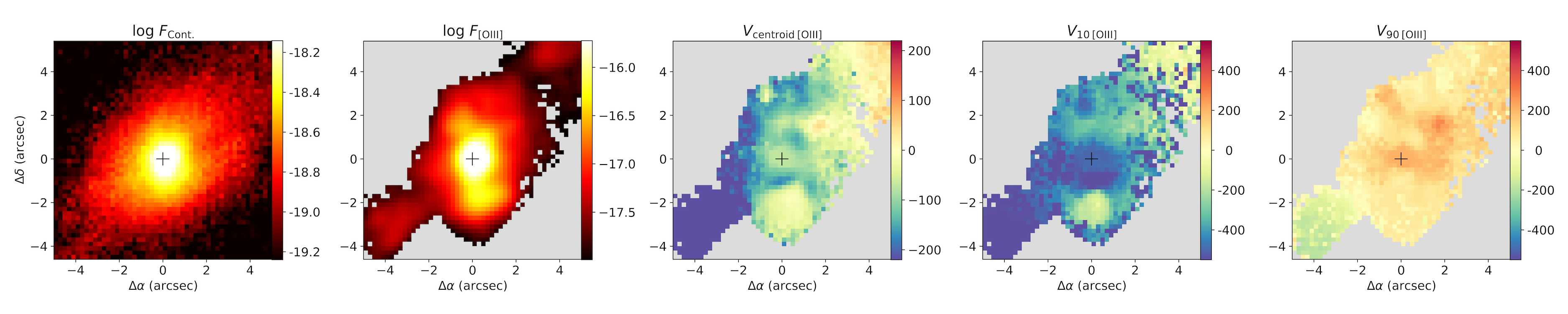}
    \caption{Same as Fig.\ref{fig:maps_J0945} for the galaxy J1114+1939.}
    \label{fig:maps_J1114}
\end{figure}

\begin{figure}
    \centering
    \includegraphics[width=0.95\textwidth]{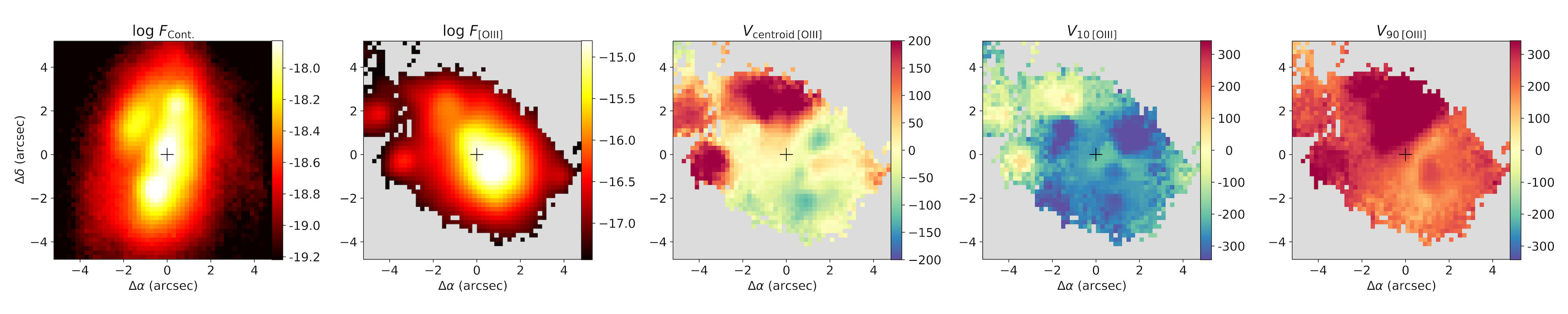}
    \caption{Same as Fig.\ref{fig:maps_J0945} for the galaxy J1116+2200.}
    \label{fig:maps_J1116}
\end{figure}

\begin{figure}
    \centering
    \includegraphics[width=0.95\textwidth]{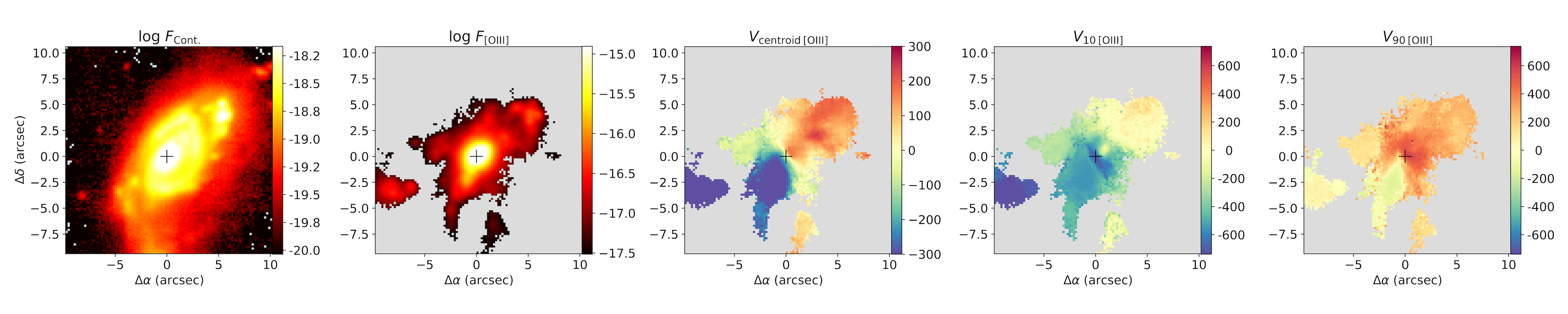}
    \caption{Same as Fig.\ref{fig:maps_J0945} for the galaxy J1316+1753.}
    \label{fig:maps_J1316}
\end{figure}

\begin{figure}
    \centering
    \includegraphics[width=0.95\textwidth]{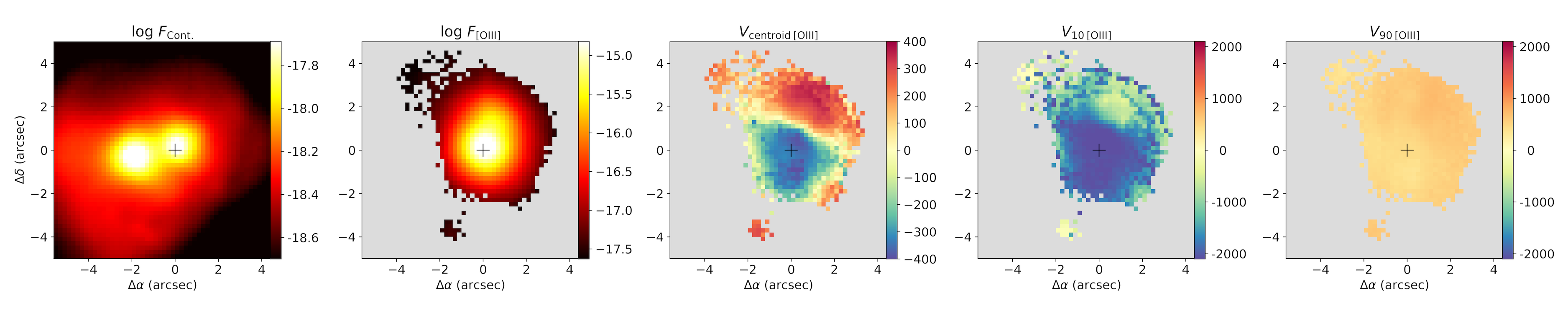}
    \caption{Same as Fig.\ref{fig:maps_J0945} for the galaxy J1347+1217.}
    \label{fig:maps_J1347}
\end{figure}

\begin{figure}
    \centering
    \includegraphics[width=0.95\textwidth]{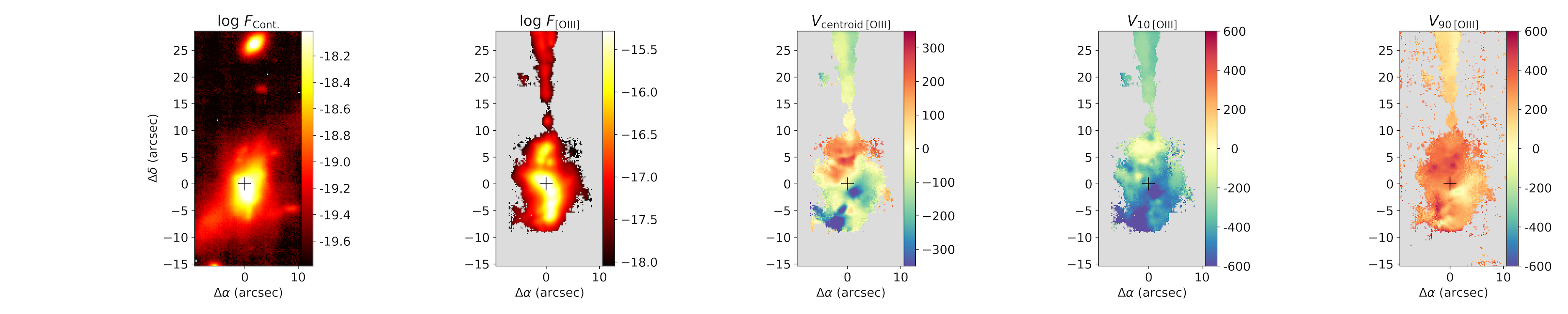}
    \caption{Same as Fig.\ref{fig:maps_J0945} for the galaxy J1356+1026.}
    \label{fig:maps_J1356}
\end{figure}

\begin{figure}
    \centering
    \includegraphics[width=0.95\textwidth]{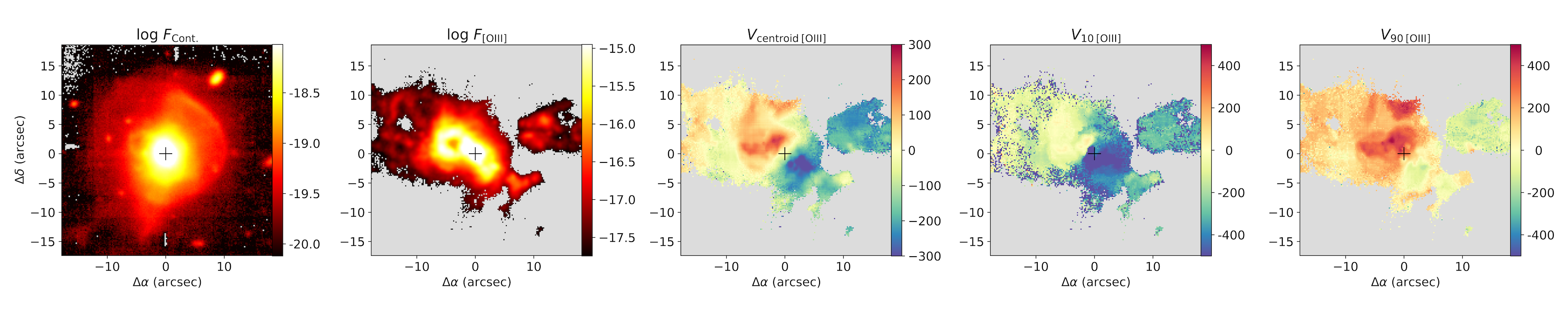}
    \caption{Same as Fig.\ref{fig:maps_J0945} for the galaxy J1430+1339.}
    \label{fig:maps_J1430}
\end{figure}

\begin{figure}
    \centering
    \includegraphics[width=0.95\textwidth]{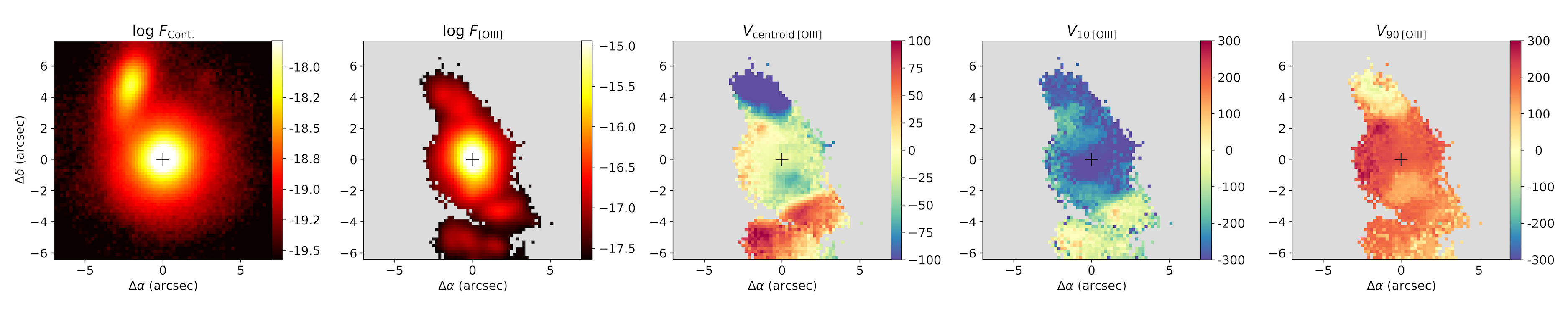}
    \caption{Same as Fig.\ref{fig:maps_J0945} for the galaxy J1518+1403.}
    \label{fig:maps_J1518}
\end{figure}

Figure \ref{fig:spec_sample} shows the extracted spectra ($r=1.5\arcsec$) for the 18 galaxies in the sample. The most prominent emission lines are labelled, and the insets show the region where the [Ar{\sc iv}] doublet, used for the electron density measurements (see Section\ref{sec:density}), is observable.  
Figures \ref{fig:maps_J0945}-\ref{fig:maps_J1518} present  $V_{\rm cen}$, $V_{10}$, and $V_{90}$ maps in addition to the \oiiir\ flux and local continuum maps for all 15 galaxies with the \oiii~line available.  Figure \ref{fig:w80+radio} shows the $W_{80}$ maps for the same galaxies. Table \ref{tab:out_103} provides mass outflow rate and kinetic power for the Global and Peak methods assuming a density of $10^3$\:\cmm~for these same galaxies.

The local continuum is the average of the fitted linear polynomial within a window of 2000\,\kms\ in a region centred at 4980\:\AA\ in the rest frame.  The local continuum maps can be linked to the distribution of stars in the galaxies, as they are the main source of continuum emission at optical wavelengths. The flux maps are simply the combination of the fluxes of the \oiiir\ Gaussian components. The flux maps presented here are extinction corrected using the values from Table \ref{tab:dens}. The grey regions indicate where the signal-to-noise ratio in any of the three components is lower than three.

The continuum and \oiii~ distributions have distinct morphologies: the former traces a more extended structure than the latter, even with the presence of spiral arms (see Fig.\,\ref{fig:maps_J1316}, for example). The orientations of the continuum and ionised gas distributions are also different, indicating that the stellar and gas have different dynamical timescales in systems heavily influenced by AGN and merger phenomena.

We note that the \oiii\ $V_{\rm cen}$ appeared shifted, with respect to the systemic velocities inferred from SDSS redshifts, in our initial maps. The discrepancy is in $\sim 100$\,\kms~ in all galaxies. A similar behaviour was also reported by \citet{jarvis19} when analysing the GMOS data for a subsample of 9 targets. To correct for this effect, we calculated the mean $V_{\rm peak}$ in a circular aperture with the diameter of the data angular resolution (last column of Table\,\ref{tab:obs}) centred at the nucleus and subtracted it from the $V_{\rm cen}$, $V_{10}$ and $V_{90}$ velocity fields presented here.  This correction does not affect our quantitative results, as the outflow velocity is purely based on the $W_{80}$ parameter and is applied to help guide the discussion and interpretation of the kinematic maps.

The velocity fields and $W_{80}$ maps are complex, especially in the galaxies with clear signatures of ongoing or previous interactions. The most notable cases are: (i) J1000+1242 with double nuclei observed in the high-resolution HST imaging \citep{jarvis19}, a pronounced tidal tail as seen on MUSE (Fig.~\ref{fig:maps_J1000}) and also reported by \citet{ulivi24}, and molecular filaments with complex velocity structures \citep{girdhar24}; (ii) J1016+0028 (Fig.~\ref{fig:maps_J1016+0028}) shows a double-peak on the continuum image, identified as dual nuclei in {\em r-band} imaging, putting this galaxy as a pre-collapse merger \citep[and references therein]{pierce23}. It also shows prominent extended features on the \oiii\ flux map resembling tidal tails; (iii) J1356+1026 (Fig.~\ref{fig:maps_J1356} has two nuclei and a companion located $\sim50$~kpc away \citep[and references therein]{greene12}. A tidal tail connecting the companion to the main galaxy is clearly visible in the \oiii\ flux map \citep[see also][]{ulivi24}, and the outflowing bubble identified by \citet{greene12} coincides with the SE region where we identify the highest $W_{80}$ values. This feature can also be partially seen in \citet{speranza24} but is limited due to the smaller FOV of their observations with GTC-MEGARA. This galaxy also shows one of the most extended coronal line-emitting regions of 13-15.5~kpc, which is excited by the AGN radiation field \citep{bianchin26}.

A remarkable mention is J1430+1339, the {\em Teacup galaxy}. This source is the most well-studied galaxy in our sample in a multitude of wavelengths, from X-ray to optical, infrared, millimetre, and radio \citep[e.g.][]{keel12b, gagne14, harrison15, ramos-almeida17, lansbury18, girdhar24, villar-martin24, audibert23, audibert25}. In our data, we observe the shell structure on the continuum, as previously reported by \citet{venturi23}, which is linked to previous merger events. The classical Teacup handle is clearly observed in the \oiii\ flux map, and it aligns with the extended radio emission \citep[and Fig. \ref{fig:flux+radio}]{harrison15}. Besides the ``handle'', we also observe another extended bubble to the east of the nucleus, each of them associated with a previous AGN outburst \citep{venturi23}. The highest $W_{80}$ is close to the nucleus, and previous works observed ionised gas AGN-driven outflows \citep{ramos-almeida17, harrison15, venturi23}, although the $V_{10}$ is not as extreme as in other sources, i.e. J1347, and $V_{\rm cen}$ resembles a rotation pattern due to its symmetrical features and not extreme velocity amplitudes ($\pm 300$~\kms).

The other well-studied galaxy in our sample is J1347+1217, also known as 4C12.50 and PKS1345+12 (see Figs.\ref{fig:w80+radio} and \ref{fig:maps_J1347}). It is the source with the most extreme kinematics, with $W_{80}$ reaching $\sim 2600$\,\kms\ \citep{holt03,bessiere24} and $V_{10}$ up to $-2000$\,\kms. Two nuclei are observed in the continuum image, and previous works associated them with an ongoing merger \citep{emonts16,tadhunter18}. The radio emission, linked to a radio AGN, appears only in the western nucleus \citep[see the radio contours in Figs.~\ref{fig:flux+radio} and \ref{fig:w80+radio}]{evans99,jarvis21}. This is also the case for \oiii\ emission (Fig.~\ref{fig:maps_J1347}) and the cold molecular gas \citep{holden24, audibert25}. There are no signatures of hot ($\geq2000$~K) molecular gas outflows in this galaxy, which is being excited by a shock induced by the radio jet \citep{villar-martin23}. Although the $V_{\rm cen}$ field could resemble rotation, it has amplitudes of $\approx400$\,\kms\ and the zero velocity line does not coincide with the peak of the \oiii\ or continuum emissions. In combination with the extreme line widths, this leads us to interpret that the \oiii\ gas emission is outflow-dominated in agreement with previous spectroscopic work \citep{holt03}. However, \citet{villar-martin23} concluded that despite the extreme outflows, there is no evidence for star formation suppression in this galaxy. 

The peak of radio emission is co-spatial with the peak of the \oiii\ flux distribution in 15 of the 18 galaxies in our sample. This behaviour has already been reported by \citet{jarvis19} for the subsample of 9 sources and in \citet{jarvis21} for the complete QFeedS-1 sample. The three exceptions to this trend are J1016+0028, J1116+2200 and J1055+1102. In J1016+0028, only the radio lobes are observed. \citet{jarvis21} postulated the possibility that the radio emission is linked to a hidden jet. This scenario is confirmed by deeper radio imaging revealing the previously hidden jet ({\color{blue}Silpa et al. in prep}), which connects the lobes and trails the extended ionised gas emission (see Figs.\:\ref{fig:flux+radio} and \ref{fig:maps_J1016+0028}). However, for J1116+2200 and J1055+1102, the peak of the radio emission is displaced from either the peak of the \oiii\ and continuum emission. For the former, the small radio knot seen in the high-resolution imaging (grey contours in Fig.\ref{fig:flux+radio}) extends along the larger \oiii~cloud \citet{jarvis21}. However, the AGN origin of the radio emission remains unconfirmed \citep{njeri25}.  In J1055+1102, the radio emission is extended along the direction of the \oiii~emission (Figs.\ref{fig:flux+radio} and \ref{fig:maps_J1055}). Deeper radio imaging from {\color{blue} Silpa et al. (in prep.)} confirms this trend, and \citet{njeri25} confirmed that there is a radio AGN in this galaxy. 
Notably,  J1116+2200 and J1055+1102 have the lowest outflow velocities in our sample (Tab. \ref{tab:out_props}).

We also note that the jet-like structures observed in radio, especially in the high-resolution images, are perpendicular to the direction with the highest $W_{80}$ values in J0958+1439, J1000+1242, J1010+1413, J1316+1753, J1430+1339, and J1518+1403. Similar structures have been identified in the nearby Seyfert galaxies \citep[e.g.][]{couto13, rogemar_n5929letter, lena15, finlez18, venturi21, bianchin22} and interpreted as equatorial outflows. \citet{rogemar_abundances} found strong evidence for high electron temperatures perpendicular to the ionisation axis of the galaxy that AGN photoionisation models cannot reproduce. The high-temperature regions are spatially coincident with regions with high emission line widths, favouring the interpretation of wide-angle shocks in the ISM.  \citet{venturi21} also analysed local Seyfert galaxies and observed the same features. In these cases, the velocity dispersion is enhanced perpendicular to the ionisation and the low-power jet axis. The authors also interpret them as shocks and a signature of jet-induced turbulence in the ISM. Recent works focused on the optical data of some of the galaxies in our sample also observed the same behaviour and attribute the enhancement in velocity to the jet-ISM interaction \citep{girdhar22,venturi23,ulivi24,speranza24}.

\onecolumn

\begin{table}
    \centering
     \caption{Mass outflow rate and kinetic power assuming $n_e=10^3$~\cmm.}

    \begin{tabular}{lcccccc}
    \hline\hline
              &  \multicolumn{2}{c}{Global}  &  \multicolumn{2}{c}{Peak}\\
          \cmidrule(l){2-3} \cmidrule(l){4-5}
    Galaxy & $\log{\dot{M}^{\rm g}_{\rm out}}$& $\log{\dot{E}^{\rm g}_{\rm kin}}$ & $\log{\dot{M}^{\rm p}_{\rm out}}$& $\log{\dot{E}^{\rm p}_{\rm kin}}$ \\
    \hline
J0945+1737 &  -0.28 $\pm$ 0.61 & 41.19 $\pm$ 0.67 & -0.22 $\pm$ 0.25 & 41.13 $\pm$ 0.25\\
J0958+1439 & 0.93 $\pm$ 0.01 & 42.25 $\pm$ 0.01 & 0.53 $\pm$ 0.01 & 41.84 $\pm$ 0.01\\
J1000+1242  & 0.29 $\pm$ 0.03 & 41.45 $\pm$ 0.03 & 0.28 $\pm$ 0.04 & 41.62 $\pm$ 0.04\\
J1010+0612  & 0.95 $\pm$ 0.09 & 42.71 $\pm$ 0.09 & 0.56 $\pm$ 0.04 & 42.30 $\pm$ 0.04\\
J1010+1413  & 1.06 $\pm$ 0.46 & 42.89 $\pm$ 0.48 & 0.76 $\pm$ 0.34 & 42.60 $\pm$ 0.34\\
J1016+0028  & -0.48 $\pm$ 0.01 & 40.55 $\pm$ 0.01 & -0.65 $\pm$ 0.02 & 40.45 $\pm$ 0.02\\
J1055+1102 & -0.14 $\pm$ 0.01 & 40.66 $\pm$ 0.01 & -0.36 $\pm$ 0.01 & 40.45 $\pm$ 0.01\\
J1100+0846  & 0.53 $\pm$ 0.15 & 42.08 $\pm$ 0.16 & 0.35 $\pm$ 0.13 & 41.90 $\pm$ 0.13\\
J1114+1939  & -0.64 $\pm$ 0.18 & 40.45 $\pm$ 0.21 & -1.01 $\pm$ 0.24 & 40.06 $\pm$ 0.24\\
J1116+2200 &  -0.18 $\pm$ 0.01 & 40.69 $\pm$ 0.01 & -0.25 $\pm$ 0.01 & 40.70 $\pm$ 0.01\\
J1316+1753 &  0.5 $\pm$ 0.06 & 41.94 $\pm$ 0.08 & 0.14 $\pm$ 0.04 & 41.51 $\pm$ 0.04\\
J1347+1217 &  0.37 $\pm$ 1.91 & 42.65 $\pm$ 2.01 & 0.1 $\pm$ 1.1 & 42.40 $\pm$ 1.10\\
J1356+1026 &  -0.17 $\pm$ 0.08 & 41.07 $\pm$ 0.09 & -0.19 $\pm$ 0.09 & 41.11 $\pm$ 0.09\\
J1430+1339 &  0.8 $\pm$ 0.05 & 42.08 $\pm$ 0.07 & 0.47 $\pm$ 0.06 & 41.81 $\pm$ 0.06\\
J1518+1403 &  -0.22 $\pm$ 0.03 & 40.67 $\pm$ 0.03 & -0.52 $\pm$ 0.02 & 40.37 $\pm$ 0.02\\
\hline
    \end{tabular}
    \tablefoot{$\dot{M}_{\rm out}$ are in $\rm M_{\odot} yr^{-1}$, and $\dot{E_{\rm kin}}$ are in erg\:s$^{-1}$.}
    \label{tab:out_103}
\end{table}

\FloatBarrier 
\twocolumn

\end{appendix}
\end{document}